\RequirePackage{fix-cm}
\documentclass[11pt]{article}
\newcounter{ourcount}
\usepackage{amsmath,amsfonts,amssymb,graphicx,epsfig,pdflscape,multirow,theorem,gensymb}
\usepackage{lscape,arydshln,float}
\usepackage[matrix,arrow,curve]{xy} 
\numberwithin{equation}{section}
\graphicspath{{./eps/}}
\usepackage[nosort]{cite}
\usepackage[usenames]{color}
\usepackage{pstricks}
\usepackage{pst-plot}
\usepackage[backref=false]{hyperref}
\usepackage[capitalise,noabbrev]{cleveref} 
\usepackage{float}
\usepackage[11pt]{moresize}
\hypersetup{
colorlinks=true,
citecolor=red,
linkcolor=darkblue,
urlcolor=darkblue
}
\definecolor{darkblue}{rgb}{0,0,.8}
\definecolor{red}{rgb}{1,0,0}

\theorembodyfont{\itshape} 
\theoremheaderfont{\scshape}
\theoremstyle{definition}
\newtheorem{Proposition}{Proposition}[section]

\newtheorem{Conjecture}{Conjecture}
\newtheorem{Lemma}[Proposition]{Lemma}
\newtheorem{Corollary}[Proposition]{Corollary}

\numberwithin{equation}{section}

\crefname{Conjecture}{Conjecture}{Conjectures}

\newcommand{\nc}{\newcommand}

\def\arxiv#1#2{\href{http://arxiv.org/abs/#1}{\textsf{arXiv:#1 #2}}}
\nc{\ir}{\mathrm{i}}
\nc{\dd}{\mathrm{d}}   
\nc{\eE}{\mathsf{e}}
\nc{\rhoj}{\rho(j)}
\nc{\bib}{\bibitem}
\nc{\be}{\begin{equation}}
\nc{\ee}{\end{equation}}
\nc{\chit}{\raisebox{0.25ex}{$\chi$}}
\nc{\dtl}{\mathsf{dTL}}
\nc{\pdtl}{\mathsf{pdTL}}
\nc{\Dbh}{\mbox{\boldmath $\widehat D$}}
\nc{\Dh}{\mbox{$\hat D$}}
\nc{\Dbb}{\mbox{\boldmath $\bar D$}}
\nc{\Dbm}{\mbox{\boldmath $\mathcal D$}}
\nc{\Dbt}{\mbox{\boldmath $\tilde{D}$}}
\nc{\Tbt}{\mbox{\boldmath $\tilde{T}$}}
\nc{\Tbh}{\mbox{\boldmath $\widehat{T}$}}

\nc{\setS}{\mathcal S}

\nc{\db}{\mbox{\boldmath $d$}}
\nc{\Ab}{\mbox{\boldmath $A$}}
\nc{\Bb}{\mbox{\boldmath $B$}}
\nc{\Cb}{\mbox{\boldmath $C$}}
\nc{\Db}{\mbox{\boldmath $D$}}
\nc{\eb}{\mbox{\boldmath $e$}}
\nc{\Fb}{\mbox{\boldmath $F$}}
\nc{\Fbt}{\mbox{\boldmath $\tilde{F}$}}
\nc{\fb}{\mbox{\boldmath $f$}}
\nc{\fbt}{\mbox{\boldmath $\tilde{f}$}}
\nc{\Gb}{\mbox{\boldmath $G$}}
\nc{\Hb}{\mbox{\boldmath $H$}}
\nc{\Ib}{\mbox{\boldmath $I$}}
\nc{\Jb}{\mbox{\boldmath $J$}}
\nc{\Kb}{\mbox{\boldmath $K$}}
\nc{\Lb}{\mbox{\boldmath $L$}}
\nc{\Mb}{\mbox{\boldmath $M$}}
\nc{\Pb}{\mbox{\boldmath $P$}}
\nc{\Qb}{\mbox{\boldmath $Q$}}
\nc{\Rb}{\mbox{\boldmath $R$}}
\nc{\Tb}{\mbox{\boldmath $T$}}
\nc{\Tbb}{\mbox{\boldmath $\bar T$}}
\nc{\Tbm}{\mbox{\boldmath $\mathcal T$}}
\nc{\tb}{\mbox{\boldmath $t$}}
\nc{\Ub}{\mbox{\boldmath $U$}}
\nc{\Vb}{\mbox{\boldmath $V$}}
\nc{\Wb}{\mbox{\boldmath $W$}}
\nc{\xb}{\mbox{\boldmath $x$}}
\nc{\yb}{\mbox{\boldmath $y$}}
\nc{\Zb}{\mbox{\boldmath $Z$}}
\nc{\Lambdab}{\boldsymbol{\Lambda}}

\nc{\even}{\textrm{ even}}
\nc{\odd}{\textrm{ odd}}

\def\qbar{\bar q}
\def\jbar{\overline j}

\nc{\Atwotwo}{\mbox{$A_2^{\textrm{\fontsize{7pt}{7pt}\selectfont $(2)$}}$}}
\nc{\Aoneone}{\mbox{$A_1^{\textrm{\fontsize{7pt}{7pt}\selectfont $(1)$}}$}}
\nc{\amf}{\mbox{$\mathfrak a$}}
\nc{\bmf}{\mbox{$\mathfrak b$}}
\nc{\cmf}{\mbox{$\mathfrak c$}}
\nc{\dmf}{\mbox{$\mathfrak d$}}
\nc{\fmf}{\mbox{$\mathfrak f$}}
\nc{\gmf}{\mbox{$\mathfrak g$}}
\nc{\Amf}{\mbox{$\mathfrak A$}}
\nc{\Bmf}{\mbox{$\mathfrak B$}}
\nc{\Cmf}{\mbox{$\mathfrak C$}}
\nc{\Dmf}{\mbox{$\mathfrak D$}}
\nc{\Fmf}{\mbox{$\mathfrak F$}}
\nc{\asf}{\mbox{$\mathsf a$}}
\nc{\bsf}{\mbox{$\mathsf b$}}
\nc{\csf}{\mbox{$\mathsf c$}}
\nc{\dsf}{\mbox{$\mathsf d$}}
\nc{\fsf}{\mbox{$\mathsf f$}}
\nc{\gsf}{\mbox{$\mathsf g$}}
\nc{\Asf}{\mbox{$\mathsf A$}}
\nc{\Bsf}{\mbox{$\mathsf B$}}
\nc{\Csf}{\mbox{$\mathsf C$}}
\nc{\Dsf}{\mbox{$\mathsf D$}}

\nc{\repV}{\mathsf{V}}
\nc{\repW}{\mathsf{W}}
\newrgbcolor{darkgreen}{0., 0.733333, 0.0621042}
\definecolor{lightblue}{rgb}{.7,.7,1}
\definecolor{lightestblue}{rgb}{.95,.95,1}
\definecolor{lightlightblue}{rgb}{.85,.85,1}
\definecolor{midblue}{rgb}{.7,.7,1}
\definecolor{lightpurple}{rgb}{1,.75,1}
\definecolor{darkpurple}{rgb}{1,.5,1}

\nc{\elegant}{1.5pt}
\nc{\moyen}{1.0pt}
\nc{\mince}{0.5pt}

\def\vvdots{\mathinner{\mkern1mu\raise1pt\vbox{\kern7pt\hbox{.}}\mkern2mu
  \raise4pt\hbox{.}\mkern2mu\raise7pt\hbox{.}\mkern1mu}}

\def\loopa{
\psframe[linewidth=.25pt](0,0)(1,1)
}
\def\loopb{
\psframe[linewidth=.25pt](0,0)(1,1)
\psarc[linewidth=1.5pt,linecolor=blue](0,1){.5}{-90}{0}
}
\def\loopc{
\psframe[linewidth=.25pt](0,0)(1,1)
\psarc[linewidth=1.5pt,linecolor=blue](1,0){.5}{90}{180}
}
\def\loopd{
\psframe[linewidth=.25pt](0,0)(1,1)
\psarc[linewidth=1.5pt,linecolor=blue](0,0){.5}{0}{90}
}
\def\loope{
\psframe[linewidth=.25pt](0,0)(1,1)
\psarc[linewidth=1.5pt,linecolor=blue](1,1){.5}{180}{270}
}
\def\loopf{
\psframe[linewidth=.25pt](0,0)(1,1)
\psline[linewidth=1.5pt,linecolor=blue](0,0.5)(1,0.5)
}
\def\loopg{
\psframe[linewidth=.25pt](0,0)(1,1)
\psline[linewidth=1.5pt,linecolor=blue](0.5,0)(0.5,1)
}
\def\looph{
\psframe[linewidth=.25pt](0,0)(1,1)
\psarc[linewidth=1.5pt,linecolor=blue](1,0){.5}{90}{180}
\psarc[linewidth=1.5pt,linecolor=blue](0,1){.5}{-90}{0}
}
\def\loopi{
\psframe[linewidth=.25pt](0,0)(1,1)
\psarc[linewidth=1.5pt,linecolor=blue](0,0){.5}{0}{90}
\psarc[linewidth=1.5pt,linecolor=blue](1,1){.5}{180}{270}
}

\def\facegrid#1#2{
\psframe[fillstyle=solid,fillcolor=lightlightblue,linewidth=0pt]#1#2
\psgrid[gridlabels=0pt,subgriddiv=1]#1#2}

\renewcommand{\ge}{\geqslant}
\renewcommand{\le}{\leqslant}

\def\GA#1#2{\Gamma_{#2,#1}}

\nc{\proof}{{\scshape Proof.\ }} 				
\nc{\eproof}{{\hfill \rule{0.5em}{0.5em}\medskip}}

\begin{document}

\topmargin -5mm
\oddsidemargin 5mm
\vspace*{-2cm}

\makeatletter 
\newcommand\Larger{\@setfontsize\semiHuge{20.00}{23.78}}
\makeatother 

\setcounter{page}{1}
\mbox{}\vspace{1cm}
\begin{center}

{\Larger \bf \mbox{
Modular covariant torus partition functions\ \ }
\\[0.18cm] 
\mbox{of dense $\boldsymbol{A_1^{(1)}}$ and dilute $\boldsymbol{A_2^{(2)}}$  loop models}
}

\end{center}

\vspace{1cm}
\begin{center}
{\vspace{-5mm}\Large Alexi Morin-Duchesne$^{a,b}$, Andreas Kl\"umper$^c$, Paul A.~Pearce$^{d,e}$}
\\[.5cm]
{\em { }$^a$Department of Applied Mathematics, Computer Science and Statistics \\ Ghent University, 9000 Ghent, Belgium}
\\[.2cm]
{\em $^{b}$Department of Mathematics, Royal Military Academy, 1000 Brussels, Belgium}
\\[.2cm]
  {\em { }$^c$Fakult\"at f\"ur Mathematik und Naturwissenschaften \\ Bergische
 Universit\"at Wuppertal, 42097 Wuppertal, Germany}
\\[.2cm]
{\em { }$^d$School of Mathematics and Statistics, University of Melbourne\\
Parkville, Victoria 3010, Australia}
\\[.2cm]
{\em { }$^e$School of Mathematics and Physics, University of Queensland}\\
{\em St Lucia, Brisbane, Queensland 4072, Australia}
\\[.2cm] 
{\tt alexi.morin.duchesne\,@\,gmail.com}
\qquad
{\tt kluemper\,@\,uni-wuppertal.de}
\qquad
{\tt papearce\,@unimelb.edu.au}
\end{center}

\vspace{12mm}
\centerline{{\bf{Abstract}}}
\vskip.3cm
\noindent 

Yang--Baxter integrable dense $\Aoneone$ and dilute $\Atwotwo$ loop models are considered on the torus in their simplest physical regimes. A combination of boundary conditions $(h,v)$ is applied in the horizontal and vertical directions with $h,v=0$ and 1 for periodic and antiperiodic boundary conditions respectively. The fugacities of non-contractible and contractible loops are denoted by $\alpha$ and $\beta$ respectively where $\beta$ is simply related to the crossing parameter $\lambda$. At roots of unity, when $\lambda/\pi\in\mathbb Q$, these models are the dense ${\cal LM}(p,p')$ and dilute ${\cal DLM}(p,p')$ logarithmic minimal models with $p,p'$ coprime integers. We conjecture the scaling limits of the transfer matrix traces in the standard modules with $d$ defects and deduce the conformal partition functions ${\cal Z}_{\textrm{dense}}^{\textrm{\tiny$(h,v)$}}(\alpha)$ and ${\cal Z}_{\textrm{dilute}}^{\textrm{\tiny$(h,v)$}}(\alpha)$ using Markov traces.  These are expressed in terms of functions ${\cal Z}_{m,m'}(g)$ known from the Coulomb gas arguments of Di Francesco, Saleur and Zuber and subsequently as sesquilinear forms in Verma characters. Crucially, we find that the partition functions are identical for the dense and dilute models. 
The coincidence of these conformal partition functions provides compelling evidence that, for given $(p,p')$, these dense and dilute theories lie in the same universality class. In root of unity cases with $\alpha=2$, the $(h,v)$ modular covariant partition functions are also expressed as sesquilinear forms in affine $u(1)$ characters involving generalized Bezout conjugates. These also give the modular covariant partition functions for the 6-vertex and Izergin--Korepin 19-vertex models in the corresponding regimes.

\newpage
\tableofcontents

\hyphenpenalty=30000

\setcounter{footnote}{0}

%
\section{Introduction}
%

The Yang--Baxter integrable~\cite{BaxBook} dense 
$\Aoneone$~\cite{Nienhuis82,BloteNienhuis89,YB95,PRZ2006,PR2007,SAPR2009,PR2011,MDPR2013,MDPR2014,MDKP17} and dilute 
$\Atwotwo$~\cite{DJS2010,SAPR2012,G12,FeherNien2015,GarbNien2017,MDP19,MDKP23} loop models provide a paradigm for studying (i) two-dimensional statistical systems with non-local degrees of freedom in the form of loop segments and (ii) geometrical phase transitions. 
In this paper, we consider only the simplest of the various physical parameter regimes in the spectral parameter $u$ and crossing parameter $\lambda$. For the so-called root of unity cases, for which the contractible loop fugacity is parameterised as $\beta=2\cos\tfrac{\pi(p'-p)}{p'}$ with $p/p'\in\mathbb{Q}$, $1\le p<p'$ and $p,p'$ coprime integers, these loop models reduce to the dense ${\cal LM}(p,p')$ and dilute ${\cal DLM}(p,p')$ logarithmic minimal models. The first few members include critical dense and dilute polymers with $(p,p')=(1,2)$ and critical bond and site percolation with $(p,p')=(2,3)$.

In fact, in the continuum scaling limit, the universal behaviour of these theories is described not by {\em rational}~\cite{FMS,MooreSeiberg} but rather by {\em logarithmic}~\cite{Gurarie,SpecialIssue} Conformal Field Theories (CFTs) with central charges $c$ and conformal weights $\Delta_{r,s}$ given by
\be
c=1-\frac{6(p-p')^2}{pp'}, 
\qquad
\Delta_{r,s}=\Delta_{r,s}^{p,p'}=\frac{(p'r-ps)^2-(p-p')^2}{4pp'}
\label{LogMinConfWts}
\ee
where, for our purposes, the Kac labels are restricted to 
$r \in \mathbb R$ and $s\in\frac 12 \mathbb{Z}$. These logarithmic theories are not unitary and do not lie in the rational universality classes represented by the RSOS minimal models ${\cal M}(p,p')$~\cite{BPZ,ABF,FB} despite the fact that they share the same formulas for the central charges and conformal weights. Instead, the universal behaviour of these dense and dilute theories are described by logarithmic universality classes.

In light of these observations, it is natural to ask: (i) Are the dense and dilute lattice models in the same logarithmic universality class? 
(ii) What are the conformal partition functions on the torus for simple periodic and antiperiodic boundary conditions and are they the same for dense and dilute theories? These are the general questions we endeavour to answer in this paper extending our previous results for critical dense polymers \cite{MDPR2013} corresponding to $\mathcal{LM}(1,2)$ and bond \cite{MDKP17} and site percolation \cite{MDKP23} corresponding to $\mathcal{LM}(2,3)$ and $\mathcal{DLM}(2,3)$ respectively.

Let us summarize the main results of the paper. We investigate the dense $\mathcal{LM}(p,p')$ and dilute $\mathcal{DLM}(p,p')$ on an $M\times N$ torus, with fugacities $\alpha=2\cos\gamma$, $\gamma\in\mathbb{R}$ and $\beta=2\cos\tfrac{\pi(p'-p)}{p'}$ for non-contractible and contractible loops respectively. We consider periodic and antiperiodic boundary conditions in the horizontal ($h\in\{0,1\}$) and vertical ($v\in\{0,1\}$) directions, leading to four lattice partition functions for each $(p,p')$ model: $Z_{\textrm{dense}}^{\textrm{\tiny$(h,\!v)$}}$ and $Z_{\textrm{dilute}}^{\textrm{\tiny$(h,\!v)$}}$. Using Markov traces, we compute the conformal scaling limit of these partition functions, denoted by $\mathcal Z_{\textrm{dense}}^{\textrm{\tiny$(h,\!v)$}}$ and $\mathcal Z_{\textrm{dilute}}^{\textrm{\tiny$(h,\!v)$}}$. Crucially, we find that they are equal for the two loop models in the regime $\frac p{p'} \in (0,1)$, namely $\mathcal Z_{\textrm{dense}}^{\textrm{\tiny$(h,\!v)$}}=\mathcal Z_{\textrm{dilute}}^{\textrm{\tiny$(h,\!v)$}}$. We express these partition functions first in terms of functions $\mathcal Z_{m,m'}(g)$ known from the Coulomb gas formalism~\cite{FSZ87npb,FSZ87}, and subsequently as series in the modular parameters $q,\bar q$ with conformal dimensions that depend on $p$ and $p'$ only through the ratio $\frac p{p'}$. The sum of the four partition functions reproduces the partition function of the $O(n)$ model computed by Di Francesco, Saleur and Zuber in~\cite{FSZ87}. Finally, for the special case $\alpha = 2$, we re-express the conformal partition functions, denoted by $\mathcal Z^{\textrm{\tiny$(h,\!v)$}}(p,p')$, in terms of sesquilinear forms in affine $u(1)$ characters $\varkappa^n(z,q)$
\begin{alignat}{2}
\mathcal Z^{\textrm{\tiny$(h,\!v)$}}(p,p')
&=\sum_{r=0}^{p-1}\sum_{s=0}^{2p'-1} (-1)^{vr} \varkappa^n_{p' r-p(s+h/2)}\big((-1)^{pv},q\big)\,\varkappa^n_{p' r+p(s+h/2)}\big((-1)^{pv},\qbar\big)
\nonumber\\&=\frac 1\kappa \displaystyle\sum_{j=0}^{P-1} (-1)^{v\rhoj}\,\varkappa^n_{j+ h'/2}(q)\varkappa^n_{\overline{j+ {h'}/2}}(\qbar)
\end{alignat}
where $\rhoj=\frac1{2p'}(j+h'/2+\overline{j+h'/2})$,
\be
n = pp', \qquad
h'=\begin{cases}
\,1& p \textrm{ odd and $h=1$},\\
\,0&\textrm{otherwise},
\end{cases}
\quad\
P=\begin{cases}
\,2n&\mbox{$pv$ even,}\\
\,4n&\mbox{$pv$ odd,}
\end{cases}
\quad\
\kappa=\frac{P}{2n}=\begin{cases}
\,1&\mbox{$pv$ even,}\\
\,2&\mbox{$pv$ odd.}
\end{cases}
\ee 
Here, $j+h'/2$ and $\overline{j+ {h'}/2}$ are integer and half-integer Bezout conjugates for $h'=0$ and $h'=1$ respectively. For the modular invariant case of periodic boundary conditions, we show that $\mathcal Z^{\textrm{\tiny$(0,0)$}}(p,p')$ is precisely the known Coulomb partition function ${\cal Z}_{\text{Coul}}(\tfrac{p}{p'})$.

The layout of the paper is as follows. In \cref{sec:model.def}, we introduce the dense and dilute loop models, discuss the four possible boundary conditions on the torus and define the lattice partition functions on the torus. In \cref{sec:Markov}, Markov traces are used to express these torus partition functions in terms of traces of the transfer matrix over standard modules of the enlarged periodic Temperley--Lieb algebras first for the dense models, and then separately for the dilute models. The scaling limits of the transfer matrix traces in the standard modules are conjectured in \cref{sec:scaling.limit}  for the two models. Using Markov traces, we compute the conformal scaling limit of the lattice partition functions. In \cref{sec:Verma}, these conformal partition functions are expressed as explicit series in the modular parameters $q,\bar q$. Showing the equivalence of our result for the $O(n)$ partition function, obtained by summing over $h$ and $v$, with that of Di Francesco, Saleur and Zuber in~\cite{FSZ87} requires an exercise in number theory which we detail in \cref{NumberTheory}. The results of \cref{sec:model.def,sec:Markov,sec:scaling.limit,sec:Verma} hold for general values of $\alpha$ and $\beta$. In \cref{sec:alpha=2}, we specialize to $\alpha=2$ and $\lambda/\pi\in\mathbb{Q}$, and re-express the corresponding partition functions in terms of sesquilinear forms in affine $u(1)$ characters involving integer and half-integer Bezout conjugates. We conclude in \cref{sec:conclusion} with some general comments. The properties of integer and half-integer Bezout conjugates are discussed in \cref{app:half.integer.Bezout}. Examples of the modular covariant partition functions for the first values of $(p,p')$ are given in \cref{sec:Zpp'.examples}.

%
\section{Dense and dilute loop models on the torus}\label{sec:model.def}
%

We define the dense and dilute loop models on a lattice made of $M$ rows and $N$ columns of square faces. A configuration $\sigma$ of the dilute loop model is a decoration of each of the $MN$ square faces of this lattice by one of nine following tiles: 
\be
\psset{unit=0.7cm}
\begin{pspicture}[shift=-.40](-0.1,-.2)(1.1,1.2)
\facegrid{(0,0)}{(1,1)}
\rput[bl](0,0){\loopa}
\end{pspicture}
\quad
\begin{pspicture}[shift=-.40](-0.1,-.2)(1.1,1.2)
\facegrid{(0,0)}{(1,1)}
\rput[bl](0,0){\loopb}
\end{pspicture}
\quad
\begin{pspicture}[shift=-.40](-0.1,-.2)(1.1,1.2)
\facegrid{(0,0)}{(1,1)}
\rput[bl](0,0){\loopc}
\end{pspicture}
\quad
\begin{pspicture}[shift=-.40](-0.1,-.2)(1.1,1.2)
\facegrid{(0,0)}{(1,1)}
\rput[bl](0,0){\loopd}
\end{pspicture}
\quad
\begin{pspicture}[shift=-.40](-0.1,-.2)(1.1,1.2)
\facegrid{(0,0)}{(1,1)}
\rput[bl](0,0){\loope}
\end{pspicture}
\quad
\begin{pspicture}[shift=-.40](-0.1,-.2)(1.1,1.2)
\facegrid{(0,0)}{(1,1)}
\rput[bl](0,0){\loopf}
\end{pspicture}
\quad
\begin{pspicture}[shift=-.40](-0.1,-.2)(1.1,1.2)
\facegrid{(0,0)}{(1,1)}
\rput[bl](0,0){\loopg}
\end{pspicture}
\quad
\begin{pspicture}[shift=-.40](-0.1,-.2)(1.1,1.2)
\facegrid{(0,0)}{(1,1)}
\rput[bl](0,0){\looph}
\end{pspicture}
\quad
\begin{pspicture}[shift=-.40](-0.1,-.2)(1.1,1.2)
\facegrid{(0,0)}{(1,1)}
\rput[bl](0,0){\loopi}
\end{pspicture}\ .
\ee
We assign to these tiles the labels $1,2, \dots, 9$. In a configuration of this model, there can be no free ends, namely any loop segment arising on a tile necessarily connects to loop segments on two adjacent tiles. For the dense loop model, only the tiles 8 and 9 are allowed. Typical configurations for the dense and dilute loop models on the torus are shown in \cref{fig:strips.and.connectivities}.

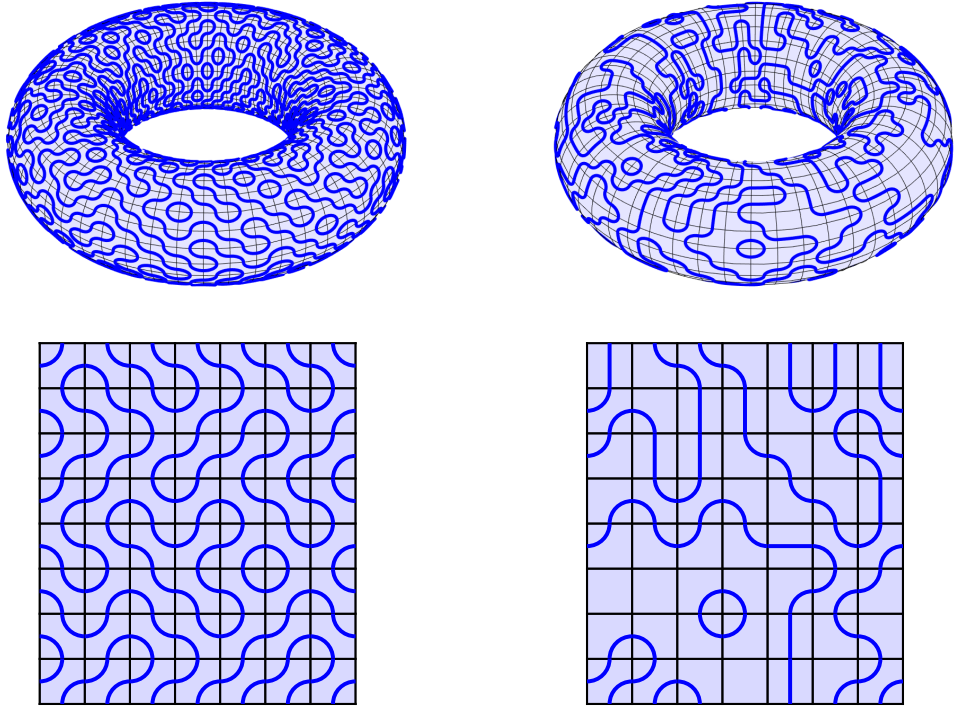
\begin{figure}[t]
\begin{equation*}
\psset{unit=0.60cm}
\begin{pspicture}[shift=-3.9](-2,-1)(5,13)
\rput(1.4,11){\includegraphics[width=3in]{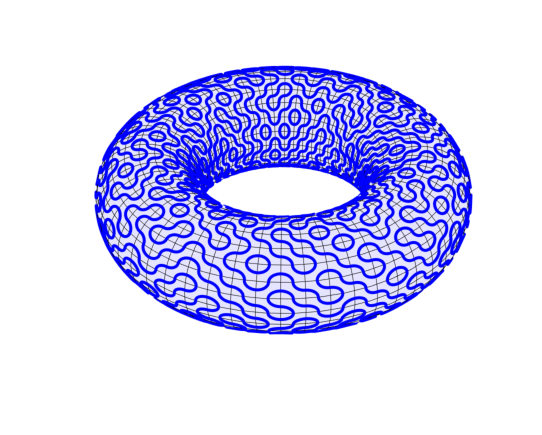}}
\facegrid{(-2,-1)}{(5,7)}
\rput(-2,6){\looph}\rput(-1,6){\loopi}\rput(0,6){\loopi}\rput(1,6){\loopi}\rput(2,6){\loopi}\rput(3,6){\loopi}\rput(4,6){\loopi}
\rput(-2,5){\loopi}\rput(-1,5){\loopi}\rput(0,5){\loopi}\rput(1,5){\looph}\rput(2,5){\looph}\rput(3,5){\loopi}\rput(4,5){\looph}
\rput(-2,4){\looph}\rput(-1,4){\looph}\rput(0,4){\looph}\rput(1,4){\looph}\rput(2,4){\looph}\rput(3,4){\looph}\rput(4,4){\looph}
\rput(-2,3){\looph}\rput(-1,3){\looph}\rput(0,3){\loopi}\rput(1,3){\looph}\rput(2,3){\loopi}\rput(3,3){\loopi}\rput(4,3){\loopi}
\rput(-2,2){\loopi}\rput(-1,2){\loopi}\rput(0,2){\loopi}\rput(1,2){\looph}\rput(2,2){\looph}\rput(3,2){\loopi}\rput(4,2){\looph}
\rput(-2,1){\loopi}\rput(-1,1){\looph}\rput(0,1){\loopi}\rput(1,1){\loopi}\rput(2,1){\loopi}\rput(3,1){\looph}\rput(4,1){\loopi}
\rput(-2,0){\loopi}\rput(-1,0){\looph}\rput(0,0){\loopi}\rput(1,0){\loopi}\rput(2,0){\looph}\rput(3,0){\looph}\rput(4,0){\loopi}
\rput(-2,-1){\looph}\rput(-1,-1){\looph}\rput(0,-1){\looph}\rput(1,-1){\looph}\rput(2,-1){\looph}\rput(3,-1){\looph}\rput(4,-1){\looph}
\end{pspicture}
\qquad\qquad\qquad\qquad
\mbox{}\vspace{-.2in}\mbox{}
\begin{pspicture}[shift=-3.9](-2,-1)(5,7)
\rput(1.4,11){\includegraphics[width=3in]{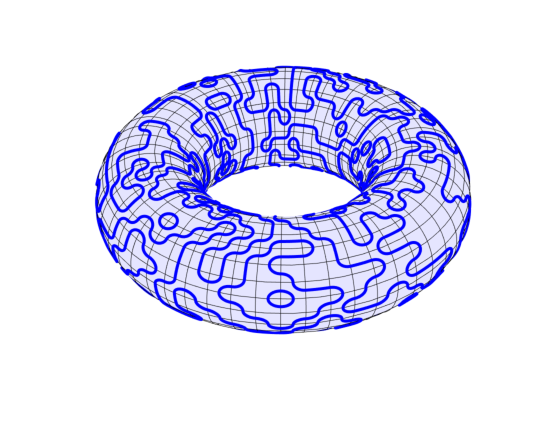}}
\facegrid{(-2,-1)}{(5,7)}
\rput(-2,6){\loopg}\rput(-1,6){\loope}\rput(0,6){\loopi}\rput(1,6){\loopd}\rput(2,6){\loopg}\rput(3,6){\loopg}\rput(4,6){\loopg}
\rput(-2,5){\looph}\rput(-1,5){\loopd}\rput(0,5){\loopg}\rput(1,5){\loopg}\rput(2,5){\loope}\rput(3,5){\looph}\rput(4,5){\loopi}
\rput(-2,4){\loopb}\rput(-1,4){\loopg}\rput(0,4){\loopg}\rput(1,4){\loope}\rput(2,4){\loopd}\rput(3,4){\loope}\rput(4,4){\loopi}
\rput(-2,3){\loopc}\rput(-1,3){\loopi}\rput(0,3){\looph}\rput(1,3){\loopd}\rput(2,3){\loope}\rput(3,3){\loopd}\rput(4,3){\loopg}
\rput(-2,2){\loopb}\rput(-1,2){\loope}\rput(0,2){\loopb}\rput(1,2){\loope}\rput(2,2){\loopf}\rput(3,2){\loopi}\rput(4,2){\looph}
\rput(-2,1){\loopa}\rput(-1,1){\loopa}\rput(0,1){\loopc}\rput(1,1){\loopd}\rput(2,1){\loopc}\rput(3,1){\looph}\rput(4,1){\loopb}
\rput(-2,0){\loopc}\rput(-1,0){\loopd}\rput(0,0){\loope}\rput(1,0){\loopb}\rput(2,0){\loopg}\rput(3,0){\loope}\rput(4,0){\loopd}
\rput(-2,-1){\looph}\rput(-1,-1){\looph}\rput(0,-1){\loopd}\rput(1,-1){\loopa}\rput(2,-1){\loopg}\rput(3,-1){\loopc}\rput(4,-1){\looph}
\end{pspicture}
\end{equation*}\\[0pt]
\caption{Typical configurations of the dense (left panels) and dilute loop model (right panels). In the upper panels, configurations are shown directly on the torus for $(M,N)=(22,52)$. Projections of typical configurations onto a doubly periodic rectangle are shown in the lower panels for $(M,N)=(8,7)$.}
\label{fig:strips.and.connectivities}
\end{figure}

The model is defined with toroidal boundary conditions, meaning that the lattice is periodic along both the horizontal and vertical axes. The loop segments form closed loops that are {\it contractible} if they can be continuously deformed to a point, and {\it non-contractible} if they encircle the torus non-trivially. The non-contractible loops may wrap the torus $i$ times around the horizontal period and $j$ times around the vertical period, with $i$ and $j$ coprime. We write this condition in terms of their greatest common divisor as $i \wedge j = 1$, with the convention $i \wedge 0 = 0 \wedge i = i$. We choose the convention $j\in \mathbb Z_{\ge 0}$ and $i \in \mathbb Z$ where, following a curve that moves upwards on the torus, we assign it a positive value of $i$ if it winds around the torus horizontally by moving to the right, and a negative integer if it winds by moving to the left. Each contractible loop is assigned a weight $\beta$, whereas each non-contractible loop is given a weight $\alpha_{i,j}$, dependent on its winding $(i,j)$. For both models, the inhomogeneous and homogeneous partition functions are defined as
\be
\label{eq:Zdef}
Z = \sum_\sigma \beta^{n_\beta(\sigma)} \prod_{i \wedge j = 1} \alpha_{i,j}^{n_{i,j}(\sigma)} \prod_{i=1}^9 \rho_i^{n_i(\sigma)}, 
\qquad Z(\alpha) = Z \big|_{\alpha_{i,j} \to \alpha}\ .
\ee
The numbers $n_\beta(\sigma)$ and $n_{i,j}(\sigma)$ respectively count the numbers of contractible loops and non-contractible loops of winding $(i,j)$ of the configuration $\sigma$, and the number $n_i(\sigma)$ counts the occurences of the $i$-th tile in $\sigma$.
The weights $\rho_i$ of the tiles are expressed in terms of the spectral parameter $u$ and the crossing parameter $\lambda$ as
\begin{subequations}
\label{eq:weights}
\begin{alignat}{2}
&\textrm{dense:}\qquad
&&\begin{array}{lll}\rho_1 = \rho_2 = \dots = \rho_7 = 0, \qquad& \rho_8 = s(\lambda-u), \qquad& \rho_9 = s(u),\end{array}
\\[0.25cm]
&\textrm{dilute:}\qquad
&&\left\{\begin{array}{ll}
\rho_1= s(2\lambda) s(3\lambda) + s(u) s(3\lambda-u), \quad
&\rho_{6}=\rho_{7}= s(u) s(3\lambda-u), \\[0.1cm]
\rho_{2} = \rho_{3} = s(2\lambda) s(3\lambda-u), \qquad
&\rho_8= s(2\lambda-u)s(3\lambda-u),\\[0.1cm]
\rho_{4}=\rho_{5}=s(2\lambda) s(u), \qquad
&\rho_9= -s(u)s(\lambda-u),
\end{array}\right.
\end{alignat}
\end{subequations}
where
\be
s(u) = \frac{\sin u}{\sin \lambda}.
\ee
The crossing parameter also parameterises the weight of contractible loops as
\be
\beta = 2\cos\frac{\pi(p'-p)}{p'}= \begin{cases}
\ 2\cos \lambda & \textrm{dense,}\\
-2\cos 4\lambda & \textrm{dilute,}
\end{cases}
\ee
where we use the parameterisation
\be
\label{eq:lambda.pp'}
\lambda =\left\{\begin{array}{ccc}
\displaystyle\frac{\pi(p'-p)}{p'} \quad& \textrm{dense,}\\[0.3cm]
\displaystyle\frac{\pi(2p'-p)}{4p'}\quad& \textrm{dilute,}
\end{array}\right.
\qquad
\frac{p}{p'} \in (0,1).
\ee
The central charge and the Kac formula for the conformal dimensions are given by \eqref{LogMinConfWts}. For the logarithmic minimal models $\mathcal{LM}(p,p')$ and dilute logarithmic minimal models $\mathcal{DLM}(p,p')$, the integers $p$ and $p'$ are coprime with $p<p'$. The dilute model admits other branches but we do not consider them here. We denote by $Z_{\textrm{dense}}$ and $Z_{\textrm{dilute}}$ the partition functions of the dense and dilute loop models, respectively.

The loop segments draw the contours of clusters. For the dense loop model on a lattice with both $M$ and $N$ even, each configuration has two dual sets of clusters, with the loop segments drawing the contours between clusters of different sets. In this case, we say that the boundary condition is periodic along both axes. For $M$ odd, we say that the model is anti-periodic in the vertical direction, meaning that by moving along this periodicity axis, the two types of clusters are interchanged. Likewise for $N$ odd, the model is anti-periodic in the horizontal direction. We therefore define the partition functions $Z^{\textrm{\tiny$(h,v)$}}$, with $h,v \in \{0,1\}$, where the clusters are constrained to have only periodic ($h,v = 0$) and antiperiodic ($h,v=1$) boundary conditions along the horizontal and vertical axes. Thus for the dense loop models, this simply translates to
\begin{subequations}
\begin{alignat}{2}
&Z_{\textrm{dense}}^{\textrm{\tiny$(0,0)$}} = Z_{\textrm{dense}}^{\textrm{\tiny $M$\,even,\,$N$\,even}},
\qquad
&&Z_{\textrm{dense}}^{\textrm{\tiny$(0,1)$}} = Z_{\textrm{dense}}^{\textrm{\tiny $M$\,odd,\,$N$\,even}},
\\[0.15cm]
&Z_{\textrm{dense}}^{\textrm{\tiny$(1,0)$}} = Z_{\textrm{dense}}^{\textrm{\tiny $M$\,even,\,$N$\,odd}},
\qquad
&&Z_{\textrm{dense}}^{\textrm{\tiny$(1,1)$}} = Z_{\textrm{dense}}^{\textrm{\tiny $M$\,odd,\,$N$\,odd}}.
\end{alignat}
\end{subequations}

\begin{figure}[htb]
\begin{equation*}
\begin{array}{cc}
\includegraphics[width=2in]{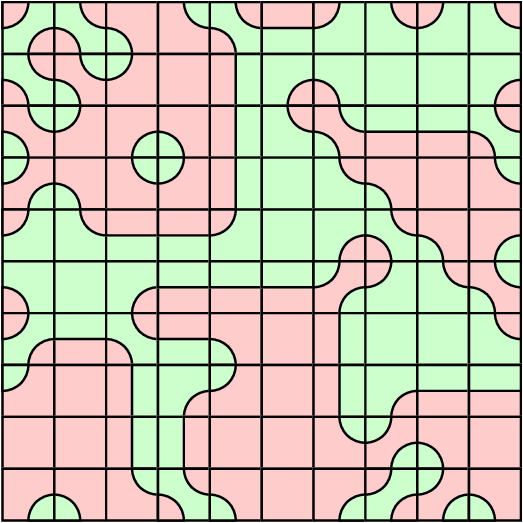} & \includegraphics[width=2in]{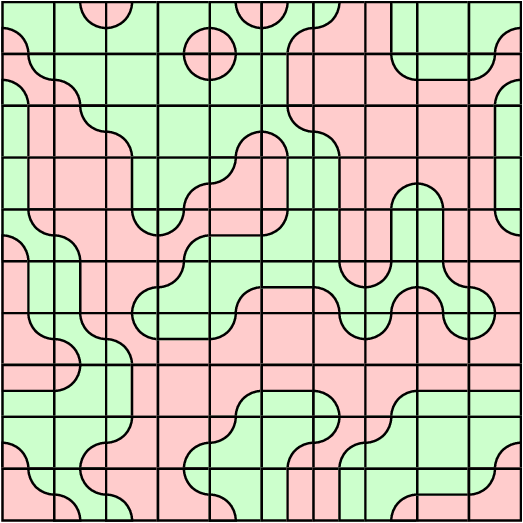}\\
(h,v)=(0,0) & (h,v)=(0,1)\\[0.3cm]
\includegraphics[width=2in]{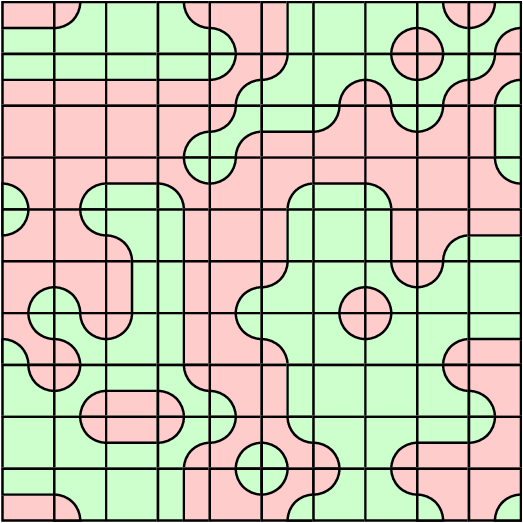} & \includegraphics[width=2in]{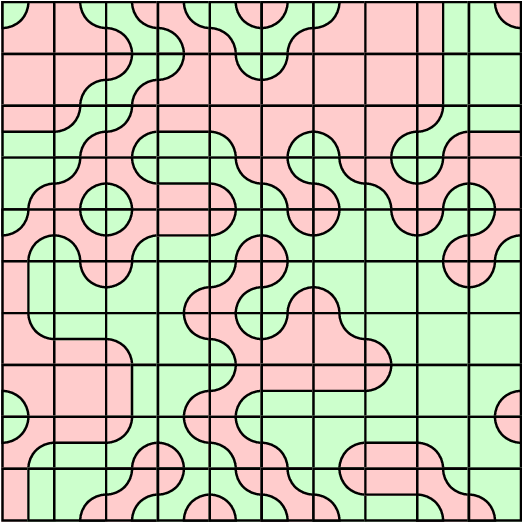}\\
(h,v)=(1,0) & (h,v)=(1,1)
\end{array}
\end{equation*}
\caption{Example loop configurations for the dilute $\Atwotwo$ loop model for the four possible $(h,v)$ boundary conditions. Generic horizontal and vertical lines cross loop segments $H$ and $V$ times, respectively, with $h=H \textrm{ mod } 2$ and $v=V \textrm{ mod } 2$. The left/right edges and top/bottom edges are identified to form a torus. Loop configurations for the dense $\Aoneone$ loop model are similar but with each square face visited by two loop segments. In this case, $h=N \textrm{ mod } 2$ and $v=M \textrm{ mod } 2$.}
\label{hvLoopConfigs}
\end{figure}

In the dilute loop model, the loop segments also draw the contours of clusters. In this case, 
independently of the parity of $M$ and $N$, there are configurations whose clusters have periodic boundary conditions and others where they have anti-periodic boundary conditions. We then define the partition functions $Z_{\textrm{dilute}}^{\textrm{\tiny$(h,v)$}}$ with $h,v\in\{0,1\}$, defined as in~\eqref{eq:Zdef}, but where the sum only runs over configurations~$\sigma$ whose clusters have the periodicities $(h,v)$. As argued in \cite{MDKP23}, the four homogeneous partition functions then correspond to assigning different weights $\alpha_{i,j}$ for the non-contractible loops entry-wise according to their winding $(i,j)$, namely\footnote{We use the notation $a\equiv b$ mod $n$ when $a$ and $b$ have the same remainder $r$ when they are divided by $n$ such that $0\le r < n$ 
and the notation $a= b$ mod $n$ when assigning to $a$ the unique value given by the remainder $r$ when $b$ is divided by $n$.}
\be
\label{eq:loop.specifications}
\hspace{-0.4cm}
\begin{array}{lll}
Z_{\textrm{dilute}}^{\textrm{\tiny$(0,0)$}}(\alpha):\hspace{0.2cm}
\alpha_{i,j} \to
\left\{\begin{array}{ll}
\alpha & (i,j) \equiv (1,0) \textrm{ mod } 2, \\
\alpha & (i,j) \equiv (0,1) \textrm{ mod } 2,\\
\alpha & (i,j)\equiv (1,1) \textrm{ mod } 2,
\end{array}\right.
\\[0.85cm]
Z_{\textrm{dilute}}^{\textrm{\tiny$(0,1)$}}(\alpha):\hspace{0.2cm}
\alpha_{i,j} \to
\left\{\begin{array}{ll}
\alpha & (i,j) \equiv(1,0) \textrm{ mod } 2, \\
0 & (i,j) \equiv (0,1) \textrm{ mod } 2,\\
0 & (i,j) \equiv (1,1) \textrm{ mod } 2,
\end{array}\right.
\\[0.85cm]
Z_{\textrm{dilute}}^{\textrm{\tiny$(1,0)$}}(\alpha):\hspace{0.2cm}
\alpha_{i,j} \to
\left\{\begin{array}{ll}
0 & (i,j) \equiv (1,0) \textrm{ mod } 2, \\
\alpha & (i,j) \equiv(0,1) \textrm{ mod } 2,\\
0 & (i,j) \equiv (1,1) \textrm{ mod } 2,
\end{array}\right.
\\[0.85cm]
Z_{\textrm{dilute}}^{\textrm{\tiny$(1,1)$}}(\alpha):\hspace{0.2cm}
\alpha_{i,j} \to
\left\{\begin{array}{ll}
0 & (i,j) \equiv (1,0) \textrm{ mod } 2, \\
0 & (i,j) \equiv (0,1) \textrm{ mod } 2,\\
\alpha & (i,j) \equiv(1,1) \textrm{ mod } 2.
\end{array}\right.\end{array}
\ee
Typical $(h,v)$ loop configurations for the dilute loop model are shown in \cref{hvLoopConfigs}. The full partition function of the dilute loop model is then obtained as
\be
\label{eq:fullZ}
Z_{\textrm{dilute}} = \sum_{h,v\,\in\, \{0,1\}}Z_{\textrm{dilute}}^{\textrm{\tiny$(h,v)$}}.
\ee

%
\section{Transfer matrices, Markov traces and partition functions}\label{sec:Markov}
%

The transfer matrix $\Tb(u)$ of the dense and dilute loop models are respectively elements of the enlarged periodic Temperley--Lieb algebra~\cite{L91,MS93,GL98,PRV10} and the enlarged dilute periodic Temperley--Lieb algebra~\cite{MDP19}:
\be
\label{eq:T(u)}
\psset{unit=0.8cm}
\Tb(u)
= \ 
\begin{pspicture}[shift=-0.9](-0.3,-0.5)(5.3,1.2)
\facegrid{(0,0)}{(5,1)}
\psarc[linewidth=0.025]{-}(0,0){0.16}{0}{90}
\psarc[linewidth=0.025]{-}(1,0){0.16}{0}{90}
\psarc[linewidth=0.025]{-}(2,0){0.16}{0}{90}
\psarc[linewidth=0.025]{-}(4,0){0.16}{0}{90}
\psline[linewidth=1.5pt,linecolor=blue,linestyle=dashed,dash=2pt 2pt]{-}(0,0.5)(-0.3,0.5)
\psline[linewidth=1.5pt,linecolor=blue,linestyle=dashed,dash=2pt 2pt]{-}(5,0.5)(5.3,0.5)
\rput(0.5,.5){$u$}
\rput(1.5,.5){$u$}
\rput(2.5,0.5){$u$}
\rput(3.5,0.5){$\ldots$}
\rput(4.5,.5){$u$}
\psline{<->}(0,-0.2)(5,-0.2)\rput(2.5,-0.45){$_N$}
\end{pspicture}
\ee
where the face operator is
\begin{alignat}{2}
\psset{unit=0.8cm}
\label{eq:face.op}
\begin{pspicture}[shift=-.40](0,0)(1,1)
\facegrid{(0,0)}{(1,1)}
\psarc[linewidth=0.025]{-}(0,0){0.16}{0}{90}
\rput(.5,.5){$u$}
\end{pspicture}
\ \ = \ &\rho_1(u)\ \
\psset{unit=0.8cm}
\begin{pspicture}[shift=-.40](0,0)(1,1)
\facegrid{(0,0)}{(1,1)}
\rput[bl](0,0){\loopa}
\end{pspicture}
\ \ + \rho_2(u)\ \
\begin{pspicture}[shift=-.40](0,0)(1,1)
\facegrid{(0,0)}{(1,1)}
\rput[bl](0,0){\loopb}
\end{pspicture}
\ \ + \rho_3(u)\ \
\begin{pspicture}[shift=-.40](0,0)(1,1)
\facegrid{(0,0)}{(1,1)}
\rput[bl](0,0){\loopc}
\end{pspicture}
\ \ + \rho_4(u)\ \
\begin{pspicture}[shift=-.40](0,0)(1,1)
\facegrid{(0,0)}{(1,1)}
\rput[bl](0,0){\loopd}
\end{pspicture}
\ \ + \rho_5(u)\ \
\begin{pspicture}[shift=-.40](0,0)(1,1)
\facegrid{(0,0)}{(1,1)}
\rput[bl](0,0){\loope}
\end{pspicture}
\nonumber\\[0.2cm] \ &\hspace{0.2cm}
+ \rho_6(u)\ \
\psset{unit=0.8cm}
\begin{pspicture}[shift=-.40](0,0)(1,1)
\facegrid{(0,0)}{(1,1)}
\rput[bl](0,0){\loopf}
\end{pspicture}
\ \ + \rho_7(u)\ \
\begin{pspicture}[shift=-.40](0,0)(1,1)
\facegrid{(0,0)}{(1,1)}
\rput[bl](0,0){\loopg}
\end{pspicture}
\ \ + \rho_8(u)\ \
\begin{pspicture}[shift=-.40](0,0)(1,1)
\facegrid{(0,0)}{(1,1)}
\rput[bl](0,0){\looph}
\end{pspicture}
\ \ + \rho_9(u)\ \
\begin{pspicture}[shift=-.40](0,0)(1,1)
\facegrid{(0,0)}{(1,1)}
\rput[bl](0,0){\loopi}
\end{pspicture}\ \, ,
\end{alignat}
with the weights $\rho_i$ as in \eqref{eq:weights}. The dashed lines at the left and right of the diagram in \eqref{eq:T(u)} indicate that the first and last face operator are connected periodically.

Let us denote by $\repW_{N,d,\omega}$ the standard module over the periodic Temperley--Lieb algebra for the dense loop model, and over the dilute periodic Temperley--Lieb algebra for the dilute loop model where $d$ is the number of defects. This defect number takes all values in $\{0,1,\dots, N\}$ for the dilute model, but only the values in this set with $d \equiv N \textrm{ mod } 2$ for the dense model. For $d\neq 0$, the parameter $\omega$ couples to the windings of the defects whereas, for $d=0$, it parameterises the weight of non-contractible loops as $\alpha = \omega + \omega^{-1}$ with $\omega=\eE^{\ir \gamma}$. We follow here the conventions in \cite{MDKP17,MDKP23} for the definition of these modules. The $M$-th power of the transfer matrix decomposes as a Laurent series
\be
\textrm{tr}_{\raisebox{-0.05cm}{\tiny$\repW_{N,d,\omega}$}} \Tb(u)^M = \sum_{j = -M}^M  \omega^{-j}C_{d,j}\, ,
\ee
where $C_{d,j}$ are certain coefficients that are independent of $\omega$. This holds for both the dense and dilute loop models, but with coefficients $C_{d,j}$ that are different for the two models. In particular, it is not difficult to see that $C_{d,j}=0$ for the dense loop model if $(-1)^{j+M}=-1$, whereas this coefficient is in general non-zero in the dilute loop model. In both cases, the coefficients can be computed as
\be
C_{d,j} = \frac1{2\pi} \int_{0}^{2\pi} \dd\gamma\, \eE^{\ir \gamma j}\, \textrm{tr}_{\raisebox{-0.05cm}{\tiny$\repW_{N,d,\eE^{\ir \gamma}}$}} \Tb(u)^M.
\ee

Using the Markov trace, it is possible to express the partition functions $Z_{\textrm{dense}}^{\textrm{\tiny$(h,v)$}}(\alpha)$ and $Z_{\textrm{dilute}}^{\textrm{\tiny$(h,v)$}}(\alpha)$ in terms of the coefficients $C_{d,j}$. For the dense loop model, this is a result that goes back to Jacobsen and Richard \cite{RJ07}. For the dilute loop model, this was argued recently in the context of site percolation \cite{MDKP23} corresponding to $\lambda = \frac \pi 3$, with the argument readily extending to the other values of $\lambda$. The result reads
\begin{subequations}
\label{eq:Cdj1}
\begin{alignat}{2}
\label{eq:ZdenseCdj}
Z_{\textrm{dense}}^{\textrm{\tiny$(h,v)$}}(\alpha) &= 
\sum_{\substack{-N\le d \le N\\ d\,\equiv\, h \,\textrm{mod}\, 2}} \sum_{\substack{-M\le j \le M\\ j\,\equiv\, v \,\textrm{mod}\, 2}} \, T_{d \wedge j} (\tfrac \alpha2)\, C_{d,j}\, ,
\\
\label{eq:ZdiluteCdj}
Z_{\textrm{dilute}}^{\textrm{\tiny$(h,v)$}}(\alpha) &= 
\sum_{\substack{-N\le d \le N\\ d\,\equiv\, h \,\textrm{mod}\, 2}} \sum_{\substack{-M\le j \le M\\ j\,\equiv\, v \,\textrm{mod}\, 2}} 2\, T_{d \wedge j} (\tfrac \alpha2)\, C_{d,j}\, ,
\end{alignat}
\end{subequations}
where the definition of $C_{d,j}$ is extended to negative $d$ by $C_{-d,j} = C_{d,-j}$ and $T_n(x)$ is the $n$-th Chebyshev polynomial of the first kind: $T_n(\cos \theta) = \cos n \theta$. There is thus an extra factor of $2$ for the dilute loop model. 

%
\section{Scaling limit and conformal partition functions}\label{sec:scaling.limit}
%

In this section, we write down conjectures for the scaling limit of the traces of the transfer matrices and use them to obtain formulas for the conformal partition functions.

\begin{Conjecture}[Scaling limits of the traces in the standard modules]
The scaling limits of the traces of the transfer matrices are given by
\be
\label{eq:traceConjectures}
\lim_{\substack{M,N \to \infty\\M/N\to \delta\\
\epsilon\, = \, M\, \text{\rm mod } 2}} \eE^{MN f_\mathrm{bulk}(u)}\,\mathrm{tr}_{\raisebox{-0.05cm}{\tiny$\repW_{N,d,\omega}$}} \Tb(u)^M =
\left\{\begin{array}{ll}
\displaystyle\frac{(q \bar q)^{-c/24}}{(q)_\infty (\bar q)_\infty}
\sum_{\ell=-\infty}^{\infty} (-1)^{\epsilon \ell}  q^{\Delta_{\gamma/\pi - \ell, d/2}} \bar q^{\Delta_{\gamma/\pi-\ell, -d/2}}
& {\rm dense,}
\\[0.5cm]
\displaystyle\frac{(q \bar q)^{-c/24}}{(q)_\infty (\bar q)_\infty}
\sum_{\ell=-\infty}^{\infty} q^{\Delta_{\gamma/\pi - 2\ell, d/2}} \bar q^{\Delta_{\gamma/\pi-2\ell, -d/2}}
& {\rm dilute,}
\end{array}\right.
\ee
where $f_\mathrm{bulk}(u)$ is the known dense or dilute bulk free energy and
\be
(q)_\infty=\prod_{n=1}^\infty (1-q^n),\qquad \omega = \eE^{\ir \gamma},\qquad q = \eE^{2 \pi \ir \tau} = \exp\big(\!-\!2\pi i\delta \eE^{-i\vartheta}\big),\qquad
\vartheta=\begin{cases}
\tfrac{\pi u}{\lambda}& {\rm dense},\\
\tfrac{\pi u}{3\lambda}& {\rm dilute}.
\end{cases}
\ee
\end{Conjecture}

The geometric anisotropy angle $\vartheta$~\cite{KP87} (and thus $q$ up to the aspect ratio $\delta$) is determined by $u$ measured in units of the
crossing parameter $\lambda$ or $3\lambda$ for a given dense or dilute model (namely for fixed $p,p'$). The isotropic points are $u=\lambda/2$ and $u=3\lambda/2$
in the dense and dilute models. In both cases, we see that the traces are invariant under the transformation $\gamma \to \gamma+2\pi$. In the dense case, the traces are also invariant under the transformation $\gamma \to \gamma+\pi$ up to a sign $(-1)^M$, a feature that is not shared by the dilute loop model. 

We defer discussion of the evidence supporting this conjecture to the conclusion and after we have fully explored the ramifications of this central assertion.

We now take the scaling limit and define the scaled coefficients and partition functions
\be
\mathcal C_{d,j} =  \lim_{\substack{M,N \to \infty\\M/N \to \delta}} \eE^{MN f_\text{bulk}(u)} C_{d,j},
\qquad
\mathcal Z^{\textrm{\tiny$(h,v)$}} = \lim_{\substack{M,N \to \infty\\M/N \to \delta}} \eE^{MN f_\text{bulk}(u)} Z^{\textrm{\tiny$(h,v)$}}(\alpha).
\ee
Treating $\frac{p}{p'}$ as a quasi-continuous variable, we can then compute $\mathcal C_{d,j}$ for all values of $\frac{p}{p'}\in (0,1)$. For the dilute model, we find
\begin{alignat}{2}
\mathcal C_{d,j} &= \frac1{2\pi}\frac{(q\bar q)^{-c/24}}{(q)_\infty (\bar q)_\infty} \sum_{\ell = -\infty}^
\infty  \int_{0}^{2\pi} \dd\gamma\, \eE^{\ir \gamma j}\, q^{\Delta_{(\gamma-2\pi \ell)/\pi,d/2}} \bar q^{\Delta_{(\gamma-2\pi \ell)/\pi,-d/2}} 
\nonumber\\&=
\frac1{2\pi} \frac{(q\bar q)^{-c/24}}{(q)_\infty (\bar q)_\infty}  \int_{-\infty}^{\infty} \dd\gamma\, \eE^{\ir \gamma j}\, q^{\Delta_{\gamma/\pi,d/2}} \bar q^{\Delta_{\gamma/\pi,-d/2}}
\nonumber\\
&= \frac1{2\pi} \frac{(q\bar q)^{-c/24}}{(q)_\infty (\bar q)_\infty}\exp\big[\tfrac{\ir \pi}{2pp'}(\tau-\bar \tau)(\tfrac{p^2 d^2}4-(p'-p)^2)\big] \int_{-\infty}^{\infty} \dd\gamma\, \exp\Big[\ir \gamma(j-\tfrac d2(\tau+\bar\tau))+\frac{\ir \gamma^2 p'}{2 p \pi} (\tau-\bar\tau)\Big]
\nonumber\\
& = \Big(\frac{p}{4p'\tau_i}\Big)^{1/2} \frac{(q\bar q)^{-1/24}}{(q)_\infty (\bar q)_\infty} \exp\Big[-\tfrac {p\pi}{4p' \tau_i} \big(j^2 + d^2 (\tau_r^2 + \tau_i^2) -2 d\, j\, \tau_r\big)\Big]
\nonumber\\&\label{eq:Cdj2}
= \mathcal Z_{d,j}\big(\tfrac p{4p'}\big)
\end{alignat}
where 
\be
\mathcal Z_{m,m'}(g) = \Big(\frac g{\tau_i}\Big)^{1/2} \frac1{\eta(q) \eta(\bar q)} \exp\Big[-\frac {\pi g}{ \tau_i} \big|m \tau - m'\big|^2\Big]
\ee
and $\eta(q)=q^{1/24} (q)_\infty$ is the Dedekind eta function.
Although the intermediate lines in \eqref{eq:Cdj2} appear to depend separately on $p$ and $p'$, it is clear from the first line that the dependence on $\lambda$ 
is in fact only through the ratio $\tfrac{p}{p'}$ appearing in the conformal weights $\Delta_{r,s}=\Delta_{r,s}^{p,p'}$. For the dense loop model, we similarly find
\begin{alignat}{2}
\mathcal C_{d,j} &= \frac1{2\pi}\frac1{(q)_\infty (\bar q)_\infty} \sum_{\ell = -\infty}^
\infty  \int_{0}^{2\pi} \dd\gamma\, \eE^{\ir \gamma j}\, \Big[q^{\Delta_{(\gamma-2\pi \ell)/\pi,d/2}} \bar q^{\Delta_{(\gamma-2\pi \ell)/\pi,-d/2}} 
\nonumber\\&\hspace{5.5cm}+ (-1)^M q^{\Delta_{(\gamma-2\pi \ell-\pi)/\pi,d/2}} \bar q^{\Delta_{(\gamma-2\pi \ell-\pi)/\pi,-d/2}}\Big] 
\nonumber\\&=
 \frac1{2\pi}\frac{1+(-1)^{M+j}}{(q)_\infty (\bar q)_\infty} \int_{-\infty}^{\infty} \dd\gamma\, \eE^{\ir \gamma j} q^{\Delta_{\gamma/\pi,d/2}} \bar q^{\Delta_{\gamma/\pi,-d/2}} = (1+(-1)^{M+j})\, \mathcal Z_{d,j}\big(\tfrac p{4p'}\big).
\end{alignat}
Thus in both cases, we have
\be
\mathcal Z_{\textrm{dense}}^{\textrm{\tiny$(h,v)$}} = \mathcal Z_{\textrm{dilute}}^{\textrm{\tiny$(h,v)$}} = 
\sum_{d \in 2\mathbb Z+h} \sum_{j \in 2\mathbb Z+v} \, 2\,T_{d \wedge j} (\tfrac \alpha2)\, \mathcal Z_{d,j}\big(\tfrac p{4p'}\big).
\ee
We note that the functions $\mathcal Z_{m,m'}(g)$ are well-known in the Coulomb gas formalism~\cite{FSZ87,RS01}. 

The equality of these conformal partition functions provides compelling evidence that the corresponding dense and dilute loop models, for $\frac p{p'} \in (0,1)$, are in the same universality classes. Moreover, under the action of the modular group with generators $T: \tau\mapsto 1+\tau$ and $S: \tau\mapsto -\tfrac 1\tau$, we find
\be
\mathcal Z_{d,j}(g,\tau+1) = \mathcal Z_{d,j-d}(g,\tau),
\qquad 
\mathcal Z_{d,j}(g,-\tfrac1 \tau) = \mathcal Z_{j,-d}(g,\tau),
\ee 
so the partition functions satisfy 
\be
\label{eq:Z.modular.props}
\begin{array}{c}
\mathcal Z^{\textrm{\tiny$(0,0)$}}(\tau+1) = \mathcal Z_{\textrm{tor}}^{\textrm{\tiny$(0,0)$}}(\tau), 
\\[0.2cm]
\mathcal Z^{\textrm{\tiny$(0,1)$}}(\tau+1) = \mathcal Z_{\textrm{tor}}^{\textrm{\tiny$(0,1)$}}(\tau), 
\\[0.2cm]
\mathcal Z^{\textrm{\tiny$(1,0)$}}(\tau+1) = \mathcal Z_{\textrm{tor}}^{\textrm{\tiny$(1,1)$}}(\tau),
\\[0.2cm]
\mathcal Z^{\textrm{\tiny$(1,1)$}}(\tau+1) = \mathcal Z_{\textrm{tor}}^{\textrm{\tiny$(1,0)$}}(\tau),
\end{array}
\qquad
\begin{array}{c}
\mathcal Z^{\textrm{\tiny$(0,0)$}}(-\tfrac1\tau) = 
\mathcal Z^{\textrm{\tiny$(0,0)$}}(\tau),
\\[0.2cm]
\mathcal Z^{\textrm{\tiny$(0,1)$}}(-\tfrac1\tau) = 
\mathcal Z^{\textrm{\tiny$(1,0)$}}(\tau),
\\[0.2cm]
\mathcal Z^{\textrm{\tiny$(1,0)$}}(-\tfrac1\tau) = 
\mathcal Z^{\textrm{\tiny$(0,1)$}}(\tau),
\\[0.2cm]
\mathcal Z^{\textrm{\tiny$(1,1)$}}(-\tfrac1\tau) = 
\mathcal Z^{\textrm{\tiny$(1,1)$}}(\tau).
\end{array}
\ee 
As a result, the partition function $\mathcal Z^{\textrm{\tiny$(0,0)$}}$ is modular invariant, whereas $\mathcal Z^{\textrm{\tiny$(0,1)$}}$, $\mathcal Z^{\textrm{\tiny$(1,0)$}}$ and $\mathcal Z^{\textrm{\tiny$(1,1)$}}$ are covariant under the modular group. More specifically, in accord with modular covariance, the action of the generators $S$ and $T$ on the ordered basis $\{\mathcal Z^{\textrm{\tiny$(0,0)$}},\mathcal Z^{\textrm{\tiny$(0,1)$}},\mathcal Z^{\textrm{\tiny$(1,0)$}},\mathcal Z^{\textrm{\tiny$(1,1)$}} \}$ yields a four-dimensional representation of the modular group
\be
\mathsf{S}=\mbox{\scriptsize $\begin{pmatrix} 1&0&0&0\\  0&0&1&0\\  0&1&0&0\\  0&0&0&1\end{pmatrix}$},\quad \mathsf{T}=\mbox{\scriptsize $\begin{pmatrix} 1&0&0&0\\  0&1&0&0\\  0&0&0&1\\  0&0&1&0\end{pmatrix}$},\qquad \mathsf{S}^2=(\mathsf{S}\mathsf{T})^3=\mathsf{I}.
\ee
Under this action, $\mathsf{T}$ is additionally an involution, namely it satisfies $\mathsf{T}^2=\mathsf{I}$. 
Lastly, we note that the full partition function of the dilute loop model defined in \eqref{eq:fullZ} is also modular invariant.

%
\section{Partition functions as sesquilinear forms in Verma characters}\label{sec:Verma}
%

In this section, we show that the four conformal partition functions can be written as sums of traces of scaled transfer matrices, for all four types of boundary conditions. This then allows us to write down character formulas for these partition functions. {Having established in \cref{sec:scaling.limit} that the partition functions of the two loop models are equal, here we choose to focus on the dilute model for simplicity.}
Its conformal partition functions are
\be
\mathcal Z^{\textrm{\tiny$(h,v)$}}_{\textrm{dilute}} = \sum_{d \in 2 \mathbb Z+h} \sum_{j \in 2 \mathbb Z+v} 2\, T_{d \wedge j} (\tfrac \alpha 2)\, \mathcal C_{d,j}\, .
\ee
For $d \neq 0$, we write $j=x+2dy$ and sum over $x \in \{0,1,\dots,2d-1\}$ and $y \in \mathbb Z$. Using the property $T_{d \wedge j} = T_{d \wedge (j+d)} = T_{d \wedge (j+2d)}$, we find
\be
\label{eq:Z.and.M}
\mathcal Z^{\textrm{\tiny$(h,v)$}}_{\textrm{dilute}} = \delta_{h,0} \sum_{j \in 2\mathbb Z+v} 2\,T_{j} (\tfrac \alpha 2)\, \mathcal C_{0,j} + \frac12 \sum_{d\in \mathbb Z^\times} \sum_{x=0}^{2d-1} \big(1+(-1)^{d+h}\big)\big(1+(-1)^{x+v}\big) T_{d\wedge x}(\tfrac \alpha 2) \mathcal M_{d,x}
\ee
where $\mathbb Z^\times=\mathbb Z\,\backslash \{0\}$ and 
\be
\mathcal M_{d,x}=\sum_{y \in \mathbb Z} \mathcal C_{d,x+2dy}\,.
\ee
{Here and below, we understand sums from $0$ to $2d-1$ for $d<0$ as simply running over the set of integers $\{0,-1,-2,\dots,2d-1\}$.}
Let us also define
\be
\mathcal Y_{m,d} = \sum_{j \in \mathbb Z} \omega_{m,d}^{-j}\, \mathcal C_{d,j} \qquad \textrm{where}\qquad \omega_{m,d} = \eE^{\ir \pi m/d}.
\ee
Because $\omega_{m,d}^{j+2d} = \omega_{m,d}^{j}$, we again write $j = x+2dy$ and sum over $x$ and $y$ to find 
\be
\label{eq:Y.and.M}
\mathcal Y_{m,d} = \sum_{x=0}^{2d-1} \omega_{m,d}^{-x} \mathcal M_{d,x}\, , \qquad 
d \neq 0.
\ee
The following proposition relates the quantities appearing in \eqref{eq:Z.and.M} and \eqref{eq:Y.and.M}.
\begin{Proposition}
\label{prop:MY}
For $d \neq 0$, we have
\be
\label{eq:MY}
\sum_{x=0}^{2d-1}  (1+(-1)^{x+v})\, T_{d \wedge x}(\tfrac \alpha 2)\, \mathcal M_{d,x} = \sum_{m=0}^{2d-1} \GA{m}{d}^{\textrm{\tiny$(v)$}} \, \mathcal Y_{m,d}\,,
\ee
where
\be
\label{eq:Lambdas}
\GA{m}{d}^{\textrm{\tiny$(v)$}} = \frac1{|2d|}
\sum_{j=0}^{d-1} \Big(1+(-1)^{j+v}+(-1)^m+(-1)^{m+j+d+v}\Big) \eE^{\ir \pi j m /d} \, T_{d\wedge j}(\tfrac \alpha 2).
\ee
\end{Proposition}
\proof
The proof is straightforward:
\begin{alignat}{2}
\sum_{m=0}^{2d-1} \GA{m}{d}^{\textrm{\tiny$(v)$}}\, \mathcal Y_{m,d}&= \frac 1{|2d|} \sum_{x=0}^{2d-1} \sum_{j=0}^{d-1} T_{d \wedge j}(\tfrac \alpha 2) \mathcal M_{d,x} \sum_{m=0}^{d-1} \eE^{\ir \pi m(j-x)/d}\big(1+(-1)^{j+v}+(-1)^m+(-1)^{m+j+v+d}\big)
\nonumber\\&= \sum_{x=0}^{2d-1} \sum_{j=0}^{d-1} T_{d \wedge j}(\tfrac \alpha 2) \mathcal M_{d,x} \big(\delta_{x,j}(1+(-1)^{j+v}) + \delta_{x,j+d}(1+(-1)^{j+v+d})\big)
\nonumber\\&= \sum_{x=0}^{2d-1}  (1+(-1)^{x+v})\, T_{d \wedge x}(\tfrac \alpha 2)\, \mathcal M_{d,x}\,.
\end{alignat}
\eproof

\noindent Because
\be
\mathcal Y_{m,d} = \lim_{\substack{M,N \to \infty\\M/N \to \delta}} \eE^{MN f_\mathrm{bulk}(u)}\,\mathrm{tr}_{\raisebox{-0.05cm}{\tiny$\repW_{N,d,\omega_{m,d}}$}} \Tb(u)^M,
\ee
the right side of \eqref{eq:traceConjectures} gives us a character formula for $\mathcal Y_{m,d}$. Using this with \eqref{eq:Z.and.M} and \eqref{eq:MY}, we obtain
\begin{alignat}{2}
\mathcal Z^{\textrm{\tiny$(h,v)$}}_{\textrm{dilute}} &= \frac{(q\bar q)^{-c/24}}{(q)_\infty(\bar q)_\infty} 
\Big[\delta_{h,0} \!\!\sum_{\ell=-\infty}^\infty \!(q\bar q)^{\Delta_{\frac \gamma \pi -2\ell,0}} 
+ 2\sum_{d=1}^\infty\, \delta_{d \equiv h\,\textrm{mod}\,2} \sum_{m=0}^{2d-1}\! \GA{m}{d}^{\textrm{\tiny$(v)$}} \!\sum_{\ell = -\infty}^\infty \!q^{\Delta_{\frac md-2\ell,\frac d2}} \bar q^{\Delta_{\frac md-2\ell,-\frac d2}}\Big],
\end{alignat}
where we used $\Gamma^{\textrm{\tiny$(v)$}}_{-d,m} = \Gamma^{\textrm{\tiny$(v)$}}_{d,m}$ and combined the identical contributions coming from $d<0$ and $d>0$.
This is the final formula for the partition functions $\mathcal Z^{\textrm{\tiny$(h,v)$}}_{\textrm{dilute}}$ as a sesquilinear form in Verma characters. The full partition function of the loop model is obtained by summing over $h,v \in \{0,1\}$: 
\begin{alignat}{2}
\mathcal Z_{\textrm{dilute}} &= \frac{(q\bar q)^{-c/24}}{(q)_\infty(\bar q)_\infty} \Big[\sum_{\ell=-\infty}^\infty (q\bar q)^{\Delta_{\frac \gamma \pi -2\ell,0}} 
+ 2\sum_{d=1}^\infty  \sum_{m=0}^{d-1} \GA{m}{d} \sum_{\ell = -\infty}^\infty q^{\Delta_{\frac {2m}d-2\ell,\frac d2}} \bar q^{\Delta_{\frac {2m}d-2\ell,-\frac d2}}\Big]
\label{fullPF}
\end{alignat}
where
\be
\GA{m}{d} =
\frac1{{d}}\sum_{j=1}^{d} \eE^{2\ir \pi j m /d} T_{d\wedge j}(\tfrac \alpha 2)=\frac1{d}\sum_{j=1}^{d} \eE^{2\ir \pi j m/d} \cos((d\wedge j)\gamma).\label{ourLambda}
\ee
In \cref{NumberTheory}, we prove that
\be
\mathcal Z_{\textrm{dilute}} = \hat Z(\tfrac p{p'},\tfrac{\gamma}\pi)\big|_{q \leftrightarrow \bar q}
\qquad \textrm{and}\qquad
\GA{m}{d} = \tfrac12 \Lambda(d,\tfrac{d}{m\wedge d}), 
\label{theidentity}
\ee
where the functions $\hat Z(g,e_0)$ and $\Lambda(M,N)$ are defined in equations (3.21) and (3.24) of \cite{FSZ87}, respectively. This implies that $\mathcal Z_{\textrm{dilute}}$ coincides with the partition function of the $O(n)$ model studied in that paper.
 
%
\section{Partition functions for $\boldsymbol{\alpha = 2}$ as affine $\boldsymbol{u(1)}$ sesquilinear forms}\label{sec:alpha=2}
%

In this section, we consider the special case $\alpha=2$. First, in \cref{sec:classicalPFs}, we express the $(h,v)$ partition functions in terms of generalized Coulomb partition functions 
$\mathcal Z_{\text{Coul}}^{\textrm{\tiny$(h,v)$}}(\tfrac p{p'})$. The rest of the section is devoted to re-expressing, in root of unity cases, these partition functions as sesquilinear forms in affine $u(1)$ characters. 

\subsection{Classical partition functions}
\label{sec:classicalPFs}
For $\alpha = 2$, we use \eqref{eq:Cdj1} and $T_n(1) = 1$ for $n \in \mathbb Z_{\ge 0}$ to simplify the partition functions of the dense and dilute loop models to
\begin{subequations}
\begin{alignat}{2}
Z_{\textrm{dense}}^{\textrm{\tiny$(h,v)$}}(\alpha = 2)&= 
\sum_{\substack{-N\le \,d\, \le N\\ d\, \equiv\, h\,\textrm{mod}\,2}} 
\textrm{tr}_{\raisebox{-0.05cm}{\tiny$\repW_{N,d,1}$}} \Tb(u)^M,
\\
Z_{\textrm{dilute}}^{\textrm{\tiny$(h,v)$}}(\alpha = 2) &= 
\sum_{\substack{-N\le \,d\, \le N\\ d\, \equiv\, h\,\textrm{mod}\,2}} 
\Big(\textrm{tr}_{\raisebox{-0.05cm}{\tiny$\repW_{N,d,1}$}} \Tb(u)^M 
+(-1)^v\,
\textrm{tr}_{\raisebox{-0.05cm}{\tiny$\repW_{N,d,-1}$}} \Tb(u)^M\Big),
\end{alignat}
\end{subequations} 
where the standard modules with negative defect numbers are defined as $\repW_{N,-d,\omega} = \repW_{N,d,\omega^{-1}}$. The same argument was used previously in \cite{MDKP17,MDKP23} for the special case $\beta = 1$. From \eqref{eq:lambda.pp'},
the corresponding conformal partition functions are identical and given by
\begin{alignat}{2}
\label{hvConfPFs}
\mathcal Z^{\textrm{\tiny$(h,v)$}}\big|_{\alpha=2}&= \frac{(q \bar q)^{-c/24}}{(q)_\infty(\bar q)_\infty}\,
\sum_{\ell\in \mathbb Z} \sum_{d \in 2\mathbb Z+h} (-1)^{v\ell} q^{\Delta_{- \ell, d/2}} \bar q^{\Delta_{-\ell, -d/2}} 
\nonumber\\
&= \frac{(q \bar q)^{-c/24}}{(q)_\infty(\bar q)_\infty} 
\sum_{r,s\in \mathbb Z} (-1)^{v r} q^{\Delta_{r, s+h/2}} \bar q^{\Delta_{r, -s-h/2}}\\
&=\frac{1}{\eta(q) \eta(\bar q)} \sum_{r,s-h/2\in\mathbb Z}  (-1)^{v r} q^{\frac{(p'r-ps)^2}{4pp'}} \bar q^{\frac{(p'r+ps)^2}{4pp'}}
={\cal Z}_{\text{Coul}}^{\textrm{\tiny$(h,v)$}}(\tfrac{p}{p'}),\nonumber
\end{alignat}
where $c$ and $\Delta_{r,s}$ are given by \eqref{LogMinConfWts} and 
\be
{\cal Z}_{\text{Coul}}^{\textrm{\tiny$(h,v)$}}(g) = \frac{1}{\eta(q)\eta(\bar q)}\sum_{r,s-h/2\in{\Bbb Z}} (-1)^{v r}q^{(r/\sqrt{g}-s\sqrt{g})^2/4}\bar q^{(r/\sqrt{g}+s\sqrt{g})^2/4}
\ee
is a generalization of the Coulomb partition function ${\cal Z}_{\text{Coul}}(g)={\cal Z}_{\text{Coul}}^{\textrm{\tiny$(0,0)$}}(g)$. 
Moreover, in the dilute case, the full conformal partition function at $\alpha = 2$ is
\be
\mathcal Z\big|_{\alpha=2} = \sum_{h,v \in\{0,1\}}\mathcal Z^{\textrm{\tiny$(h,v)$}}\big|_{\alpha=2}
= {\frac{2}{\eta(q) \eta(\bar q)}} 
\sum_{r \in 2 \mathbb Z} \sum_{s \in \mathbb Z/2} q^{\frac{(p'r-ps)^2}{4pp'}} \bar q^{\frac{(p'r+ps)^2}{4pp'}} 
= 2\,{\cal Z}_{\text{Coul}}(\tfrac{p}{4p'}).
\ee

\subsection[Partition functions as sesquilinear forms in affine $u(1)$ characters]{Partition functions as  sesquilinear forms in affine $\boldsymbol{u(1)}$ characters}

For $\alpha=2$ and $\lambda/\pi\in\mathbb Q$, that is $p$ and $p'$ coprime integers, the conformal spectra of the dense and dilute logarithmic minimal models 
coincide with that of the 6-vertex and Izergin--Korepin 19-vertex models respectively. At these points, the 6-vertex model exhibits an $s\ell_2$ loop algebra symmetry~\cite{DFM2001}. These latter vertex models also admit an extended affine $u(1)$ symmetry. In this way, this inherited symmetry is reflected in the partition functions of the loop models. 

As a consequence, the partition functions written as sesquilinear forms in Verma characters can be reorganized into finite sesquilinear forms in affine $u(1)$ characters. These are characters of $u(1)$ coset CFTs~\cite{PR2011} with effective central charge $c_\text{eff}=1$, defined as
\begin{equation}
{\varkappa^n_j(q)} = \varkappa_j^{n}(1,q), \qquad \varkappa_j^{n}(z,q)=\frac{\Theta_{j,n}(q,z)}{q^{1/24}(q)_\infty}=\frac{q^{-1/24}}{(q)_\infty} \sum_{k\in{\mathbb Z}} z^kq^{(j+2kn)^2/4n},
\label{eq:u1chars}
\end{equation}
at level $n=p p'$ with associated conformal weights
\be
\Delta^n_j=\mbox{min}\big[\tfrac{j^2}{4n},\tfrac{(2n-j)^2}{4n}\big],\qquad j=0,1,\ldots,2n.
\ee
These affine characters satisfy the symmetries
\be
\varkappa_{j+2n}^n(z,q)=z^{-1}\varkappa_{j}^n(z,q), \qquad 
\varkappa_{2n-j}^n(z,q)=z^{-1}\varkappa_{j}^n(z^{-1},q).
\label{folding1}
\ee
The specializations of the affine $u(1)$ characters relevant here have $z=\pm 1$ 
with periodicity and folding relations
\begin{subequations}
\begin{alignat}{3}
&
\varkappa_{j+2n}^n(1,q) = \varkappa_{j}^n(1,q),\qquad &&
\varkappa_{j+4n}^n(-1,q) = \varkappa_{j}^n(-1,q),\label{periods}\\[4pt]
&
\varkappa_{2n-j}^n(1,q)=\varkappa_j^n(1,q),\qquad&&
\varkappa_{2n-j}^n(-1,q)=-\varkappa_j^n(-1,q),\qquad&
\varkappa_{4n-j}^n(-1,q)=\varkappa_j^n(-1,q),
\label{folding2}
\end{alignat}
\end{subequations}
as well as the relation
\be
\varkappa_n^n(-1,q)=0.
\ee
Following from a ${\mathbb Z}_2$ {folding}, the affine $u(1)$ characters also satisfy the intertwining relations 
\begin{alignat}{2}
\varkappa_j^n(\pm1,q)&=\frac{q^{-1/24}}{(q)_\infty} \sum_{k\in{\mathbb Z}} (\pm 1)^k q^{\frac{(j+2kn)^2}{4n}}
=\frac{q^{-1/24}}{(q)_\infty} \Big[\sum_{k\in{\mathbb Z}} q^{\frac{(2j+8kn)^2}{16n}}\pm \sum_{k\in{\mathbb Z}} q^{\frac{(2j-4n+8kn)^2}{16n}}\Big]\quad\nonumber\\[0.1cm]
&=\varkappa_{2j}^{4n}(q)\pm \varkappa_{4n-2j}^{4n}(q)= \varkappa_{2j}^{4n,\pm}(q)\label{varkappapm}
\end{alignat}
where we introduce the notation
\be
\label{eq:varkappa.n.4n}
\varkappa_j^{n,\pm}(q)=\varkappa_{j}^{n}(q)\pm\varkappa_{n-j}^{n}(q).
\ee

Writing $q=\eE^{2\pi \ir \tau}$, the action of the modular group on affine $u(1)$ characters yields a $2n$-dimensional representation
 \begin{subequations}
\begin{alignat}{3}
{\cal T}:&\qquad 
\varkappa^n_j(\eE^{2 \pi \ir (\tau+1)}) &&=
\exp\!\big[2\pi\ir\,(\Delta_j^n-\tfrac1{24})
\big] \varkappa^n_j(\eE^{2\pi \ir\tau}),\\[0.1cm]
{\cal S}:&\qquad \ \ \varkappa^n_j(\eE^{-2\pi \ir/\tau})
&&=\frac{1}{\sqrt{2n}}\sum_{k=0}^{2n-1} \eE^{-\pi \ir jk/n} \varkappa^n_k(\eE^{2\pi \ir\tau}),
\end{alignat}
\end{subequations}
with
\be
{\cal S}^2=({\cal S}{\cal T})^3={\cal I}.
\ee
The action of the modular transformation $\cal T$ on sesquilinear forms in affine $u(1)$ characters is given by
\be
{\cal T}: 
\quad
\left\{\ 
\begin{array}{l}
\varkappa^n_j(q)\varkappa^n_k(\qbar) \mapsto \exp\!\big[2\pi\ir(\Delta_j^n-\Delta_k^n)\big] \varkappa^n      _j(q) \varkappa^n_k(\qbar),\\[6pt]
\varkappa^{4n,\pm}_{2j}(q)\varkappa^{4n,\pm}_{2k}(\qbar) \mapsto \exp\!\big[2\pi\ir(\Delta_{2j}^{4n}-\Delta_{2k}^{4n})\big] \varkappa^{4n,\pm}_{2j}(q) \varkappa^{4n,\pm}_{2k}(\qbar),\\[6pt]
\varkappa^{4n,\pm}_{2j+1}(q)\varkappa^{4n,\pm}_{2k+1}(\qbar) \mapsto \exp\!\big[2\pi\ir(\Delta_{2j+1}^{4n}-\Delta_{2k+1}^{4n})\big] \varkappa^{4n,\mp}_{2j+1}(q) \varkappa^{4n,\mp}_{2k+1}(\qbar),
\end{array}\right.
\qquad j,k \in \mathbb Z,
\label{Tsign}
\ee
since 
$\exp\!\big[\!\pm\! 2\pi\ir(\Delta_{j}^{4n}-\Delta_{4n-j}^{4n})\big]=(-1)^j$ for $j\in \mathbb{Z}$.

For $\alpha=2$, we now express the modular covariant partition functions \eqref{hvConfPFs} as sesquilinear forms in affine $u(1)$ characters at level $n=p p'$. These partition functions  are denoted by 
\be
\mathcal Z^{\textrm{\tiny$(h,\!v)$}}(p,p') = {\cal Z}^{\textrm{\tiny$(h,\!v)$}}\big|_{\alpha = 2},\  \quad h,v \in \{0,1\}; \qquad 
\mathcal Z(p,p') = \sum_{h,v\in \{0,1\}}\mathcal Z^{\textrm{\tiny$(h,\!v)$}}(p,p').
\ee
By splitting the infinite sum in \eqref{hvConfPFs} into two equal copies and shifting $r$ by $p$ in the second copy, we obtain
\be
\label{eq:ZhvZrs}
\mathcal Z^{\textrm{\tiny$(h,\!v)$}}(p,p')
=\frac 12\sum_{r=0}^{2p-1}\sum_{s=0}^{2p'-1} (-1)^{vr} {\cal Z}_{r,s}^{\textrm{\tiny$(h,\!v)$}}
\ee
where
\be
{\cal Z}_{r,s}^{\textrm{\tiny$(h,\!v)$}}=\frac1{\eta(q)\eta(\bar q)}\sum_{r'\,\in\, 2p\mathbb Z+r} \,  \sum_{s'\,\in\, 2p'\mathbb Z+s+h/2} \!\! q^{\frac{(p' r'-p s')^2}{4p p'}}
 \Big[\bar q^{\frac{(p' r'+p s')^2}{4pp'}}+(-1)^{pv}\bar q^{\frac{(p' r'+p s'+2pp')^2}{4p p'}}\Big].\label{rshvCoulIdentities}
\ee
\begin{Proposition}
\label{prop:Zrs}
The functions ${\cal Z}_{r,s}^{\textrm{\tiny$(h,\!v)$}}$, given by \eqref{rshvCoulIdentities}, are products of affine $u(1)$ characters
\be
\label{eq:Zrshv.kappa}
{\cal Z}_{r,s}^{\textrm{\tiny$(h,\!v)$}}=\varkappa^n_{p' r-p(s+h/2)}\big((-1)^{pv},q\big)\,\varkappa^n_{p' r+p(s+h/2)}\big((-1)^{pv},\qbar\big), \qquad n = pp'.
\ee 
\end{Proposition}
\proof
We evaluate \eqref{rshvCoulIdentities} by substituting $r' = 2p j+ r$ and $s' = 2p' k+ s$ and summing over $j,k\in \mathbb Z$:
\begin{alignat}{2}
\eta(q)\eta(\bar q) {\cal Z}_{r,s}^{\textrm{\tiny$(h,\!v)$}}
&= \sum_{j,k \in \mathbb Z} q^{(p'r-p(s+h/2)+2pp'(j-k))^2/4pp'}\Big(\bar q^{(p'r+p(s+h/2)+2pp'(j+k))^2/4pp'}
\nonumber\\[-0.2cm]&\hspace{6cm}
+(-1)^{p v}\bar q^{(p'r+p(s+h/2)+2pp'(j+k+1))^2/4pp'}\Big)
\nonumber\\&
= \sum_{j',k \in \mathbb Z} q^{(p'r-p(s+h/2)+2pp'j')^2/4pp'}\Big(\bar q^{(p'r+p(s+h/2)+2pp'(j'+2k))^2/4pp'}
\nonumber\\[-0.2cm]&\hspace{6cm}
+(-1)^{p v}\bar q^{(p'r+p(s+h/2)+2pp'(j'+2k+1))^2/4pp'}\Big)
\nonumber\\&
= \sum_{j',k \in \mathbb Z} (-1)^{p v k}q^{(p'r-p(s+h/2)+2pp'j')^2/4pp'}\bar q^{(p'r+p(s+h/2)+2pp'(j'+k))^2/4pp'}
\nonumber\\&
= \sum_{j',k' \in \mathbb Z} \Big((-1)^{p v j'}q^{(p'r-p(s+h/2)+2pp'j')^2/4pp'}\Big)\Big((-1)^{p v k'}\bar q^{(p'r+p(s+h/2)+2pp'k')^2/4pp'}\Big)
\nonumber\\&
=\eta(q)\eta(\bar q)\,\varkappa^n_{p' r-p(s+h/2)}\big((-1)^{pv},q\big)\,\varkappa^n_{p' r+p(s+h/2)}\big((-1)^{pv},\qbar\big).
\end{alignat}
Here, we changed the summation index $j$ to $j'=j-k$ after the second equality sign and $k$ to $k'=j'+k$ after the fourth equality sign.\eproof

The functions ${\cal Z}_{r,s}^{\textrm{\tiny$(h,v)$}}$ satisfy the symmetries
\be
\label{eq:Zrs.symmetries}
{\cal Z}_{r,s}^{\textrm{\tiny$(h,v)$}} = {\cal Z}_{r+2p,s}^{\textrm{\tiny$(h,v)$}}
= {\cal Z}_{r,s+2p'}^{\textrm{\tiny$(h,v)$}} = {\cal Z}_{-r,-s-h}^{\textrm{\tiny$(h,v)$}}\ ,
\qquad
{\cal Z}_{r,2p'-s-h}^{\textrm{\tiny$(h,v)$}} = \big({\cal Z}_{r,s}^{\textrm{\tiny$(h,v)$}}\big)^*,
\ee
where in the last identity, complex conjugation consists in interchanging $q$ and $\bar q$.
The four torus partition functions are then written as sesquilinear forms in affine $u(1)$ characters as
\begin{alignat}{2}
\mathcal Z^{\textrm{\tiny$(h,\!v)$}}(p,p')
&=\sum_{r=0}^{p-1}\sum_{s=0}^{2p'-1} (-1)^{vr} {\cal Z}_{r,s}^{\textrm{\tiny$(h,\!v)$}}
\nonumber\\&=\sum_{r=0}^{p-1}\sum_{s=0}^{2p'-1} (-1)^{vr} \varkappa^n_{p' r-p(s+h/2)}\big((-1)^{pv},q\big)\,\varkappa^n_{p' r+p(s+h/2)}\big((-1)^{pv},\qbar\big).
\label{eq:Zhv.u1char}
\end{alignat}

Using the above results, we also rewrite the full dilute partition function \eqref{eq:fullZ} in terms of the characters $\varkappa^n_j(\pm 1,q)$ as 
\begin{alignat}{2}
\label{eq:Z.u1char}
\mathcal Z(p,p')&=\sum_{v=0,1}\sum_{r=0}^{p-1}\sum_{s=0,\frac 12,1,\ldots}^{2p'-\frac 12} (-1)^{vr} 
 \varkappa^n_{p'r-ps}\big((-1)^{pv},q\big)\varkappa^n_{p'r+ps}\big((-1)^{pv},\qbar\big).
\end{alignat}
For $p$ even, this simplifies to
\begin{alignat}{2}
\label{eq:Z.u1char.peven}
\mathcal Z(p,p')\big|_{\text{$p$ even}}&=2\sum_{r=0,2,4.\ldots}^{p-1}\sum_{s=0,\frac 12,1,\ldots}^{2p'-\frac 12} 
 \varkappa^n_{p'r-ps}(q)\varkappa^n_{p'r+ps}(\qbar).
\end{alignat}
The expressions \eqref{eq:Zhv.u1char}, \eqref{eq:Z.u1char} and \eqref{eq:Z.u1char.peven} can be equivalently rewritten in terms of the characters $\varkappa^{4n,\pm}_{j}(q)$ using~\eqref{eq:varkappa.n.4n}. In the next section, we introduce Bezout conjugates and use them to write these sesquilinear forms in affine $u(1)$ characters using affine $u(1)$ indices.

\subsection[Bezout conjugates and sesquilinear forms using affine $u(1)$ indices]{Bezout conjugates and sesquilinear forms using affine $\boldsymbol{u(1)}$ indices}

As is the case for the minimal models ${\cal M}(p,p')$~\cite{FMS}, the CFT description of coset theories typically involves Bezout conjugate integer pairs $j, \overline j\in \mathbb{Z}$.  
However, the sesquilinear forms \eqref{eq:Zhv.u1char} are sums of terms of the form $\varkappa^n_{p' r-p(s+h/2)}\big((-1)^{pv},q\big)\,\varkappa^n_{p' r+p(s+h/2)}\big((-1)^{pv},\qbar\big)$, whose labels are half-integers if $p$ is odd and $h=1$. We now re-express these  in terms of Bezout conjugates $j,\overline j$ and their generalisations to half-integers $j+1/2,\overline {j+1/2}$. We define 
\be
h'=\begin{cases}
\,1& p \textrm{ odd and\ } h=1,\\
\,0&\textrm{otherwise}.
\end{cases}
\ee
From \eqref{periods}, the affine characters have the periodicity property
\be
\varkappa^n_{j+P}\big((-1)^{pv},q\big) = \varkappa^n_{j}\big((-1)^{pv},q\big), 
\qquad
P=\begin{cases}
\, 2n&\mbox{$pv$ even,}\\
\, 4n&\mbox{$pv$ odd.}
\end{cases}
\ee

The Bezout conjugates $j+h'/2$ and $\overline{j+h'/2}$ are integers for $h' =
0$ and half-integers for $h'=1$. They are defined by
\be
\label{eq:Bezout.def}
\begin{array}{c}
j+\tfrac {h'}2 = p' r-p(s+\tfrac h2) \textrm{ mod } P,\\[0.25cm]
\overline{j+\tfrac {h'}2} = p' r+p(s+\tfrac h2)\textrm{ mod } P,\\
\end{array}
\qquad 0\le j+\tfrac{h'}2, \overline{j+\tfrac{h'}2} <P,
\qquad h'=0,1.
\ee
In \cref{app:half.integer.Bezout}, we discuss the properties of these Bezout
conjugates and give a proof of the following proposition. For the integer
values in $\mathbb U_{h'}$ corresponding to $h'=0$, these properties are discussed in \cite{FMS}. 
\begin{Proposition}
\label{prop:Bijection}
This Bezout construction gives a bijection, for $h'\in \{0,1\}$, between the set of Kac labels $\mathbb K$ and the set of $u(1)$ indices $\mathbb U_{h'}$
\begin{subequations}
\label{Bijection}
\begin{alignat}{2}
&\mathbb{K}=\{(r,s)\in\mathbb{Z}^2\,\big|\, 0\le r\le p-1, 0\le s\le \tfrac{P}{n}\, p'-1\}\,,
\\  
&\mathbb{U}_{h'}=\{j+\tfrac {h'}2\!\in \mathbb{Z}+\tfrac {h'}2\,\big|\, 0\le j+\tfrac {h'}2<P\}\,,
\end{alignat}
\end{subequations}
with $r$, $s$ and $j$ related as in \eqref{eq:Bezout.def}.
\end{Proposition}

The Bezout number (or conjugator) $\omega_0\in \mathbb{N}_0$ is defined by
\be
\omega_0=2^{h'}\big(p'r_0+p(s_0+\tfrac h2)\big)
=\begin{cases}
\,\overline 1&h'=0,\\ 
\,2\, \overline{\tfrac12}&h'=1,\label{Bezoutconjugator}
\end{cases}
\ee
where $(r_0,s_0)\in \mathbb{K}$ is uniquely determined by the condition
\be
(\tfrac12)^{h'}\equiv p'r_0-p(s_0+\tfrac h2) \textrm{ mod } P.\label{omegamultpj}
\ee
It is straightforward to show that $\omega_0^2 \equiv 1 \textrm{ mod } P$ and that $\omega_0$ is odd in all cases.
Bezout conjugation is implemented by multiplication with the Bezout conjugator up to shifts of $\tfrac P2$
\be
\overline{j+\tfrac{h'}2} \equiv \omega_0(j+\tfrac{h'}2+\mu \tfrac P2) \equiv \omega_0(j+\tfrac{h'}2)+\mu\tfrac P2\textrm{ mod } P,
\label{shiftconj}
\ee
where $\mu \in \{0,1\}$ is given by
\begin{subequations}
\label{eq:muValues}
\begin{alignat}{2}
p \textrm{ odd}&:\qquad 
\begin{cases}
\,(h,v)=(h,0)\!:
&\mu=0,\\[4pt]
\,(h,v)=(0,1)\!:
&\mu=
\begin{cases}
\,0& r s_0 -r_0 s \textrm{ even,}\\
\,1& r s_0 -r_0 s \textrm{ odd,}
\end{cases}\\[12pt]
\,(h,v)=(1,1)\!:\ \ \   
&\mu=
\begin{cases}
\,0& r-r_0 \textrm{ even},\\
\,1& r-r_0 \textrm{ odd,}
\end{cases}
\end{cases}\label{muValues1}
\\[0.3cm]
p \textrm{ even} &:\qquad \begin{cases}
\,(h,v)=(0,v)\!:
&\mu=0,\quad\\[4pt]
\,(h,v)=(1,v)\!:\ \ \ 
&\mu=\begin{cases} 
\,0& r-r_0 \textrm{ even,}\\
\,1&r-r_0 \textrm{ odd}.\end{cases}\\[14pt]
\end{cases}\ \ \label{muValues2}
\end{alignat}
\end{subequations}
It is easily verified that Bezout conjugation is an involution
\be
\overline{\Big(\,\overline{j+\tfrac{h'}2}\,\Big)}\equiv\omega_0(\omega_0(j+\tfrac{h'}2+\mu \tfrac P2)+\mu \tfrac P2)\equiv j+\tfrac{h'}2 +\mu(1+\omega_0)\tfrac P2\equiv j+\tfrac{h'}2 \textrm{ mod } P
\ee
where we used the fact that $\omega_0$ is odd at the last step.

It follows that, once $(p,p')$ and $(h,v)$ are fixed, the half-period shift $\mu$ is either not needed ($\mu=0$) or applied with a $\mathbb{Z}_2$ column- or checkerboard-alternation in the Kac tables. Example Kac tables of Bezout conjugates $\{j+h'/2,\overline{j+h'/2}\}=
\{p' r\!-\!p(s\!+\!\tfrac h2),p' r\!+\!p(s\!+\!\tfrac h2)\} \textrm{ mod } P$ are shown in \cref{pOddBezoutConjugates}  for $(p,p')=(3,4)$ and $(p,p')=(3,5)$, and in \cref{pEvenBezoutConjugates} for $(p,p')=(4,5)$.

\begin{figure}[htbp] 
\begin{center}
\psset{unit=0.95cm}
\begin{pspicture}(-0.5,-.3)(7.5,8.5)
\psframe[linewidth=0pt,fillstyle=solid,fillcolor=lightlightblue](0,0)(7,9)
\multirput(0,0)(0,2){5}{\multirput[bl](0,0)(2,0){4}{\psframe[linewidth=0pt,fillstyle=solid,fillcolor=lightpurple](0,0)(1,1)}}
\multirput[bl](1,1)(0,2){4}{\psframe[linewidth=0pt,fillstyle=solid,fillcolor=darkpurple](0,0)(1,1)}
\multirput[bl](3,1)(0,2){4}{\psframe[linewidth=0pt,fillstyle=solid,fillcolor=darkpurple](0,0)(1,1)}
\multirput[bl](5,1)(0,2){4}{\psframe[linewidth=0pt,fillstyle=solid,fillcolor=darkpurple](0,0)(1,1)}
 \psgrid[gridlabels=0pt,subgriddiv=1](0,0)(7,9)

\rput(.5,.5){\small $0,\!0$}
\rput(1.5,.5){\small $4,\!4$}
\rput(2.5,.5){\small $8,\!8$}
\rput(3.5,.5){\small $12,\!12$}
\rput(4.5,.5){\small $16,\!16$}
\rput(5.5,.5){\small $20,\!20$}
\rput(6.5,.5){\small $0,\!0$}

\rput(.5,1.5){\small $21,\!3$}
\rput(1.5,1.5){\small $1,\!7$}
\rput(2.5,1.5){\small $5,\!11$}
\rput(3.5,1.5){\small $9,\!15$}
\rput(4.5,1.5){\small $13,\!19$}
\rput(5.5,1.5){\small $17,\!23$}
\rput(6.5,1.5){\small $21,\!3$}

\rput(.5,2.5){\small $18,\!6$}
\rput(1.5,2.5){\small $22,\!10$}
\rput(2.5,2.5){\small $2,\!14$}
\rput(3.5,2.5){\small $6,\!18$}
\rput(4.5,2.5){\small $10,\!22$}
\rput(5.5,2.5){\small $14,\!2$}
\rput(6.5,2.5){\small $18,\!6$}

\rput(  .5,3.5){\small $15,\!9$}
\rput(1.5,3.5){\small $19,\!13$}
\rput(2.5,3.5){\small $23,\!17$}
\rput(3.5,3.5){\small $3,\!21$}
\rput(4.5,3.5){\small $7,\!1$}
\rput(5.5,3.5){\small $11,\!5$}
\rput(6.5,3.5){\small $15,\!9$}

\rput(  .5,4.5){\small $12,\!12$}
\rput(1.5,4.5){\small $16,\!16$}
\rput(2.5,4.5){\small $20,\!20$}
\rput(3.5,4.5){\small $0,\!0$}
\rput(4.5,4.5){\small $4,\!4$}
\rput(5.5,4.5){\small $8,\!8$}
\rput(6.5,4.5){\small $12,\!12$}

\rput(  .5,5.5){\small $9,\!15$}
\rput(1.5,5.5){\small $13,\!19$}
\rput(2.5,5.5){\small $17,\!23$}
\rput(3.5,5.5){\small $21,\!3$}
\rput(4.5,5.5){\small $1,\!7$}
\rput(5.5,5.5){\small $5,\!11$}
\rput(6.5,5.5){\small $9,\!15$}

\rput(  .5,6.5){\small $6,\!18$}
\rput(1.5,6.5){\small $10,\!22$}
\rput(2.5,6.5){\small $14,\!2$}
\rput(3.5,6.5){\small $18,\!6$}
\rput(4.5,6.5){\small $22,\!10$}
\rput(5.5,6.5){\small $2,\!14$}
\rput(6.5,6.5){\small $6,\!18$}

\rput(  .5,7.5){\small $3,\!21$}
\rput(1.5,7.5){\small $7,\!1$}
\rput(2.5,7.5){\small $11,\!5$}
\rput(3.5,7.5){\small $15,\!9$}
\rput(4.5,7.5){\small $19,\!13$}
\rput(5.5,7.5){\small $23,\!7$}
\rput(6.5,7.5){\small $3,\!21$}

\rput(  .5,8.5){\small $0,\!0$}
\rput(1.5,8.5){\small $4,\!4$}
\rput(2.5,8.5){\small $8,\!8$}
\rput(3.5,8.5){\small $12,\!12$}
\rput(4.5,8.5){\small $16,\!16$}
\rput(5.5,8.5){\small $20,\!20$}
\rput(6.5,8.5){\small $0,\!0$}

{\color{blue}
 \rput(.5,-.5){$0$}
 \rput(1.5,-.5){$1$}
 \rput(2.5,-.5){$2$}
 \rput(3.5,-.5){$3$}
 \rput(4.5,-.5){$4$}
 \rput(5.5,-.5){$5$}
 \rput(6.5,-.5){$6$}
 \rput(7.5,-.5){$r$}
 \rput(-.5,.5){$0$}
 \rput(-.5,1.5){$1$}
 \rput(-.5,2.5){$2$}
 \rput(-.5,3.5){$3$}
 \rput(-.5,4.5){$4$}
 \rput(-.5,5.5){$5$}
 \rput(-.5,6.5){$6$}
 \rput(-.5,7.5){$7$}
 \rput(-.5,8.5){$8$}
 \rput(-.5,9.25){$s$}}
 \psline[linewidth=2pt,linecolor=blue](0,0)(3,0)(3,8)(0,8)(0,0)
\end{pspicture}\hspace{.5in}
\begin{pspicture}(-0.5,-.3)(7.5,8.5)
\psframe[linewidth=0pt,fillstyle=solid,fillcolor=lightlightblue](0,0)(7,9)
\multirput(0,0)(0,2){5}{\multirput[bl](0,0)(2,0){4}{\psframe[linewidth=0pt,fillstyle=solid,fillcolor=lightpurple](0,0)(1,1)}}
\multirput[bl](1,1)(0,2){4}{\psframe[linewidth=0pt,fillstyle=solid,fillcolor=darkpurple](0,0)(1,1)}
\multirput[bl](3,1)(0,2){4}{\psframe[linewidth=0pt,fillstyle=solid,fillcolor=darkpurple](0,0)(1,1)}
\multirput[bl](5,1)(0,2){4}{\psframe[linewidth=0pt,fillstyle=solid,fillcolor=darkpurple](0,0)(1,1)}
 \psgrid[gridlabels=0pt,subgriddiv=1](0,0)(7,9)
 
\rput(  .5,.5){\small $\tfrac {93}{2},\!\tfrac {3}{2}$}
\rput(1.5,.5){\small $\tfrac {5}{2},\!\tfrac {11}{2}$}
\rput(2.5,.5){\small $\tfrac {13}{2},\!\tfrac {19}{2}$}
\rput(3.5,.5){\small $\tfrac {21}{2},\!\tfrac {27}{2}$}
\rput(4.5,.5){\small $\tfrac {29}{2},\!\tfrac {35}{2}$}
\rput(5.5,.5){\small $\tfrac {37}{2},\!\tfrac {43}{2}$}
\rput(6.5,.5){\small $\tfrac {45}{2},\!\tfrac {51}{2}$}

\rput(  .5,1.5){\small $\tfrac {87}{2},\!\tfrac {9}{2}$}
\rput(1.5,1.5){\small $\tfrac {95}{2},\!\tfrac {17}{2}$}
\rput(2.5,1.5){\small $\tfrac {7}{2},\!\tfrac {25}{2}$}
\rput(3.5,1.5){\small $\tfrac {15}{2},\!\tfrac {33}{2}$}
\rput(4.5,1.5){\small $\tfrac {23}{2},\!\tfrac {41}{2}$}
\rput(5.5,1.5){\small $\tfrac {31}{2},\!\tfrac {49}{2}$}
\rput(6.5,1.5){\small $\tfrac {39}{2},\!\tfrac {57}{2}$}

\rput(  .5,2.5){\small $\tfrac {81}{2},\!\tfrac {15}{2}$}
\rput(1.5,2.5){\small $\tfrac {89}{2},\!\tfrac {23}{2}$}
\rput(2.5,2.5){\small $\tfrac {1}{2},\!\tfrac {31}{2}$}
\rput(3.5,2.5){\small $\tfrac {9}{2},\!\tfrac {39}{2}$}
\rput(4.5,2.5){\small $\tfrac {17}{2},\!\tfrac {47}{2}$}
\rput(5.5,2.5){\small $\tfrac {25}{2},\!\tfrac {55}{2}$}
\rput(6.5,2.5){\small $\tfrac {33}{2},\!\tfrac {63}{2}$}

\rput(  .5,3.5){\small $\tfrac {75}{2},\!\tfrac {21}{2}$}
\rput(1.5,3.5){\small $\tfrac {83}{2},\!\tfrac {29}{2}$}
\rput(2.5,3.5){\small $\tfrac {91}{2},\!\tfrac {37}{2}$}
\rput(3.5,3.5){\small $\tfrac {3}{2},\!\tfrac {45}{2}$}
\rput(4.5,3.5){\small $\tfrac {11}{2},\!\tfrac {53}{2}$}
\rput(5.5,3.5){\small $\tfrac {19}{2},\!\tfrac {61}{2}$}
\rput(6.5,3.5){\small $\tfrac {27}{2},\!\tfrac {69}{2}$}

\rput(  .5,4.5){\small $\tfrac {69}{2},\!\tfrac {27}{2}$}
\rput(1.5,4.5){\small $\tfrac {77}{2},\!\tfrac {35}{2}$}
\rput(2.5,4.5){\small $\tfrac {85}{2},\!\tfrac {43}{2}$}
\rput(3.5,4.5){\small $\tfrac {93}{2},\!\tfrac {51}{2}$}
\rput(4.5,4.5){\small $\tfrac {5}{2},\!\tfrac {59}{2}$}
\rput(5.5,4.5){\small $\tfrac {13}{2},\!\tfrac {67}{2}$}
\rput(6.5,4.5){\small $\tfrac {21}{2},\!\tfrac {75}{2}$}

\rput(  .5,5.5){\small $\tfrac {63}{2},\!\tfrac {33}{2}$}
\rput(1.5,5.5){\small $\tfrac {71}{2},\!\tfrac {41}{2}$}
\rput(2.5,5.5){\small $\tfrac {79}{2},\!\tfrac {49}{2}$}
\rput(3.5,5.5){\small $\tfrac {87}{2},\!\tfrac {57}{2}$}
\rput(4.5,5.5){\small $\tfrac {95}{2},\!\tfrac {65}{2}$}
\rput(5.5,5.5){\small $\tfrac {7}{2},\!\tfrac {73}{2}$}
\rput(6.5,5.5){\small $\tfrac {15}{2},\!\tfrac {81}{2}$}

\rput(  .5,6.5){\small $\tfrac {57}{2},\!\tfrac {39}{2}$}
\rput(1.5,6.5){\small $\tfrac {65}{2},\!\tfrac {47}{2}$}
\rput(2.5,6.5){\small $\tfrac {73}{2},\!\tfrac {55}{2}$}
\rput(3.5,6.5){\small $\tfrac {81}{2},\!\tfrac {63}{2}$}
\rput(4.5,6.5){\small $\tfrac {89}{2},\!\tfrac {71}{2}$}
\rput(5.5,6.5){\small $\tfrac {1}{2},\!\tfrac {79}{2}$}
\rput(6.5,6.5){\small $\tfrac {9}{2},\!\tfrac {87}{2}$}

\rput(  .5,7.5){\small $\tfrac {51}{2},\!\tfrac {45}{2}$}
\rput(1.5,7.5){\small $\tfrac {59}{2},\!\tfrac {53}{2}$}
\rput(2.5,7.5){\small $\tfrac {67}{2},\!\tfrac {61}{2}$}
\rput(3.5,7.5){\small $\tfrac {75}{2},\!\tfrac {69}{2}$}
\rput(4.5,7.5){\small $\tfrac {83}{2},\!\tfrac {77}{2}$}
\rput(5.5,7.5){\small $\tfrac {91}{2},\!\tfrac {85}{2}$}
\rput(6.5,7.5){\small $\tfrac {3}{2},\!\tfrac {93}{2}$}

\rput(  .5,8.5){\small $\tfrac {45}{2},\!\tfrac {51}{2}$}
\rput(1.5,8.5){\small $\tfrac {53}{2},\!\tfrac {59}{2}$}
\rput(2.5,8.5){\small $\tfrac {61}{2},\!\tfrac {67}{2}$}
\rput(3.5,8.5){\small $\tfrac {69}{2},\!\tfrac {75}{2}$}
\rput(4.5,8.5){\small $\tfrac {77}{2},\!\tfrac {83}{2}$}
\rput(5.5,8.5){\small $\tfrac {85}{2},\!\tfrac {91}{2}$}
\rput(6.5,8.5){\small $\tfrac {93}{2},\!\tfrac {3}{2}$}

{\color{blue}
 \rput(.5,-.5){$0$}
 \rput(1.5,-.5){$1$}
 \rput(2.5,-.5){$2$}
 \rput(3.5,-.5){$3$}
 \rput(4.5,-.5){$4$}
 \rput(5.5,-.5){$5$}
 \rput(6.5,-.5){$6$}
 \rput(7.5,-.5){$r$}
\rput(-.5,.5){$\tfrac 12$}
 \rput(-.5,1.5){$\tfrac 32$}
 \rput(-.5,2.5){$\tfrac 52$}
 \rput(-.5,3.5){$\tfrac 72$}
 \rput(-.5,4.5){$\tfrac 92$}
 \rput(-.5,5.5){$\tfrac {11}2$}
 \rput(-.5,6.5){$\tfrac {13}2$}
 \rput(-.5,7.5){$\tfrac {15}2$}
 \rput(-.5,8.5){$\tfrac {17}2$}
 \rput(-.5,9.25){$s\!+\!\tfrac 12$}}

\psline[linewidth=2pt,linecolor=blue](0,0)(3,0)(3,8)(0,8)(0,0)
\end{pspicture}\\[20pt]
\begin{pspicture}(-0.5,-.3)(7.5,11.2)
\psframe[linewidth=0pt,fillstyle=solid,fillcolor=lightlightblue](0,0)(7,11)
\multirput(0,0)(0,2){6}{\multirput[bl](0,0)(2,0){4}{\psframe[linewidth=0pt,fillstyle=solid,fillcolor=lightpurple](0,0)(1,1)}}
\multirput[bl](1,1)(0,2){5}{\psframe[linewidth=0pt,fillstyle=solid,fillcolor=darkpurple](0,0)(1,1)}
\multirput[bl](3,1)(0,2){5}{\psframe[linewidth=0pt,fillstyle=solid,fillcolor=darkpurple](0,0)(1,1)}
\multirput[bl](5,1)(0,2){5}{\psframe[linewidth=0pt,fillstyle=solid,fillcolor=darkpurple](0,0)(1,1)}
 \psgrid[gridlabels=0pt,subgriddiv=1](0,0)(7,11)

\rput(.5,.5){\small $0,\!0$}
\rput(1.5,.5){\small $5,\!5$}
\rput(2.5,.5){\small $10,\!10$}
\rput(3.5,.5){\small $15,\!15$}
\rput(4.5,.5){\small $20,\!20$}
\rput(5.5,.5){\small $25,\!25$}
\rput(6.5,.5){\small $0,\!0$}

\rput(.5,1.5){\small $27,\!3$}
\rput(1.5,1.5){\small $2,\!8$}
\rput(2.5,1.5){\small $7,\!13$}
\rput(3.5,1.5){\small $12,\!18$}
\rput(4.5,1.5){\small $17,\!23$}
\rput(5.5,1.5){\small $22,\!28$}
\rput(6.5,1.5){\small $27,\!3$}

\rput(.5,2.5){\small $24,\!6$}
\rput(1.5,2.5){\small $29,\!11$}
\rput(2.5,2.5){\small $4,\!16$}
\rput(3.5,2.5){\small $9,\!21$}
\rput(4.5,2.5){\small $14,\!26$}
\rput(5.5,2.5){\small $19,\!1$}
\rput(6.5,2.5){\small $24,\!6$}

\rput(  .5,3.5){\small $21,\!9$}
\rput(1.5,3.5){\small $26,\!14$}
\rput(2.5,3.5){\small $1,\!19$}
\rput(3.5,3.5){\small $6,\!24$}
\rput(4.5,3.5){\small $11,\!29$}
\rput(5.5,3.5){\small $16,\!4$}
\rput(6.5,3.5){\small $21,\!9$}

\rput(  .5,4.5){\small $18,\!12$}
\rput(1.5,4.5){\small $23,\!17$}
\rput(2.5,4.5){\small $28,\!22$}
\rput(3.5,4.5){\small $3,\!27$}
\rput(4.5,4.5){\small $8,\!2$}
\rput(5.5,4.5){\small $13,\!7$}
\rput(6.5,4.5){\small $18,\!12$}

\rput(  .5,5.5){\small $15,\!15$}
\rput(1.5,5.5){\small $20,\!20$}
\rput(2.5,5.5){\small $25,\!25$}
\rput(3.5,5.5){\small $0,\!0$}
\rput(4.5,5.5){\small $5,\!5$}
\rput(5.5,5.5){\small $10,\!10$}
\rput(6.5,5.5){\small $15,\!15$}

\rput(  .5,6.5){\small $12,\!18$}
\rput(1.5,6.5){\small $17,\!23$}
\rput(2.5,6.5){\small $22,\!28$}
\rput(3.5,6.5){\small $27,\!3$}
\rput(4.5,6.5){\small $2,\!8$}
\rput(5.5,6.5){\small $7,\!13$}
\rput(6.5,6.5){\small $12,\!18$}

\rput(  .5,7.5){\small $9,\!21$}
\rput(1.5,7.5){\small $14,\!26$}
\rput(2.5,7.5){\small $19,\!1$}
\rput(3.5,7.5){\small $24,\!6$}
\rput(4.5,7.5){\small $29,\!11$}
\rput(5.5,7.5){\small $4,\!16$}
\rput(6.5,7.5){\small $9,\!21$}

\rput(  .5,8.5){\small $6,\!24$}
\rput(1.5,8.5){\small $11,\!29$}
\rput(2.5,8.5){\small $16,\!4$}
\rput(3.5,8.5){\small $21,\!9$}
\rput(4.5,8.5){\small $26,\!14$}
\rput(5.5,8.5){\small $1,\!19$}
\rput(6.5,8.5){\small $6,\!24$}

\rput(  .5,9.5){\small $3,\!27$}
\rput(1.5,9.5){\small $8,\!2$}
\rput(2.5,9.5){\small $13,\!7$}
\rput(3.5,9.5){\small $18,\!12$}
\rput(4.5,9.5){\small $23,\!17$}
\rput(5.5,9.5){\small $28,\!22$}
\rput(6.5,9.5){\small $3,\!27$}

\rput(  .5,10.5){\small $0,\!0$}
\rput(1.5,10.5){\small $5,\!5$}
\rput(2.5,10.5){\small $10,\!10$}
\rput(3.5,10.5){\small $15,\!15$}
\rput(4.5,10.5){\small $20,\!20$}
\rput(5.5,10.5){\small $25,\!25$}
\rput(6.5,10.5){\small $0,\!0$}

{\color{blue}
 \rput(.5,-.5){$0$}
 \rput(1.5,-.5){$1$}
 \rput(2.5,-.5){$2$}
 \rput(3.5,-.5){$3$}
 \rput(4.5,-.5){$4$}
 \rput(5.5,-.5){$5$}
 \rput(6.5,-.5){$6$}
 \rput(7.5,-.5){$r$}
 \rput(-.5,.5){$0$}
 \rput(-.5,1.5){$1$}
 \rput(-.5,2.5){$2$}
 \rput(-.5,3.5){$3$}
 \rput(-.5,4.5){$4$}
 \rput(-.5,5.5){$5$}
 \rput(-.5,6.5){$6$}
 \rput(-.5,7.5){$7$}
 \rput(-.5,8.5){$8$}
 \rput(-.5,9.5){$9$}
 \rput(-.5,10.5){$10$}
 \rput(-.5,11.25){$s$}}
 
\psline[linewidth=2pt,linecolor=blue](0,0)(3,0)(3,10)(0,10)(0,0)
\end{pspicture}\hspace{.5in}
\begin{pspicture}(-0.5,-.3)(7.5,11.2)
\psframe[linewidth=0pt,fillstyle=solid,fillcolor=lightlightblue](0,0)(7,11)
\multirput(0,0)(0,2){6}{\multirput[bl](0,0)(2,0){4}{\psframe[linewidth=0pt,fillstyle=solid,fillcolor=lightpurple](0,0)(1,1)}}
\multirput[bl](1,1)(0,2){5}{\psframe[linewidth=0pt,fillstyle=solid,fillcolor=darkpurple](0,0)(1,1)}
\multirput[bl](3,1)(0,2){5}{\psframe[linewidth=0pt,fillstyle=solid,fillcolor=darkpurple](0,0)(1,1)}
\multirput[bl](5,1)(0,2){5}{\psframe[linewidth=0pt,fillstyle=solid,fillcolor=darkpurple](0,0)(1,1)}
 \psgrid[gridlabels=0pt,subgriddiv=1](0,0)(7,11)
 
\rput(  .5,.5){\small $\tfrac {117}{2},\!\tfrac {3}{2}$}
\rput(1.5,.5){\small $\tfrac {7}{2},\!\tfrac {13}{2}$}
\rput(2.5,.5){\small $\tfrac {17}{2},\!\tfrac {23}{2}$}
\rput(3.5,.5){\small $\tfrac {27}{2},\!\tfrac {33}{2}$}
\rput(4.5,.5){\small $\tfrac {37}{2},\!\tfrac {43}{2}$}
\rput(5.5,.5){\small $\tfrac {47}{2},\!\tfrac {53}{2}$}
\rput(6.5,.5){\small $\tfrac {57}{2},\!\tfrac {63}{2}$}

\rput(  .5,1.5){\small $\tfrac {111}{2},\!\tfrac {9}{2}$}
\rput(1.5,1.5){\small $\tfrac {1}{2},\!\tfrac {19}{2}$}
\rput(2.5,1.5){\small $\tfrac {11}{2},\!\tfrac {29}{2}$}
\rput(3.5,1.5){\small $\tfrac {21}{2},\!\tfrac {39}{2}$}
\rput(4.5,1.5){\small $\tfrac {31}{2},\!\tfrac {49}{2}$}
\rput(5.5,1.5){\small $\tfrac {41}{2},\!\tfrac {59}{2}$}
\rput(6.5,1.5){\small $\tfrac {51}{2},\!\tfrac {69}{2}$}

\rput(  .5,2.5){\small $\tfrac {105}{2}\!,\!\tfrac {15}{2}$}
\rput(1.5,2.5){\small $\tfrac {115}{2}\!,\!\tfrac {25}{2}$}
\rput(2.5,2.5){\small $\tfrac {5}{2},\!\tfrac {35}{2}$}
\rput(3.5,2.5){\small $\tfrac {15}{2},\!\tfrac {45}{2}$}
\rput(4.5,2.5){\small $\tfrac {25}{2},\!\tfrac {55}{2}$}
\rput(5.5,2.5){\small $\tfrac {35}{2},\!\tfrac {65}{2}$}
\rput(6.5,2.5){\small $\tfrac {45}{2},\!\tfrac {75}{2}$}

\rput(  .5,3.5){\small $\tfrac {99}{2},\!\tfrac {21}{2}$}
\rput(1.5,3.5){\small $\tfrac {109}{2}\!,\!\tfrac {31}{2}$}
\rput(2.5,3.5){\small $\tfrac {119}{2}\!,\!\tfrac {41}{2}$}
\rput(3.5,3.5){\small $\tfrac {9}{2},\!\tfrac {51}{2}$}
\rput(4.5,3.5){\small $\tfrac {19}{2},\!\tfrac {61}{2}$}
\rput(5.5,3.5){\small $\tfrac {29}{2},\!\tfrac {71}{2}$}
\rput(6.5,3.5){\small $\tfrac {39}{2},\!\tfrac {81}{2}$}

\rput(  .5,4.5){\small $\tfrac {93}{2},\!\tfrac {27}{2}$}
\rput(1.5,4.5){\small $\tfrac {103}{2}\!,\!\tfrac {37}{2}$}
\rput(2.5,4.5){\small $\tfrac {113}{2}\!,\!\tfrac {47}{2}$}
\rput(3.5,4.5){\small $\tfrac {3}{2},\!\tfrac {57}{2}$}
\rput(4.5,4.5){\small $\tfrac {13}{2},\!\tfrac {67}{2}$}
\rput(5.5,4.5){\small $\tfrac {23}{2},\!\tfrac {77}{2}$}
\rput(6.5,4.5){\small $\tfrac {33}{2},\!\tfrac {87}{2}$}

\rput(  .5,5.5){\small $\tfrac {87}{2},\!\tfrac {33}{2}$}
\rput(1.5,5.5){\small $\tfrac {97}{2},\!\tfrac {43}{2}$}
\rput(2.5,5.5){\small $\tfrac {107}{2}\!,\!\tfrac {53}{2}$}
\rput(3.5,5.5){\small $\tfrac {117}{2}\!,\!\tfrac {63}{2}$}
\rput(4.5,5.5){\small $\tfrac {7}{2},\!\tfrac {73}{2}$}
\rput(5.5,5.5){\small $\tfrac {17}{2},\!\tfrac {83}{2}$}
\rput(6.5,5.5){\small $\tfrac {27}{2},\!\tfrac {93}{2}$}

\rput(  .5,6.5){\small $\tfrac {81}{2},\!\tfrac {39}{2}$}
\rput(1.5,6.5){\small $\tfrac {91}{2},\!\tfrac {49}{2}$}
\rput(2.5,6.5){\small $\tfrac {101}{2}\!,\!\tfrac {59}{2}$}
\rput(3.5,6.5){\small $\tfrac {111}{2}\!,\!\tfrac {69}{2}$}
\rput(4.5,6.5){\small $\tfrac {1}{2},\!\tfrac {79}{2}$}
\rput(5.5,6.5){\small $\tfrac {11}{2},\!\tfrac {89}{2}$}
\rput(6.5,6.5){\small $\tfrac {21}{2},\!\tfrac {99}{2}$}

\rput(  .5,7.5){\small $\tfrac {75}{2},\!\tfrac {45}{2}$}
\rput(1.5,7.5){\small $\tfrac {85}{2},\!\tfrac {55}{2}$}
\rput(2.5,7.5){\small $\tfrac {95}{2},\!\tfrac {65}{2}$}
\rput(3.5,7.5){\small $\tfrac {105}{2}\!,\!\tfrac {75}{2}$}
\rput(4.5,7.5){\small $\tfrac {115}{2}\!,\!\tfrac {85}{2}$}
\rput(5.5,7.5){\small $\tfrac {5}{2},\!\tfrac {95}{2}$}
\rput(6.5,7.5){\small $\tfrac {15}{2}\!,\!\tfrac {105}{2}$}

\rput(  .5,8.5){\small $\tfrac {69}{2},\!\tfrac {51}{2}$}
\rput(1.5,8.5){\small $\tfrac {79}{2},\!\tfrac {61}{2}$}
\rput(2.5,8.5){\small $\tfrac {89}{2},\!\tfrac {71}{2}$}
\rput(3.5,8.5){\small $\tfrac {99}{2},\!\tfrac {81}{2}$}
\rput(4.5,8.5){\small $\tfrac {109}{2}\!,\!\tfrac {91}{2}$}
\rput(5.5,8.5){\small $\tfrac {119}{2}\!,\!\!\tfrac {101}{2}$}
\rput(6.5,8.5){\small $\tfrac {9}{2},\!\tfrac {111}{2}$}

\rput(  .5,9.5){\small $\tfrac {63}{2},\!\tfrac {57}{2}$}
\rput(1.5,9.5){\small $\tfrac {73}{2},\!\tfrac {67}{2}$}
\rput(2.5,9.5){\small $\tfrac {83}{2},\!\tfrac {77}{2}$}
\rput(3.5,9.5){\small $\tfrac {93}{2},\!\tfrac {87}{2}$}
\rput(4.5,9.5){\small $\tfrac {103}{2}\!,\!\tfrac {97}{2}$}
\rput(5.5,9.5){\small $\tfrac {113}{2}\!,\!\!\tfrac {107}{2}$}
\rput(6.5,9.5){\small $\tfrac {3}{2},\!\tfrac {117}{2}$}

\rput(  .5,10.5){\small $\tfrac {57}{2},\!\tfrac {63}{2}$}
\rput(1.5,10.5){\small $\tfrac {67}{2},\!\tfrac {73}{2}$}
\rput(2.5,10.5){\small $\tfrac {77}{2},\!\tfrac {83}{2}$}
\rput(3.5,10.5){\small $\tfrac {87}{2},\!\tfrac {93}{2}$}
\rput(4.5,10.5){\small $\tfrac {97}{2}\!,\!\tfrac {103}{2}$}
\rput(5.5,10.5){\small $\tfrac {107}{2}\!,\!\!\tfrac {113}{2}$}
\rput(6.5,10.5){\small $\tfrac {117}{2},\!\tfrac {3}{2}$}

{\color{blue}
 \rput(.5,-.5){$0$}
 \rput(1.5,-.5){$1$}
 \rput(2.5,-.5){$2$}
 \rput(3.5,-.5){$3$}
 \rput(4.5,-.5){$4$}
 \rput(5.5,-.5){$5$}
 \rput(6.5,-.5){$6$}
 \rput(7.5,-.5){$r$}
 \rput(-.5,.5){$\tfrac 12$}
 \rput(-.5,1.5){$\tfrac 32$}
 \rput(-.5,2.5){$\tfrac 52$}
 \rput(-.5,3.5){$\tfrac 72$}
 \rput(-.5,4.5){$\tfrac 92$}
 \rput(-.5,5.5){$\tfrac {11}2$}
 \rput(-.5,6.5){$\tfrac {13}2$}
 \rput(-.5,7.5){$\tfrac {15}2$}
 \rput(-.5,8.5){$\tfrac {17}2$}
 \rput(-.5,9.5){$\tfrac {19}2$}
 \rput(-.5,10.5){$\tfrac {21}2$}
 \rput(-.5,11.25){$s\!+\!\tfrac 12$}}
 
\psline[linewidth=2pt,linecolor=blue](0,0)(3,0)(3,10)(0,10)(0,0)
\end{pspicture}
\end{center}
\mbox{}\vspace{-20pt}\mbox{}
\caption{
Kac tables of Bezout conjugates $\{j,\overline j\}\big|_{h=0}$ and $\{j\!+\!\tfrac 12,\overline{j\!+\!\tfrac 12}\}\big|_{h=1}$ for $(p,p')=(3,4)$ in the upper panels and $(p,p')=(3,5)$ in the lower panels. The Bezout conjugators are $\omega_0|_{h=0}=7$ and $\omega_0|_{h=1}=31$ for $(p,p')=(3,4)$, and $\omega_0|_{h=0}=\omega_0|_{h=1}=19$ for $(p,p')=(3,5)$. The periodicity is $P=2n$ in the left panels and $P=4n$ in the right panels. Only Bezout conjugates within the framed box in the lower left contribute to the modular covariant partition functions.}
   \label{pOddBezoutConjugates}
\end{figure}

\begin{figure}[tb] 
\begin{center}
\psset{unit=0.95cm}
\begin{pspicture}(-0.5,-.3)(5.5,11.2)
\psframe[linewidth=0pt,fillstyle=solid,fillcolor=lightlightblue](0,0)(5,11)
\multirput(0,0)(0,2){6}{\multirput[bl](0,0)(2,0){3}{\psframe[linewidth=0pt,fillstyle=solid,fillcolor=lightpurple](0,0)(1,1)}}
\multirput[bl](1,1)(0,2){5}{\psframe[linewidth=0pt,fillstyle=solid,fillcolor=darkpurple](0,0)(1,1)}
\multirput[bl](3,1)(0,2){5}{\psframe[linewidth=0pt,fillstyle=solid,fillcolor=darkpurple](0,0)(1,1)}
\psgrid[gridlabels=0pt,subgriddiv=1](0,0)(5,11)

\rput(.5,.5){\small $0,\!0$}
\rput(1.5,.5){\small $5,\!5$}
\rput(2.5,.5){\small $10,\!10$}
\rput(3.5,.5){\small $15,\!15$}
\rput(4.5,.5){\small $20,\!20$}

\rput(.5,1.5){\small $36,\!4$}
\rput(1.5,1.5){\small $1,\!9$}
\rput(2.5,1.5){\small $6,\!14$}
\rput(3.5,1.5){\small $11,\!19$}
\rput(4.5,1.5){\small $16,\!24$}

\rput(.5,2.5){\small $32,\!8$}
\rput(1.5,2.5){\small $37,\!13$}
\rput(2.5,2.5){\small $2,\!18$}
\rput(3.5,2.5){\small $7,\!23$}
\rput(4.5,2.5){\small $12,\!28$}

\rput(  .5,3.5){\small $28,\!12$}
\rput(1.5,3.5){\small $33,\!17$}
\rput(2.5,3.5){\small $38,\!22$}
\rput(3.5,3.5){\small $3,\!27$}
\rput(4.5,3.5){\small $8,\!32$}

\rput(  .5,4.5){\small $24,\!16$}
\rput(1.5,4.5){\small $29,\!21$}
\rput(2.5,4.5){\small $34,\!26$}
\rput(3.5,4.5){\small $39,\!31$}
\rput(4.5,4.5){\small $4,\!36$}

\rput(  .5,5.5){\small $20,\!20$}
\rput(1.5,5.5){\small $25,\!25$}
\rput(2.5,5.5){\small $30,\!30$}
\rput(3.5,5.5){\small $35,\!35$}
\rput(4.5,5.5){\small $0,\!0$}

\rput(  .5,6.5){\small $16,\!24$}
\rput(1.5,6.5){\small $21,\!29$}
\rput(2.5,6.5){\small $26,\!34$}
\rput(3.5,6.5){\small $31,\!39$}
\rput(4.5,6.5){\small $36,\!4$}

\rput(  .5,7.5){\small $12,\!28$}
\rput(1.5,7.5){\small $17,\!33$}
\rput(2.5,7.5){\small $22,\!38$}
\rput(3.5,7.5){\small $27,\!3$}
\rput(4.5,7.5){\small $32,\!8$}

\rput(  .5,8.5){\small $8,\!32$}
\rput(1.5,8.5){\small $13,\!37$}
\rput(2.5,8.5){\small $18,\!2$}
\rput(3.5,8.5){\small $23,\!7$}
\rput(4.5,8.5){\small $28,\!12$}

\rput(  .5,9.5){\small $4,\!36$}
\rput(1.5,9.5){\small $9,\!1$}
\rput(2.5,9.5){\small $14,\!6$}
\rput(3.5,9.5){\small $19,\!11$}
\rput(4.5,9.5){\small $24,\!16$}

\rput(  .5,10.5){\small $0,\!0$}
\rput(1.5,10.5){\small $5,\!5$}
\rput(2.5,10.5){\small $10,\!10$}
\rput(3.5,10.5){\small $15,\!15$}
\rput(4.5,10.5){\small $20,\!20$}

{\color{blue}
 \rput(.5,-.5){$0$}
 \rput(1.5,-.5){$1$}
 \rput(2.5,-.5){$2$}
 \rput(3.5,-.5){$3$}
 \rput(4.5,-.5){$4$}
 \rput(5.5,-.5){$r$}
 \rput(-.5,.5){$0$}
 \rput(-.5,1.5){$1$}
 \rput(-.5,2.5){$2$}
 \rput(-.5,3.5){$3$}
 \rput(-.5,4.5){$4$}
 \rput(-.5,5.5){$5$}
 \rput(-.5,6.5){$6$}
 \rput(-.5,7.5){$7$}
 \rput(-.5,8.5){$8$}
 \rput(-.5,9.5){$9$}
 \rput(-.5,10.5){$10$}
 \rput(-.5,11.25){$s$}}
 
\psline[linewidth=2pt,linecolor=blue](0,0)(4,0)(4,10)(0,10)(0,0)
\end{pspicture}\hspace{.75in}
\begin{pspicture}(-0.5,-.3)(5.5,11.2)
\psframe[linewidth=0pt,fillstyle=solid,fillcolor=lightlightblue](0,0)(5,11)
\multirput(0,0)(0,2){6}{\multirput[bl](0,0)(2,0){3}{\psframe[linewidth=0pt,fillstyle=solid,fillcolor=lightpurple](0,0)(1,1)}}
\multirput[bl](1,1)(0,2){5}{\psframe[linewidth=0pt,fillstyle=solid,fillcolor=darkpurple](0,0)(1,1)}
\multirput[bl](3,1)(0,2){5}{\psframe[linewidth=0pt,fillstyle=solid,fillcolor=darkpurple](0,0)(1,1)}
\psgrid[gridlabels=0pt,subgriddiv=1](0,0)(5,11)
 
\rput(.5,.5){\small $38,\!2$}
\rput(1.5,.5){\small $3,\!7$}
\rput(2.5,.5){\small $8,\!12$}
\rput(3.5,.5){\small $13,\!17$}
\rput(4.5,.5){\small $18,\!22$}

\rput(.5,1.5){\small $34,\!6$}
\rput(1.5,1.5){\small $39,\!11$}
\rput(2.5,1.5){\small $4,\!16$}
\rput(3.5,1.5){\small $9,\!21$}
\rput(4.5,1.5){\small $14,\!26$}

\rput(.5,2.5){\small $33,\!10$}
\rput(1.5,2.5){\small $35,\!15$}
\rput(2.5,2.5){\small $0,\!20$}
\rput(3.5,2.5){\small $5,\!25$}
\rput(4.5,2.5){\small $10,\!30$}

\rput(  .5,3.5){\small $26,\!14$}
\rput(1.5,3.5){\small $31,\!19$}
\rput(2.5,3.5){\small $36,\!24$}
\rput(3.5,3.5){\small $1,\!29$}
\rput(4.5,3.5){\small $6,\!34$}

\rput(  .5,4.5){\small $22,\!18$}
\rput(1.5,4.5){\small $27,\!23$}
\rput(2.5,4.5){\small $32,\!28$}
\rput(3.5,4.5){\small $37,\!33$}
\rput(4.5,4.5){\small $2,\!38$}

\rput(  .5,5.5){\small $18,\!22$}
\rput(1.5,5.5){\small $23,\!27$}
\rput(2.5,5.5){\small $28,\!32$}
\rput(3.5,5.5){\small $33,\!37$}
\rput(4.5,5.5){\small $38,\!2$}

\rput(  .5,6.5){\small $14,\!26$}
\rput(1.5,6.5){\small $19,\!31$}
\rput(2.5,6.5){\small $24,\!36$}
\rput(3.5,6.5){\small $29,\!1$}
\rput(4.5,6.5){\small $34,\!6$}

\rput(  .5,7.5){\small $10,\!30$}
\rput(1.5,7.5){\small $15,\!35$}
\rput(2.5,7.5){\small $20,\!0$}
\rput(3.5,7.5){\small $25,\!5$}
\rput(4.5,7.5){\small $30,\!10$}

\rput(  .5,8.5){\small $6,\!34$}
\rput(1.5,8.5){\small $11,\!39$}
\rput(2.5,8.5){\small $16,\!4$}
\rput(3.5,8.5){\small $21,\!9$}
\rput(4.5,8.5){\small $26,\!14$}

\rput(  .5,9.5){\small $2,\!38$}
\rput(1.5,9.5){\small $7,\!3$}
\rput(2.5,9.5){\small $12,\!8$}
\rput(3.5,9.5){\small $17,\!13$}
\rput(4.5,9.5){\small $22,\!18$}

\rput(  .5,10.5){\small $38,\!2$}
\rput(1.5,10.5){\small $3,\!7$}
\rput(2.5,10.5){\small $8,\!12$}
\rput(3.5,10.5){\small $13,\!17$}
\rput(4.5,10.5){\small $18,\!22$}

{\color{blue}
 \rput(.5,-.5){$0$}
 \rput(1.5,-.5){$1$}
 \rput(2.5,-.5){$2$}
 \rput(3.5,-.5){$3$}
 \rput(4.5,-.5){$4$}
 \rput(5.5,-.5){$r$}
 \rput(-.5,.5){$\tfrac 12$}
 \rput(-.5,1.5){$\tfrac 32$}
 \rput(-.5,2.5){$\tfrac 52$}
 \rput(-.5,3.5){$\tfrac 72$}
 \rput(-.5,4.5){$\tfrac 92$}
 \rput(-.5,5.5){$\tfrac {11}2$}
 \rput(-.5,6.5){$\tfrac {13}2$}
 \rput(-.5,7.5){$\tfrac {15}2$}
 \rput(-.5,8.5){$\tfrac {17}2$}
 \rput(-.5,9.5){$\tfrac {19}2$}
 \rput(-.5,10.5){$\tfrac {21}2$}
 \rput(-.5,11.25){$s\!+\!\tfrac 12$}}
 
\psline[linewidth=2pt,linecolor=blue](0,0)(4,0)(4,10)(0,10)(0,0)
\end{pspicture}
\end{center}
\mbox{}\vspace{-16pt}\mbox{}
\caption{Kac tables of Bezout conjugates $\{j,\overline j\}$ for $(p,p')=(4,5)$, for $h=0$ in the left panel and $h=1$ in the right panel, with $\omega_0\big|_{h=0}=9$, $\omega_0\big|_{h=1}=29$ and $P=2n$. Only Bezout conjugates within the framed box in the lower left contribute to the modular covariant partition functions.}
\label{pEvenBezoutConjugates}
\end{figure}

Having defined the Bezout conjugate pairs, we can now return to the partition functions computed in \cref{sec:Verma}. Indeed, using \cref{prop:Zrs} and \cref{prop:Bijection}, we readily obtain \cref{prop:Zrs.factorisation}.
\begin{Proposition}
\label{prop:Zrs.factorisation}
We have
\be
{\cal Z}_{r,s}^{\textrm{\tiny$(h,\!v)$}}=\varkappa^n_{j+h'/2}\big((-1)^{pv},q\big)\,\varkappa^n_{\overline{j+ {h'}/2}}\big((-1)^{pv},\qbar\big)\label{hvCoulIdentities}
\ee
where $j+h'/2$ and $\overline{j+ {h'}/2}$ are the Bezout conjugate pairs in \eqref{eq:Bezout.def}.
\end{Proposition}
From \eqref{eq:Zhv.u1char}, we find
\be
\mathcal Z^{\textrm{\tiny$(h,\!v)$}}(p,p')
=\sum_{r=0}^{p-1}\sum_{s=0}^{2p'-1} (-1)^{vr} {\cal Z}_{r,s}^{\textrm{\tiny$(h,\!v)$}}
=\tfrac12 \sum_{r=0}^{p-1}\sum_{s=0}^{4p'-1} (-1)^{vr} {\cal Z}_{r,s}^{\textrm{\tiny$(h,\!v)$}},
\ee
where we used \eqref{eq:Zrs.symmetries} at the last equality. We use the first of these expressions for $pv$ even and the second for $pv$ odd. Using \eqref{hvCoulIdentities} and the bijection \eqref{Bijection}, these partition functions are expressed in terms of the Bezout conjugates as
\begin{alignat}{2}
\mathcal Z^{\textrm{\tiny$(h,\!v)$}}(p,p')
&=\frac{1}{\kappa} \displaystyle\sum_{j=0}^{P-1} (-1)^{v\rhoj}\,\varkappa^n_{j+ h'/2}(q)\varkappa^n_{\overline{j+ {h'}/2}}(\qbar)\\
&=\left\{\begin{array}{ll}
\displaystyle\sum_{j=0}^{2n-1} (-1)^{v\rhoj}\,
\varkappa^n_{j}(q)\varkappa^n_{\overline{j}}(\qbar)
\quad&pv \textrm{ even},\\[16pt]
\displaystyle\tfrac 12\sum_{j=0}^{4n-1} (-1)^{v\rhoj}\, \varkappa^n_{j+ {h}/2}(-1,q)\varkappa^n_{\overline{j+ {h}/2}}(-1,\qbar)\quad&pv \textrm{ odd},
\end{array}\right.
\label{CovariantPFs}
\end{alignat}
where
\be
\label{eq:kappa}
\kappa=\frac{P}{2n}=
\begin{cases}
\,1&pv \textrm{ even,}\\
\,2&pv \textrm{ odd,}
\end{cases}
\qquad 
\rhoj=\tfrac1{2p'}\Big(j+\tfrac{h'}2+\overline{j+\tfrac{h'}2}\,\Big).
\ee
Indeed, using the definition \eqref{eq:Bezout.def} of the the Bezout numbers, it is simple to show that $r\equiv \rhoj$ mod~$p\kappa$. If $v=0$, then $(-1)^{vr}=(-1)^{v\rhoj}=1$ independent of the value of $\rhoj$. Otherwise, we have $v=1$ and $(-1)^r=(-1)^{\rhoj}$ since either $p$ or $\kappa$ must be even.

One might have expected these modular covariant partition functions to be expressible in terms of the set of affine $u(1)$ characters $\{\varkappa^n_j(q): j=0,1,2,\ldots,2n\!-\!1\}$. This is indeed possible for critical bond and site percolation with $(p,p')=(2,3)$~\cite{MDKP17,MDKP23}, and more generally for even $p$. But to include all $(h,v)$ and $(p,p')$, one needs an extended family of characters. Two choices are possible: (i) the functions $\{\varkappa^n_j(\pm 1,q): j=0,\frac12,1, \dots, 2n-\frac12\}$, or (ii) the functions $\{\varkappa^{4n}_j(q): j=0,1,2,\ldots,4n-1\}$, obtained from the first set using \eqref{varkappapm}. It is therefore possible to write all the modular covariant partition functions as sesquilinear forms in either of these families of characters. Some example modular covariant partition functions for small $(p,p')$ values are given explicitly in \cref{sec:Zpp'.examples} as sesquilinear forms in these families of characters.

We note that the partition functions $\mathcal Z^{\textrm{\tiny$(h,0)$}}(p,p')$ with $h\in \{0,1\}$ have only positive coefficients in the $q$-expansions. In constrast, the other two partition functions $\mathcal Z^{\textrm{\tiny$(h,1)$}}(p,p')$ have both positive and negative coefficients. These two sets of partition functions are obtained from each other by applying the modular transformations $\cal S$ and $\cal T$ respectively to $\mathcal Z^{\textrm{\tiny$(1,0)$}}(p,p')$. In the case of $\cal T$ applied to $\mathcal Z^{\textrm{\tiny$(1,0)$}}(p,p')$ to obtain $\mathcal Z^{\textrm{\tiny$(1,1)$}}(p,p')$, the sign factors introduced are simply encoded in the action of $\cal T$ via \eqref{Tsign} giving
\be
(h,v)=(1,1):\qquad \exp\!\Big[2\pi\ir\big(\Delta_{2(j+h'/2)}^{4n}-\Delta_{2(\overline{j+h'/2})}^{4n}\big)\Big]=(-1)^{r},\qquad 0\le r\le p-1,
\ee
in accord with $\cal T$ acting as an involution.

%
\section{Conclusion}\label{sec:conclusion}
%

In this paper, we studied the Yang--Baxter integrable dense $\Aoneone$ and dilute $\Atwotwo$ loop models on the torus with non-contractible and contractible loop fugacities $\alpha$ and $\beta$ respectively. We restricted our attention to the simplest regimes and
applied four combinations of periodic and anti-periodic boundary conditions in the horizontal and vertical directions given by $h,v\in\{0,1\}$. For the root of unity cases, that is $\lambda/\pi\in\mathbb Q$, these loop models reduce to the dense ${\cal LM}(p,p')$ and dilute ${\cal DLM}(p,p')$ logarithmic minimal models with $p,p'$ coprime integers. 

For both dense and dilute models in the general setting, we conjectured explicit forms \eqref{eq:traceConjectures}, in the continuum scaling limit, for the traces of the periodic transfer matrices in the standard modules with an arbitrary number of defects. 
Using Markov traces, we were thus able to deduce the four conformal partition functions $\mathcal Z^{\textrm{\tiny$(h,v)$}}$ and verify that they satisfy modular covariance under the action of the modular group.
Remarkably, despite their very different microscopic description, precisely the same conformal partition functions were obtained for both dense and dilute loop models. The expressions we obtained are in fact related to the Coulomb gas formulation and are written as sesquilinear forms in Verma characters. 
The concurrence of all this conformal data provides compelling evidence for a strong form of universality, as logarithmic CFTs, between these extensive families of dense and dilute loop models which include critical dense and dilute polymers as well as critical bond and site percolation.

At the special value $\alpha=2$, for root of unity cases, the conformal spectra of the dense and dilute loop models precisely coincides with that of the 6-vertex and Izergin--Korepin 19-vertex models respectively where $\omega=\eE^{\ir\gamma}$ and the twist $\gamma$ vanishes. At these points, the conformal partition functions are expressed as sesquilinear forms in the characters of $u(1)$ coset CFTs~\cite{PR2011}. In these special cases, the modular covariant conformal partition functions were shown to be given by Coulomb partition functions~\cite{FSZ87npb,FSZ87} and generalisations thereof, which involve half-integer values of $s$ and sign factors $(-1)^r$, where $(r,s)$ are the Kac labels. 

Perhaps surprisingly, {for $p$ odd,} the standard set of affine $u(1)$ characters $\{\varkappa^n_j(q)\!: j=0,1,2,\ldots,2n-1\}$ with $n=pp'$ that appear in the modular invariant partition functions $\mathcal Z^{\textrm{\tiny$(0,0)$}}$ do not suffice to describe the other three modular covariant partition functions. Instead, to describe all four partition functions, one can use either (i) the functions $\{\varkappa^n_j(\pm 1,q): j=0,\frac12,1, \dots, 2n-\frac12\}$, or (ii) the functions $\{\varkappa^{4n}_j(q): j=0,1,2,\ldots,4n-1\}$. 
These theories are not rational, so there is no reason to expect a fixed finite basic set of irreducible representations and associated characters for all boundary conditions. It is conceivable that the introduction of more complicated seams on the torus would involve other fractional values for $s$ and necessitate further extensions to the basic set of affine $u(1)$ characters.

The results of this paper hinge on our key conjecture \eqref{eq:traceConjectures} for the scaling limits of the transfer matrix traces in the standard modules for the dense and dilute models. So let us conclude by reviewing the evidence, both direct and indirect, supporting this conjecture. 
For general $\beta$ and $\gamma$ but $M$ restricted to even values, the dense case agrees with (2.69) of \cite{PS90} in the context of the XXZ spin chain where we note that the representation of the periodic Temperley--Lieb algebra at a given fixed value of the magnetisation $S^z = \frac d2$ has the same spectra as the standard modules with $d$ defects. 
Similarly, for general $\beta$ and $\gamma$, we have shown that the full partition function \eqref{fullPF} deduced from this conjecture is an infinite sesquilinear Verma form that agrees with the result of \cite{FSZ87}. This follows from the nontrivial number theory results in Appendix~\ref{NumberTheory}. 
Next let us consider cases with $\gamma=0$ and $\lambda/\pi\in\mathbb{Q}$ starting with critical percolation given by $(p,p')=(2,3)$. 
For the dense case, corresponding to critical bond percolation, our conjecture agrees with the analytic results of \cite{MDKP17}. For the dilute case, corresponding to critical site percolation, 
our conjecture agrees with the 162 leading eigenvalues obtained numerically \cite{MDKP23} by solving the logarithmic form of the Bethe ansatz equations. 
Indeed, for $(p,p')=(2,3)$ our conjecture is in agreement, in both dense and dilute cases, with our previous results for the four twisted conformal partition functions.
Since the sesquilinear form in \eqref{eq:traceConjectures} only depends on $\beta = 2 \cos \frac {\pi(p'-p)}{p'}$
and $(p,p')$ through the conformal weights, it is natural to conjecture that the general result is obtained by the simple replacement $\Delta_{r,s}^{2,3}\mapsto\Delta_{r,s}^{p,p'}$.
In principle, individual $(p,p')$ cases could be confirmed numerically by solving the associated logarithmic Bethe ansatz equations as in \cite{MDKP23}.
Finally we observe that, for all $(p,p')$, our conjecture leads to twisted partition functions with the expected modular covariance. 
Moreover, for $\gamma=0$, the deduced modular invariant partition functions $\mathcal Z^{\textrm{\tiny$(0,0)$}}(p,p')$ agree with the conjectured affine $u(1)$ sesquilinear forms of Pearce-Rasmussen~\cite{PR2011} with  $n_{p,p'}=-1$ for all $p,p'$. This appears to rule out the identification of critical percolation with the $c=0$ triplet model ${\cal WLM}(2,3)$~\cite{GabRW2011} for which $n_{2,3}=2$.

\subsection*{Acknowledgments}

AMD was supported by the EOS Research Project, project number 30889451. AK is supported by DFG through the program FOG 2316. AMD and AK acknowledge the hospitality of the SwissMAP Research Station in Les Diablerets during the conference {\it Integrability in Condensed Matter Physics and Quantum Field Theory}.
Part of this work was carried out while PAP was visiting the APCTP as an ICTP Visiting Scholar. The writing of this paper was completed while AK and PAP participated in the MATRIX workshop MPI2024.

\appendix

\section{Partition functions of the $\boldsymbol{O(n)}$ model}
\label{NumberTheory}

Let us recall from \cite{FSZ87} that the conformal partition function for the critical $O(n)$ model on a torus reads
\be
\label{eq:Zge0}
\hat Z(g,e_0) = \frac1{\eta(q)\eta(\bar q)} \bigg(\sum_{P\in \mathbb Z} (q \bar q)^{h_{e_0+2P,0}} + \underset{P \wedge N = 1,\, N | M}{\sum_{M \in \mathbb N^\times}\sum_{P \in \mathbb Z} \sum_{N \in \mathbb N^{\times}}} \Lambda(M,N) q^{h_{2P/N,M/2}} \bar q^{\bar h_{2P/N,M/2}}\bigg)
\ee
where
\be
\label{eq:LambdaMN}
\Lambda(M,N)= \sum_{\gamma_i | \beta_i \le \gamma_i \le \alpha_i} \frac2{p_1^{\gamma_1}p_2^{\gamma_2} \cdots p_k^{\gamma_k}} \sum_{\delta_i = 0}^{\min(\gamma_i,1)} (-1)^{\sum_i \delta_i} \cos(\pi e_0 p_1^{\gamma_1-\delta_1} \cdots p_k^{\gamma_k-\delta_k})
\ee
and
\be
\label{eq:hrs}
h_{r,s} = \frac{(r+g s)^2}{4g}, \qquad 
\bar h_{r,s} = \frac{(r-g s)^2}{4g}.
\ee
Here, the integers $M$ and $N$ in the definition of $\Lambda(M,N)$ are always such that $N$ divides $M$, and they decompose as products of primes $p_i$ as
\be
M = p_1^{\alpha_1} \cdots p_k^{\alpha_k}, \qquad 
N = p_1^{\beta_1} \cdots p_k^{\beta_k},
\ee
for some $\alpha_i$ and $\beta_i$ satisfying $\beta_i \le \alpha_i$ for all $i$. In this appendix, we show that the conformal partition function $\mathcal Z_{\textrm{dilute}}$ for the dilute loop model is identical to the partition function of the $O(n)$ loop model.

\begin{Proposition}
The partition functions of the dilute loop model and the $O(n)$ model satisfy
\be
\label{eq:Zdil=ZOn}
\mathcal Z_{\textrm{dilute}} = \hat Z(\tfrac p{p'},\tfrac{\gamma}\pi) \Big|_{q \leftrightarrow \bar q}.
\ee
\end{Proposition}
\noindent The expressions \eqref{fullPF} and \eqref{eq:Zge0} are already very similar. We note that the exchange $q \leftrightarrow \bar q$ in~\eqref{eq:Zdil=ZOn} is necessary because of the conventions used in defining the conformal weights, whereby $h_{r,s}$ and $\bar h_{r,s}$ are respectively related to $\bar\Delta_{r,s}$ and $\Delta_{r,s}$:
\be
h_{r,s}\big|_{g = \frac p{p'}} = \frac{(rp'+s p)^2}{4 pp'} = \bar\Delta_{r,s}-\frac{c-1}{24}, \qquad 
\bar h_{r,s}\big|_{g = \frac p{p'}} = \frac{(rp'-s p)^2}{4 pp'} = \Delta_{r,s}-\frac{c-1}{24}.
\ee
 The proof of \eqref{eq:Zdil=ZOn} follows from the following two propositions.
\begin{Proposition}
The double sum over $m$ and $\ell$ in \eqref{fullPF} is identical to the double sum over $P$ and~$N$ in \eqref{eq:Zge0}, namely the two sets of $2$-tuples
\begin{subequations}
\begin{alignat}{2}
S_1 &= \bigg\{ \bigg(\frac{m-\ell d}{m\wedge d}, \frac d{m\wedge d}\bigg) \,\Big| \,\ell \in \mathbb Z, m = 1, 2, \dots, d\bigg\},
\\[0.15cm]
S_2 &= \big\{ ( P,N)\,|\,P \in \mathbb Z, N \in \mathbb N^\times, P\wedge N = 1, N | d \big\},
\end{alignat}
\end{subequations}
are identical for all $d \in \mathbb N^\times$. 
\end{Proposition}
\proof
To prove this result, we show that $S_1 \subseteq S_2$ and $S_2 \subseteq S_1$, and also that both sets are free of degeneracies. First, it is easy to see that $S_2$ is free of degeneracies, and that all the elements of $S_1$ appear in
$S_2$, since one can check that all the conditions in $S_2$ are satisfied: 
\be
\frac{m-\ell d}{m\wedge d} \in \mathbb Z, 
\qquad 
\frac d{m\wedge d} \in \mathbb N^\times, 
\qquad 
\frac d{m \wedge d} \Big| d,
\qquad
\frac{m-\ell d}{m\wedge d} \wedge  \frac d{m\wedge d}=
\frac{m}{m\wedge d} \wedge \frac d{m\wedge d} =1.
\ee
Next, to show that $S_2 \subseteq S_1$, we construct a unique pair $(m,\ell)$ for each $(P,N) \in S_2$. Let us note that 
\begin{equation}
\label{eq:mapS1S2}
P=\frac{m-\ell d}{m\wedge d}, \qquad N=\frac d{m\wedge d}.
\end{equation}
defines a map from $S_1$ to $S_2$. We define the inverse map from $S_2$ to $S_1$ using the relation
\be
\label{eq:mapS2S1}
\frac {Pd}N = m - \ell d.
\ee
We view this as defining the integers $m$ and $\ell$ from $P$ and $N$. Indeed, because $N$ divides $d$, it is clear that the left side is an integer. Then $\ell$ is the unique integer such that $m$ is in $\{1,2, \dots, d\}$. In other words, we have
\be
m-1 = \frac{Pd}N-1  \textrm{ mod } d, \qquad \ell = \frac md-\frac PN.
\ee
Finally, we show that the map $S_2 \to S_1$ defined from \eqref{eq:mapS2S1} is the inverse of the map \eqref{eq:mapS1S2}. Indeed, we have
\begin{equation}
\frac d{m \wedge d}
=\frac d{\left[\left(\frac PN+\ell\right) d\right]\wedge d} 
=\frac d{\left(\frac PN d\right)\wedge d}=\frac d{\left(P\frac dN\right)\wedge\left(N\frac dN\right)}= \frac d{(P\wedge N)  \frac dN}=N
\end{equation}
and
\be
\frac{m-\ell d}{m \wedge d} = (m-\ell d) \frac N d = P.
\ee
The two maps are thus inverses of one another. This also confirms that $S_1$ has no degeneracies.\eproof
\begin{Proposition}
For all positive integers $d$ and $m$, the functions $\GA{m}{d}$ and $\Lambda(M,N)$ satisfy
\be
\GA{m}{d} = \tfrac12 \Lambda(d,n)\qquad \textrm{for} \qquad n=\tfrac{d}{m\wedge d}.
\label{eq:GA=Ldn}
\ee
\end{Proposition}
\proof Here and below, we use the Moebius function $\mu(n)$ and the Euler totient function $\phi(n)$ defined, for $n\in\mathbb{N}$, by
\begin{subequations}
\begin{alignat}{2}
\mu(n) &= \left\{
\begin{array}{cl}
0& \text{$n$ has a squared factor}, \\[0.1cm]
+1 & \text{$n$ has an even number of prime factors and no squared factors}, \\[0.1cm]
-1 & \text{$n$ has an odd number of prime factors and no squared factors},
\end{array}
\right. \\
\phi(n) &= \text{[the number of integers coprime to $n$ in the set $\{1,2,\dots, n\}$].}
\end{alignat}
\end{subequations}
From \eqref{ourLambda} and \eqref{eq:LambdaMN} with $\pi e_0=\gamma$, we respectively have
\begin{subequations}
\begin{alignat}{2}
\GA{m}{d} &= 
\frac1{d}\sum_{j=1}^{d} \eE^{2\ir \pi j m /d} \cos\big(({d\wedge j})\gamma\big),\label{ourLambda2}
\\[0.2cm]
\tfrac12 \Lambda(d,n)&=
\sum_{r|\frac dn}\frac1 {nr}\sum_{a|nr}\mu(a)\cos\left(\frac{nr}a\gamma\right)\!,\qquad \mbox{where $n$ divides $d$}.\label{FSZLambda}
\end{alignat}
\end{subequations}
The identification of \eqref{eq:LambdaMN} and \eqref{FSZLambda} with $M=d$ and $N=n$ follows from the decompositions into primes $p_i$ (which we only need for $a$ in the case where $\mu(a)\not=0$)
\begin{subequations}
\begin{alignat}{3}
d&=p_1^{\alpha_1}\cdots p_k^{\alpha_k},\quad 
&&n=p_1^{\beta_1}\cdots p_k^{\beta_k},\qquad
r=p_1^{\gamma_1-\beta_1}\cdots p_k^{\gamma_k-\beta_k},\qquad a=p_1^{\delta_1}\cdots p_k^{\delta_k},\\[0.2cm]
\tfrac dn&=p_1^{\alpha_1-\beta_1}\cdots p_k^{\alpha_k-\beta_k},\quad 
&&\tfrac{nr}a=p_1^{\gamma_1-\delta_1}\cdots p_k^{\gamma_k-\delta_k},
\end{alignat}
\end{subequations}
with
\be
\mu(a)=(-1)^{\sum_i \delta_i},\qquad \beta_i\le\gamma_i\le\alpha_i,\qquad 0\le\delta_i\le\min[\gamma_i,1],\qquad 1\le i\le k,
\ee
so that $n|d$, $r|\tfrac d n$ and $a|n r$.

To prove \eqref{eq:GA=Ldn}, we proceed in four steps. In the first step, we prove
\be
\GA{m}{d} = \GA{m\wedge d}{d}.\label{mostimportant}
\ee
We factorize $m={m'}(m\wedge d)$, with $d={n}(m\wedge d)$ so that $\tfrac md=\tfrac{m'}n$ where $m'$ and $n$ are coprime, and show that
\be
\sum_{j=1}^d  \eE^{2\ir \pi j m'/n} \cos\big(({d\wedge j})\gamma\big)=\sum_{j=1}^d  \eE^{2\ir \pi j/n} \cos\big(({d\wedge j})\gamma\big).\label{A9}
\ee
The result \eqref{A9} follows from the following intermediate result. For $m'$ and $n$ coprime (with $n|d$), the following two sets are equal
\be
\big\{\big(j m'\ \mathrm{mod}\ n,d\wedge j\big)\big| 1\le j\le d\big\} = \big\{(j\ \mathrm{mod}\ n,d\wedge j)\big| 1\le j\le d\big\}.\label{identsets}
\ee
To prove this, we first show the existence of an integer $r$ such that
\be
\left(m'+r n\right)\wedge\frac dn=1.
\label{coprimerelation}
\ee
To establish this, we factorize $d/n=p\cdot q\cdot r$ where $p$ contains all prime factors $p_i^{\mu_i}$ for which $p_i$ is also a factor of $m'$, $q$ contains all prime factors $p_j^{\nu_j}$ for which $p_j$ also appears in $n$, and $r$ is the remaining factor. Note
that $p$, $q$ and $r$ are mutually 
coprime because $m'$ and $n$ are coprime. With this choice of~$r$, we see that
\eqref{coprimerelation} holds because any (prime) factor of $d/n$
is a divisor of one of the summands in $m'+rn$, but not of both:
(i) any divisor of $p$ is also a divisor of $m'$, but it is not a divisor of 
$r$ and not of $n$,
(ii) a divisor of $q$ is also a divisor of $n$, but not of $m'$,
(iii) a divisor of $r$ is not a divisor of $m'$.
We abbreviate
\be
m''=m'+r n,
\ee
which is coprime to $d/n$ from \eqref{coprimerelation} and also coprime to $n$ (as
$m'$ is so),
hence
\be
m''\wedge d=1.
\ee
Next, we replace $m'$ in \eqref{identsets} by $m''$ and prove the stronger identification
\be
\{\big(j m''\ \mbox{mod $d$},d\wedge j\big)\big| 1\le j\le d\}= \{(j\ \mbox{mod $d$},d\wedge j)\big| 1\le j\le d\},\label{identsetsmodd}
\ee
where mod $n$ has been replaced by mod $d$. In the left-hand side, we substitute
\be
d\wedge j=d\wedge (jm''),
\ee
using the fact that $m''$ and $d$ are coprime. But now, for $j=1, 2,\dots,d$, the set of all $jm''$ mod $d$ is identical to the set of all $j$ mod $d$. Hence
\eqref{identsetsmodd} and \eqref{identsets} are proved.

In the second step, we use \eqref{ourLambda2} to obtain another expression for $\GA{m}{d}$. To do this, we derive
certain linear equations for the constants $\GA{m}{d}$, restricting ourselves to $m|d$ using \eqref{mostimportant}. For $n$ a divisor of $d$, we have
\be
\sum_{k=1}^n\GA{k\frac dn}{d}
=\frac1{d}\sum_{j=1}^{d} \Big(\sum_{k=1}^n\eE^{2\ir \pi j k /n}\Big) \cos\big(({d\wedge j})\gamma\big)
\label{sumLambda1}
\ee
as well as
\be
\sum_{k=1}^n\GA{k\frac dn}{d}=\sum_{r|n}\phi(n/r)\GA{r\frac dn}{d}.
\label{sumLambda2}
\ee
The first identity follows readily from \eqref{ourLambda2}. To show the second identity, we use the fact that for any divisor $r$ of $n$ there are $\phi(n/r)$ integers $k$ between
$1$ and $n$ that have $r$ as greatest common divisor with $n$ ($k\wedge n=r \Leftrightarrow
(k/r)\wedge (n/r)=1$).
Moreover,
\be
\Big(k\frac dn\Big) \wedge d=(k\wedge n)\frac dn =r\frac dn=\Big(r\frac dn\Big)\wedge d.
\ee
Combining this with \eqref{mostimportant} completes the proof of \eqref{sumLambda2}.

The sum over exponentials in \eqref{sumLambda1} evaluates to $n$ if $n|j$.
Otherwise it vanishes. This reduces the summation over $j$ to $j=n\cdot
\ell$ with $\ell=1, 2,\dots,d/n$, and yields
\be
\sum_{r|n}\phi(n/r)\GA{\frac rnd}{d}
=\frac n{d}\sum_{\ell=1}^{d/n} \cos\bigg(\Big({\frac dn\wedge
  \ell}\Big)n\gamma\bigg).
\ee
This equation, which holds for every divisor $n$ of $d$, is readily solved using the M\"obius inversion formula to give
\be
\phi(n)\GA{\frac dn}{d}
=\sum_{a|n}\mu(a)\frac n{ad}\sum_{\ell=1}^{ad/n} \cos\bigg(\Big({\frac {ad}n\wedge
  \ell}\Big)\frac na\gamma\bigg),\label{someeq}
\ee
for every divisor $n$ of $d$. Similar to the proof of \eqref{sumLambda2}, we rewrite the second sum in \eqref{someeq} and arrive at
\be
\GA{\frac dn}{d}
=\frac n{\phi(n)d}\sum_{a|n}\frac{\mu(a)}{a}
\sum_{r|\frac{ad}n}\phi\Big(\frac{ad}{nr}\Big)\cos\Big(\frac{nr}a\gamma\Big).
\label{ourformulared}
\ee

In the third step, we obtain new expressions for both $\GA{\frac dn}{d}$ and $\tfrac12\Lambda(d,n)$. 
In the summation \eqref{ourformulared}, $nr/a$ is a divisor of $d$, hence we may write in a
redundant manner
\be
\GA{\frac dn}{d}
=\frac n{\phi(n)d}\sum_{b|d}\sum_{a|n}\frac{\mu(a)}{a}
\sum_{r|\frac{ad}n}\delta_{b,\frac{nr}a}\,\phi\Big(\frac{ad}{nr}\Big)\cos\Big(\frac{nr}a\gamma\Big).
\label{ourformulared1}
\ee
Next, we carry out the summation over $r$ and have to respect that $a$ is a
  divisor of $n$, namely $a=n/m$ with $m|n$. Because $r=ab/n=b/m$, $m$ is also a
divisor of $b$. Hence $m|n\wedge b$ so we have $m=n\wedge b / k$ with $k|n\wedge
b$. We find that
\be
\GA{\frac dn}{d}
=\frac n{\phi(n)d}\sum_{b|d}\sum_{k|n\wedge b}\frac{\mu(\frac{nk}{n\wedge b})}{\frac{nk}{n\wedge b}}
\phi(\tfrac{d}{b})\cos\left(b\gamma\right).
\label{ourformulared2}
\ee
We rewrite \eqref{FSZLambda} in a similar way
\be
\tfrac12\Lambda(d,n)
=\sum_{b|d}\sum_{r|\frac dn}\frac1 {nr}\sum_{a|nr}\mu(a)\delta_{b,\frac{nr}a}\cos\left(\frac{nr}a\gamma\right)\!.
\ee
We sum over $r$ and have to respect that $r$ is a divisor of $d/n$,
  $a=\frac{nr}b=\frac n{n\wedge b}\frac{n\wedge b}b r$ is an integer, and $\frac
  n{n\wedge b}$ and $\frac b{n\wedge b}$ are coprime. Hence $r$
  is a multiple of $\frac b{n\wedge b}$, so we write $r=\frac b{n\wedge b}k$ with
  $k|\frac{n\wedge b}{nb}d$.
We find that
\be
\tfrac12\Lambda(d,n)
=\sum_{b|d}\sum_{k|\frac{n\wedge b}{n b}d}\frac{\mu(\tfrac{n k}{n\wedge b})} {\tfrac{n k b}{n\wedge b}}\cos\left(b\gamma\right).
\label{LambdaFSZ2}
\ee

In the fourth and final step, we see that the right hand sides of \eqref{ourformulared2} and
\eqref{LambdaFSZ2} are identical if for all $b|d$
\be
\frac n{\phi(n)}\sum_{k|n\wedge b}\frac{\mu\left(\frac {nk}{n\wedge
    b}\right)}{\frac {nk}{n\wedge b}}
=
\frac {d/b}{\phi(d/b)}\sum_{k|\frac{n\wedge b}{nb}d}\frac{\mu\left(\frac
  {nk}{n\wedge b}\right)}{\frac {nk}{n\wedge b}}.\label{posited}
\ee
This holds because we have more generally for integers $a$ and $\ell$
\be
\frac {a\ell}{\phi(a\ell)}\sum_{k|\ell}\frac{\mu(ak)}{ak}=\frac{\mu(a)}{\phi(a)},\label{master}
\ee
which is independent of $\ell$. Here, we inserted $a=\frac n{n\wedge b}$ in both sides, $\ell=n\wedge b$ for
the left-hand side of~\eqref{posited} and $\ell=\frac{n\wedge b}{nb}d$ for the right-hand side of \eqref{posited}.
To prove \eqref{master}, we factorize the integer $\ell$ into $s\cdot t$ where all
prime factors of $s$ appear in $a$, and $t$ is coprime to $a$ (and $s$). Then
\be
\sum_{k|\ell}\frac{\mu(ak)}{ak}=\sum_{k|t}\frac{\mu(ak)}{ak}
=\frac{\mu(a)}{a}\sum_{k|t}\frac{\mu(k)}{k}
=\frac{\mu(a)}{a}\frac{\phi\left(t\right)}{t},
\ee
where we used a known identity arising from the M\"obius inversion formula applied to the divisor sum of Euler's totient function.
The left-hand side of \eqref{master} becomes
\be
\frac{\mu(a)}{a}\frac {a\ell}{\phi(a\ell)}\frac{\phi\left(t\right)}t
=
\frac{\mu(a)}{a}\frac {at}{\phi(at)}\frac{\phi\left(t\right)}t\,,
\label{almostthere}
\ee
where we used $\phi(ast)/ast=\phi(at)/at$ (multiply appearing primes drop out).
Finally, we use the multiplicativity of the combination $\phi(at)/at$ for coprime integers $a, t$
to arrive at the right-hand side of~\eqref{master}.\eproof

\section{Integer and half-integer Bezout conjugates}\label{app:half.integer.Bezout}

In this appendix, we discuss integer and half-integer Bezout conjugates $(j+h'/2,\overline{j+h'/2})$ and their properties.
In the case $h'=0$, the affine $u(1)$ indices $j,\jbar$ are integers and the usual Bezout Lemma applies. This lemma and its corollary below are common knowledge, and we refer to \cite{FMS,WikiBezout} for more information.
\begin{Lemma}[Bezout's Lemma]
For all non-zero integers $p, p'$, there exist $r_0, s_0\in\mathbb Z$ such that
\be
p\wedge p'=r_0 p'-s_0 p.
\ee
The solution $(r_0,s_0)$ is not unique since $\big(r_0+\tfrac{kp}{p\wedge p'},s_0+\tfrac{kp'}{p\wedge p'}\big)$ is also a solution for any $k\in\mathbb Z$. Then $p\wedge p'$ is the smallest positive integer of the form $r_0 p'+s_0 p$. The linear combinations $r_0 p'+s_0 p$ with $r_0,s_0 \in \mathbb Z$
are precisely the multiples of $p\wedge p'$. 
\end{Lemma}

\begin{Corollary}
\label{cor:Bezout.sets}
For all non-zero coprime integers $p,p'$, the sets of Bezout conjugates 
\be
\mathbb{B}_{\pm}=\{rp'\pm sp\ \mathrm{ mod }\ 2pp'\,\big|\, 0\le r\le p-1, 0\le s \le 2p'-1\}\label{Bconjugates}
\ee
are each in bijection with the set $\{0,1,2,\ldots,2pp'-1\}$. 
\end{Corollary}

If $1\le p<p'$ and $p,p'$ are coprime so that $p\wedge p'=1$, then 
\be
1=r_0 p'-s_0 p=(r_0+kp) p'-(s_0+kp') p,\qquad k\in\mathbb Z.
\label{repone}
\ee
We can choose $k$ so that $1\le s_0\le p'-1$ since $s_0$ cannot be zero. It follows that $1\le r_0\le p-1$ and $s_0 p<r_0 p'$. With these constraints, the Bezout pair $(r_0,s_0)$ is uniquely determined 
and $r p-s p'\in\mathbb Z$ for any $r,s\in\mathbb Z$. In this case, the conjugator is defined in terms of $(r_0,s_0)$ by $\omega_0=r_0p'+s_0p$ mod $2pp'$ and satisfies 
\be
\omega_0(rp'\pm sp)\equiv rp'\mp sp \textrm{ mod } 2pp',
\ee
which follows from
\begin{alignat}{2}
\omega_0\cdot(rp'\pm sp)-1\cdot(rp'\mp sp)
&=(r_0p'+s_0p)(rp'\pm sp)-(r_0p'-s_0p)(rp'\mp sp)
\nonumber\\[0.1cm]  
&=(r s_0\pm r_0 s)2pp',
\end{alignat}    
where we used the definition of $\omega_0$ and
\eqref{repone}.

For $h'=1$, the affine $u(1)$ indices $j+\frac12$ and $\overline{j+\frac12}$ are half-integers and Corollary \ref{cor:Bezout.sets} needs to be generalised. 
\begin{Proposition}
For $p,p'$ coprime and $\kappa \in\{1, 2\}$, the sets
\be
\mathbb{K}^\pm=\{p'r\pm ps \ \mathrm{ mod }\, P \, \big|\, 0\le r\le p-1,\ 0\le s\le2\kappa p'-1\}
\label{setrs}
\ee
are each in bijection with the set $\{0, 1, \dots, P-1\}$ 
with $P=2\kappa p p'$.
\end{Proposition}
\proof 
We note that the set \eqref{setrs} contains $P$ integer elements. Pairwise, these elements
are not identical. To see this, let us assume that there are pairs
$(r_1,s_1)$ and $(r_2,s_2)$ in \eqref{setrs} with $p'r_1\pm ps_1\equiv p'r_2\pm ps_2 \textrm{ mod } P$,
then there must exist an integer $t$ such that
\be
p'(r_1-r_2)=\mp p(s_1-s_2+2t\kappa p').
\label{eqwithcoprimes}
\ee
Since $p$ has no common prime factor with $p'$, $p$ has to be a divisor of
$r_1-r_2$. But this implies $r_1-r_2=0$ since $r_1$ and $r_2$ are integers
between 0 and $p-1$. From \eqref{eqwithcoprimes}, we also conclude that 
$s_1-s_2+2t\kappa p'=0$. But again under the condition for the ranges of
$s_1$ and $s_2$, this implies $s_1=s_2$.
\eproof

This shows that the map from Kac labels $(r,s)$ to the affine $u(1)$ indices $j+h'/2$ and $\overline{j+h'/2}$ in \cref{prop:Bijection} is bijective, ending the proof of this proposition. This also implies that the map from $j+h'/2$ to $\overline{j+h'/2}$ is a bijection. As this map is also an involution, it is a generalized Bezout conjugation extended to half integers. 

Using these bijections, it follows that there is a unique pair $(r_0,s_0)\in \mathbb{K}$ satisfying \eqref{omegamultpj}. Taking \eqref{Bezoutconjugator} as a definition of $\omega_0$ and multiplying with $j+\tfrac {h'}2 =p'r-p(s+\tfrac h2) \textrm{ mod } P$, we obtain
\begin{equation}
\omega_0\big(j+\tfrac {h'}2\big)
\equiv p' r+p(s+\tfrac h2)+2^{h'}2 p p'
\big(s_0r-r_0s+(r-r_0)\tfrac h2\big) \textrm{ mod } P,
\label{omegamultpj1}
\end{equation}
which we find from
\begin{alignat}{2}
  &\omega_0\cdot(p'r-p(s+\tfrac h2))-1\cdot(p'r+p(s+\tfrac h2))\nonumber\\[0.1cm]
  &  \equiv 2^{h'}\big(p'r_0+p(s_0+\tfrac h2)\big)\big(p'r-p(s+\tfrac h2)\big)
    -  2^{h'}\big(p'r_0-p(s_0+\tfrac h2)\big)\big(p'r+p(s+\tfrac h2)\big) \textrm{ mod } P\nonumber\\[0.1cm]
  &\equiv 2^{h'}2 p p'
\big(rs_0-r_0s+(r-r_0)\tfrac h2\big) \textrm{ mod } P,
\end{alignat}    
where we used the definition \eqref{Bezoutconjugator} of $\omega_0$ and \eqref{omegamultpj} in the second line. One can check case by case that \eqref{omegamultpj1} agrees with \eqref{shiftconj} with the identification $\overline{j+h'/2}=p' r+p(s+\frac h2)\textrm{ mod } P$ and the values~\eqref{eq:muValues} for $\mu$.

\section{Example sesquilinear forms}
\label{sec:Zpp'.examples}
\allowdisplaybreaks

In this appendix, we list some example modular covariant partition functions for various simple values of $(p,p')$. Each term in these sesquilinear forms, symbolically 
$\varkappa_j(q) \varkappa_{\jbar}(\qbar)$ or $\varkappa_{j+1/2}(q) \varkappa_{\overline{j+1/2}}(\qbar)$, corresponds to an entry in the Kac table of Bezout conjugates, such as those shown in \cref{pOddBezoutConjugates,pEvenBezoutConjugates}. For these terms, complex conjugation is equivalent to Bezout conjugation. The resulting sesquilinear forms are simplified by reducing the range of the $u(1)$ indices using the folding relations \eqref{folding2}.

As an example, let us consider the case $(p,p')=(3,4)$, $(h,v)=(1,1)$, for which $n=12$, $pv$ is odd, and the top-right panel in \cref{pOddBezoutConjugates} is relevant. We start by taking the entries row-by-row from the bottom and reducing each entry by applying the folding 
$\varkappa^n_{j}(-1,q)\mapsto \varkappa^n_{4n-j}(-1,q)$ for $j>2n$ to obtain
\begin{subequations}
\begin{alignat}{2}
\mathcal Z^{\textrm{\tiny$(1,1)$}}(3,4)
&={\tfrac12\sum_{j=0}^{4n-1} (-1)^{r}\, \varkappa^{n}_{j+1/2}\big(\!-\!1,q\big)\varkappa^{n}_{\overline{j+1/2}}\big(\!-\!1,\qbar\big) \qquad \textrm{with} \qquad n = pp'=12}
\nonumber\\&
=\tfrac12\Big[|\varkappa_{3/2}^{{12}}(-1,q)|^2-\varkappa_{5/2}^{{12}}(-1,q)\varkappa_{11/2}^{{12}}(-1,\qbar)+\varkappa_{13/2}^{{12}}(-1,q)\varkappa_{19/2}^{{12}}(-1,\qbar)
\nonumber\\&
+|\varkappa_{9/2}^{{12}}(-1,q)|^2-\varkappa_{1/2}^{{12}}(-1,q)\varkappa_{17/2}^{{12}}(-1,\qbar)+\varkappa_{7/2}^{{12}}(-1,q)\varkappa_{25/2}^{{12}}(-1,\qbar)
\nonumber\\&
+|\varkappa_{15/2}^{{12}}(-1,q)|^2-\varkappa_{7/2}^{{12}}(-1,q)\varkappa_{23/2}^{{12}}(-1,\qbar)+\varkappa_{1/2}^{{12}}(-1,q)\varkappa_{31/2}^{{12}}(-1,\qbar)
\nonumber\\&
+|\varkappa_{21/2}^{{12}}(-1,q)|^2-\varkappa_{13/2}^{{12}}(-1,q)\varkappa_{29/2}^{{12}}(-1,\qbar)+\varkappa_{5/2}^{{12}}(-1,q)\varkappa_{37/2}^{{12}}(-1,\qbar)
\nonumber\\&
+|\varkappa_{27/2}^{{12}}(-1,q)|^2-\varkappa_{19/2}^{{12}}(-1,q)\varkappa_{35/2}^{{12}}(-1,\qbar)+\varkappa_{11/2}^{{12}}(-1,q)\varkappa_{43/2}^{{12}}(-1,\qbar)
\nonumber\\&
+|\varkappa_{33/2}^{{12}}(-1,q)|^2-\varkappa_{25/2}^{{12}}(-1,q)\varkappa_{41/2}^{{12}}(-1,\qbar)+\varkappa_{17/2}^{{12}}(-1,q)\varkappa_{47/2}^{{12}}(-1,\qbar)
\nonumber\\&
+|\varkappa_{39/2}^{{12}}(-1,q)|^2-\varkappa_{31/2}^{{12}}(-1,q)\varkappa_{47/2}^{{12}}(-1,\qbar)+\varkappa_{23/2}^{{12}}(-1,q)\varkappa_{41/2}^{{12}}(-1,\qbar)
\nonumber\\&
+|\varkappa_{45/2}^{{12}}(-1,q)|^2-\varkappa_{37/2}^{{12}}(-1,q)\varkappa_{43/2}^{{12}}(-1,\qbar)+\varkappa_{29/2}^{{12}}(-1,q)\varkappa_{35/2}^{{12}}(-1,\qbar)+\mbox{c.c.}\Big]\label{eq:Z3411}\\
&=-2\varkappa_{1/2}^{{12}}(-1,q)\varkappa_{17/2}^{{12}}(-1,\qbar)+2|\varkappa_{3/2}^{12}(-1,q)|^2-2\varkappa_{5/2}^{{12}}(-1,q)\varkappa_{11/2}^{{12}}(-1,\qbar)
\nonumber\\&
-2\varkappa_{7/2}^{{12}}(-1,q)\varkappa_{23/2}^{{12}}(-1,\qbar)+2|\varkappa_{9/2}^{12}(-1,q)|^2-2\varkappa_{11/2}^{{12}}(-1,q)\varkappa_{5/2}^{{12}}(-1,\qbar)
\nonumber\\&
+2\varkappa_{13/2}^{{12}}(-1,q)\varkappa_{19/2}^{{12}}(-1,\qbar)+2|\varkappa_{15/2}^{12}(-1,q)|^2-2\varkappa_{17/2}^{{12}}(-1,q)\varkappa_{1/2}^{{12}}(-1,\qbar)
\nonumber\\&
+2\varkappa_{19/2}^{{12}}(-1,q)\varkappa_{13/2}^{{12}}(-1,\qbar)+2|\varkappa_{21/2}^{12}(-1,q)|^2-2\varkappa_{23/2}^{{12}}(-1,q)\varkappa_{7/2}^{{12}}(-1,\qbar).\nonumber
\end{alignat}
\end{subequations}
In the second step, sign factors are introduced by applying the folding 
{$\varkappa^{n}_{j}(-1,q)= -\varkappa^n_{2n-j}(-1,q)$ for $j>n$.} Similarly, for $\mathcal Z^{\textrm{\tiny$(1,0)$}}(3,4)$, the same steps produce a partition function involving the characters $\varkappa_{j+1/2}^{12}(q)$.
In this case, all coefficients are positive since $(-1)^{vr}=1$ and $\varkappa^n_{j}(q)= \varkappa^n_{2n-j}(q)$ for $j>n$ in the second folding.

\subsection[$(p,p')=(1,2)$ with $c = -2$]{$\boldsymbol{(p,p')=(1,2)}$ with $\boldsymbol{c = -2}$}

This is the case of dense~\cite{PR2007,MDPR2013} and dilute critical polymers. The expression \eqref{CovariantPFs} for the four conformal partition functions as sesquilinear forms in affine $u(1)$ characters gives
\begin{subequations}
\begin{alignat}{2}
\mathcal Z^{\textrm{\tiny$(0,0)$}}(1,2)&=|\varkappa_0^{2}(q)|^2+2|\varkappa_1^2(q)|^{2}+|\varkappa_2^{2}(q)|^2
=|\varkappa_0^{8,+}(q)|^2+2|\varkappa_2^{8,+}(q)|^2+|\varkappa_4^{8,+}(q)|^2,\\[4pt]
\mathcal Z^{\textrm{\tiny$(0,1)$}}(1,2)&=|\varkappa_0^{2}(-1,q)|^2+2|\varkappa_1^{2}(-1,q)|^2+|\varkappa_2^{2}(-1,q)|^2
\nonumber\\&
=|\varkappa_0^{8,-}(q)|^2+2|\varkappa_2^{8,-}(q)|^2+|\varkappa_4^{8,-}(q)|^2,\\[4pt]
\mathcal Z^{\textrm{\tiny$(1,0)$}}(1,2)&=2|\varkappa_{1/2}^{2}(q)|^2+2|\varkappa_{3/2}^{2}(q)|^2
=2|\varkappa_1^{8,+}(q)|^2+2|\varkappa_3^{8,+}(q)|^2,\\[4pt]
\mathcal Z^{\textrm{\tiny$(1,1)$}}(1,2)&=2|\varkappa_{1/2}^{2}(-1,q)|^2+2|\varkappa_{3/2}^{2}(-1,q)|^2
=2|\varkappa_1^{8,-}(q)|^2+2|\varkappa_3^{8,-}(q)|^2.
\end{alignat}
\end{subequations}
These expressions are in agreement\footnote{Equations (C.4a) and (C.4b) in \cite{MDKP17} have typos, as the second argument of each $\varkappa_j^2(q,z)$ function should be $(-1)^M$.}
with Appendix~C of \cite{MDKP17}.

\subsection[$(p,p')=(1,3)$ with $c = -7$]{$\boldsymbol{(p,p')=(1,3)}$ with $\boldsymbol{c = -7}$}

The expression \eqref{CovariantPFs} for the four conformal partition functions as sesquilinear forms in affine $u(1)$ characters gives
\begin{subequations}
\begin{alignat}{2}
\mathcal Z^{\textrm{\tiny$(0,0)$}}(1,3)
&=|\varkappa_0^{3}(q)|^2+2|\varkappa_1^{3}(q)|^2+2|\varkappa_2^{3}(q)|^2+|\varkappa_3^{3}(q)|^2
\nonumber\\&
=|\varkappa_0^{12,+}(q)|^2+2|\varkappa_2^{12,+}(q)|^2+2|\varkappa_4^{12,+}(q))|^2+|\varkappa_6^{12,+}(q)|^2,\\[4pt]
\mathcal Z^{\textrm{\tiny$(0,1)$}}(1,3)
&=|\varkappa_0^{3}(-1,q)|^2+2|\varkappa_1^{3}(-1,q)|^2+2|\varkappa_2^{3}(-1,q)|^2+|\varkappa_3^{3}(-1,q)|^2
\nonumber\\&
=|\varkappa_0^{12,-}(q)|^2+2|\varkappa_2^{12,-}(q)|^2+2|\varkappa_4^{12,-}(q))|^2+|\varkappa_6^{12,-}(q)|^2,\\[4pt]
\mathcal Z^{\textrm{\tiny$(1,0)$}}(1,3)
&=2|\varkappa_{1/2}^{3}(q)|^2+2|\varkappa_{3/2}^{3}(q)|^2+2|\varkappa_{5/2}^{3}(q)|^2
\nonumber\\&
=2|\varkappa_1^{12,+}(q)|^2+2|\varkappa_3^{12,+}(q)^2+2|\varkappa_5^{12,+}(q)|^2,\\[4pt]
\mathcal Z^{\textrm{\tiny$(1,1)$}}(1,3)
&=2|\varkappa_{1/2}^{3}(-1,q)|^2+2|\varkappa_{3/2}^{3}(-1,q)|^2+2|\varkappa_{5/2}^{3}(-1,q)|^2
\nonumber\\&
=2|\varkappa_1^{12,-}(q)|^2+2|\varkappa_3^{12,-}(q)^2+2|\varkappa_5^{12,-}(q)|^2.
\end{alignat}
\end{subequations}

\subsection[$(p,p')=(2,3)$ with $c = 0$]{$\boldsymbol{(p,p')=(2,3)}$ with $\boldsymbol{c = 0}$}

This is the case of critical bond and site percolation~\cite{MDKP17,MDKP23}. The expression \eqref{CovariantPFs} for the four conformal partition functions as sesquilinear forms in affine $u(1)$ characters gives
\begin{subequations}
\begin{alignat}{2}
\mathcal Z^{\textrm{\tiny$(0,0)$}}(2,3)
&=|\varkappa_0^6(q)|^2+2\varkappa_1^6(q)\varkappa_5^6(\qbar)+2|\varkappa_2^6(q)|^2+2|\varkappa_3^6(q)|^2+2|\varkappa_4^6(q)|^2+2\varkappa_5^6(q)\varkappa_1^6(\qbar)+|\varkappa_6^6(q)|^2
\nonumber\\&
=|\varkappa_0^{24,+}(q)|^2+2\varkappa_2^{24,+}(q)\varkappa_{10}^{24,+}(\qbar)+2|\varkappa_4^{24,+}(q)|^2+2|\varkappa_6^{24,+}(q)|^2+2|\varkappa_8^{24,+}(q)|^2
\nonumber\\&
+2\varkappa_{10}^{24,+}(q)\varkappa_2^{24,+}(\qbar)+|\varkappa_{12}^{24,+}(q)|^2,\\[4pt]
\mathcal Z^{\textrm{\tiny$(0,1)$}}(2,3)
&=|\varkappa_0^6(q)|^2-2\varkappa_1^6(q)\varkappa_5^6(\qbar)+2|\varkappa_2^6(q)|^2-2|\varkappa_3^6(q)|^2+2|\varkappa_4^6(q)|^2-2\varkappa_5^6(q)\varkappa_1^6(\qbar)+|\varkappa_6^6(q)|^2
\nonumber\\&
=|\varkappa_0^{24,+}(q)|^2-2\varkappa_2^{24,+}(q)\varkappa_{10}^{24,+}(\qbar)+2|\varkappa_4^{24,+}(q)|^2-2|\varkappa_6^{24,+}(q)|^2+2|\varkappa_8^{24,+}(q)|^2
\nonumber\\&
-2\varkappa_{10}^{24,+}(q)\varkappa_2^{24,+}(\qbar)+|\varkappa_{12}^{24,+}(q)|^2,\\[4pt]
\mathcal Z^{\textrm{\tiny$(1,0)$}}(2,3)
&=\varkappa_0^6(q)\varkappa_6^6(\qbar)\!+\!2|\varkappa_1^6(q)|^2\!+\!2\varkappa_2^6(q)\varkappa_4^6(\qbar)+2|\varkappa_3^6(q)|^2\!+\!2\varkappa_4^6(q)\varkappa_2^6(\qbar)\!+\!2|\varkappa_5^6(q)|^2\!+\!\varkappa_6^6(q)\varkappa_0^6(\qbar)
\nonumber\\&
=\varkappa_0^{24,+}(q)\varkappa_{12}^{24,+}(\qbar)+2|\varkappa_2^{24,+}(q)|^2+2\varkappa_4^{24,+}(q)\varkappa_8^{24,+}(\qbar)+2|\varkappa_6^{24,+}(q)|^2
+2\varkappa_8^{24,+}(q)\varkappa_4^{24,+}(\qbar)
\nonumber\\&
+2|\varkappa_{10}^{24,+}(q)|^2+\varkappa_{12}^{24,+}(q)\varkappa_0^{24,+}(\qbar),\\[4pt]
\mathcal Z^{\textrm{\tiny$(1,1)$}}(2,3)
&=-\varkappa_0^6(q)\varkappa_6^6(\qbar)\!+\!2|\varkappa_1^6(q)|^2\!-\!2\varkappa_2^6(q)\varkappa_4^6(\qbar)\!+\!2|\varkappa_3^6(q)|^2\!-\!2\varkappa_4^6(q)\varkappa_2^6(\qbar)\!+\!2|\varkappa_5^6(q)|^2\!-\!\varkappa_6^6(q)\varkappa_0^6(\qbar)
\nonumber\\&
=-\varkappa_0^{24,+}(q)\varkappa_{12}^{24,+}(\qbar)+2|\varkappa_2^{24,+}(q)|^2-2\varkappa_4^{24,+}(q)\varkappa_8^{24,+}(\qbar)+2|\varkappa_6^{24,+}(q)|^2
-2\varkappa_8^{24,+}(q)\varkappa_4^{24,+}(\qbar)
\nonumber\\&
+2|\varkappa_{10}^{24,+}(q)|^2-\varkappa_{12}^{24,+}(q)\varkappa_0^{24,+}(\qbar).
\end{alignat}
\end{subequations}

\subsection[$(p,p')=(3,4)$ with $c = \frac12$]{$\boldsymbol{(p,p')=(3,4)}$ with $\boldsymbol{c = \frac12}$}

This is the case of the dense and dilute logarithmic Ising model. 
{The expression \eqref{CovariantPFs} for the four conformal partition functions as sesquilinear forms in affine $u(1)$ characters gives}
\begin{subequations}
\begin{alignat}{2}
\mathcal Z^{\textrm{\tiny$(0,0)$}}(3,4)
&=|\varkappa_0^{12}(q)|^2+2\varkappa_1^{12}(q)\varkappa_7^{12}(\qbar)+2\varkappa_2^{12}(q)\varkappa_{10}^{12}(\qbar)+2|\varkappa_3^{12}(q)|^2
+2|\varkappa_4^{12}(q)|^2+2\varkappa_5^{12}(q)\varkappa_{11}^{12}(\qbar)
\nonumber\\&
+2|\varkappa_6^{12}(q)|^2+2\varkappa_7^{12}(q)\varkappa_{1}^{12}(\qbar)+2|\varkappa_8^{12}(q)|^2+2|\varkappa_9^{12}(q)|^2+2\varkappa_{10}^{12}(q)\varkappa_{2}^{12}(\qbar)
+2\varkappa_{11}^{12}(q)\varkappa_{5}^{12}(\qbar)
\nonumber\\&
+|\varkappa_{12}^{12}(q)|^2
\nonumber\\&
=|\varkappa_0^{48,+}(q)|^2+2\varkappa_2^{48,+}(q)\varkappa_{14}^{48,+}(\qbar)+2\varkappa_4^{48,+}(q)\varkappa_{20}^{48,+}(\qbar)+2|\varkappa_6^{48,+}(q)|^2
+2|\varkappa_8^{48,+}(q)|^2
\nonumber\\&
+2\varkappa_{10}^{48,+}(q)\varkappa_{22}^{48,+}(\qbar)
+2|\varkappa_{12}^{48,+}(q)|^2+2\varkappa_{14}^{48,+}(q)\varkappa_{2}^{48,+}(\qbar)+2|\varkappa_{16}^{48,+}(q)|^2+2|\varkappa_{18}^{48,+}(q)|^2(\qbar)
\nonumber\\&
+2\varkappa_{20}^{48,+}(q)\varkappa_{4}^{48,+}(\qbar)+2\varkappa_{22}^{48,+}(q)\varkappa_{10}^{48,+}(\qbar)+|\varkappa_{24}^{48,+}(q)|^2,&\\[4pt]
\mathcal Z^{\textrm{\tiny$(0,1)$}}(3,4)
&=|\varkappa_0^{{12}}(-1,q)|^2\!-\!2\varkappa_1^{{12}}(-1,q)\varkappa_{7}^{{12}}(\qbar)\!-\!2\varkappa_2^{{12}}(-1,q)\varkappa_{10}^{{12}}(-1,\qbar)\!+\!2|\varkappa_3^{{12}}(-1,q)|^2
\!-\!2|\varkappa_4^{{12}}(-1,q)|^2
\nonumber\\&
+2\varkappa_{5}^{{12}}(-1,q)\varkappa_{11}^{{12}}(-1,\qbar)+2|\varkappa_{6}^{{12}}(-1,q)|^2-2\varkappa_{7}^{{12}}(-1,q)\varkappa_{1}^{{12}}(-1,\qbar)+2|\varkappa_{8}^{{12}}(-1,q)|^2
\nonumber\\&
+2|\varkappa_{9}^{{12}}(-1,q)|^2-2\varkappa_{10}^{{12}}(-1,q)\varkappa_{2}^{{12}}(-1,\qbar)+2\varkappa_{11}^{{12}}(-1,q)\varkappa_{5}^{{12}}(-1,\qbar)+|\varkappa_{12}^{{12}}(-1,q)|^2
\nonumber\\&
=|\varkappa_0^{{48,-}}(q)|^2-2\varkappa_2^{{48,-}}(q)\varkappa_{14}^{{48,-}}(\qbar)-2\varkappa_4^{{48,-}}(q)\varkappa_{20}^{{48,-}}(\qbar)+2|\varkappa_6^{{48,-}}(q)|^2
-2|\varkappa_8^{{48,-}}(q)|^2
\nonumber\\&
+2\varkappa_{10}^{{48,-}}(q)\varkappa_{22}^{{48,-}}(\qbar)+2|\varkappa_{12}^{{48,-}}(q)|^2-2\varkappa_{14}^{{48,-}}(q)\varkappa_{2}^{{48,-}}(\qbar)+2|\varkappa_{16}^{{48,-}}(q)|^2
+2|\varkappa_{18}^{{48,-}}(q)|^2
\nonumber\\&
-2\varkappa_{20}^{{48,-}}(q)\varkappa_{4}^{{48,-}}(\qbar)+2\varkappa_{22}^{{48,-}}(q)\varkappa_{10}^{{48,-}}(\qbar)+|\varkappa_{24}^{{48,-}}(q)|^2,&\\[4pt]
\mathcal Z^{\textrm{\tiny$(1,0)$}}(3,4)
&=2\varkappa_{1/2}^{12}(q)\varkappa_{17/2}^{12}(\qbar)+2|\varkappa_{3/2}^{12}(q)|^2+2\varkappa_{5/2}^{12}(q)\varkappa_{11/2}^{12}(\qbar)
+2\varkappa_{7/2}^{12}(q)\varkappa_{23/2}^{12}(\qbar)+2|\varkappa_{9/2}^{12}(q)|^2
\nonumber\\&
+2\varkappa_{11/2}^{12}(q)\varkappa_{5/2}^{12}(\qbar)+2\varkappa_{13/2}^{12}(q)\varkappa_{19/2}^{12}(\qbar)+2|\varkappa_{15/2}^{12}(q)|^2
+2\varkappa_{17/2}^{12}(q)\varkappa_{1/2}^{12}(\qbar)
\nonumber\\&
+2\varkappa_{19/2}^{12}(q)\varkappa_{13/2}^{12}(\qbar)+2|\varkappa_{21/2}^{12}(q)|^2+2\varkappa_{23/2}^{12}(q)\varkappa_{7/2}^{12}(\qbar)
\nonumber\\&
=2\varkappa_1^{{48,+}}(q)\varkappa_{17}^{{48,+}}(\qbar)+2|\varkappa_3^{48,+}(q)|^2+2\varkappa_5^{{48,+}}(q)\varkappa_{11}^{{48,+}}(\qbar)
+2\varkappa_7^{{48,+}}(q)\varkappa_{23}^{{48,+}}(\qbar)+2|\varkappa_9^{48,+}(q)|^2
\nonumber\\&
+2\varkappa_{11}^{{48,+}}(q)\varkappa_{5}^{{48,+}}(\qbar)+2\varkappa_{13}^{{48,+}}(q)\varkappa_{19}^{{48,+}}(\qbar)+2|\varkappa_{15}^{48,+}(q)|^2
+2\varkappa_{17}^{{48,+}}(q)\varkappa_{1}^{{48,+}}(\qbar)
\nonumber\\&
+2\varkappa_{19}^{{48,+}}(q)\varkappa_{13}^{{48,+}}(\qbar)+2|\varkappa_{21}^{48,+}(q)|^2+2\varkappa_{23}^{{48,+}}(q)\varkappa_{7}^{{48,+}}(\qbar),&\\[4pt]
\mathcal Z^{\textrm{\tiny$(1,1)$}}(3,4)
&=-2\varkappa_{1/2}^{{12}}(-1,q)\varkappa_{17/2}^{{12}}(-1,\qbar)+2|\varkappa_{3/2}^{12}(-1,q)|^2-2\varkappa_{5/2}^{{12}}(-1,q)\varkappa_{11/2}^{{12}}(-1,\qbar)\nonumber\\&
-2\varkappa_{7/2}^{{12}}(-1,q)\varkappa_{23/2}^{{12}}(-1,\qbar)
+2|\varkappa_{9/2}^{12}(-1,q)|^2-2\varkappa_{11/2}^{{12}}(-1,q)\varkappa_{5/2}^{{12}}(-1,\qbar)\nonumber\\&
+2\varkappa_{13/2}^{{12}}(-1,q)\varkappa_{19/2}^{{12}}(-1,\qbar)
+2|\varkappa_{15/2}^{12}(-1,q)|^2
-2\varkappa_{17/2}^{{12}}(-1,q)\varkappa_{1/2}^{{12}}(-1,\qbar)
\nonumber\\&
+2\varkappa_{19/2}^{{12}}(-1,q)\varkappa_{13/2}^{{12}}(-1,\qbar)
+2|\varkappa_{21/2}^{12}(-1,q)|^2-2\varkappa_{23/2}^{{12}}(-1,q)\varkappa_{7/2}^{{12}}(-1,\qbar)
\nonumber\\&
=-2\varkappa_1^{{48,-}}(q)\varkappa_{17}^{{48,-}}(\qbar)\!+\!2|\varkappa_3^{48,-}(q)|^2\!-\!2\varkappa_5^{{48,-}}(q)\varkappa_{11}^{{48,-}}(\qbar)
\!-\!2\varkappa_7^{{48,-}}(q)\varkappa_{23}^{{48,-}}(\qbar)\!+\!2|\varkappa_9^{48,-}(q)|^2
\nonumber\\&
-2\varkappa_{11}^{{48,-}}(q)\varkappa_{5}^{{48,-}}(\qbar)+2\varkappa_{13}^{{48,-}}(q)\varkappa_{19}^{{48,-}}(\qbar)+2|\varkappa_{15}^{48,-}(q)|^2
-2\varkappa_{17}^{{48,-}}(q)\varkappa_{1}^{{48,-}}(\qbar)
\nonumber\\&
+2\varkappa_{19}^{{48,-}}(q)\varkappa_{13}^{{48,-}}(\qbar)+2|\varkappa_{21}^{48,-}(q)|^2-2\varkappa_{23}^{{48,-}}(q)\varkappa_{7}^{{48,-}}(\qbar).&
\end{alignat}
\end{subequations}

\subsection[$(p,p')=(3,5)$ with $c = -\frac35$]{$\boldsymbol{(p,p')=(3,5)}$ with $\boldsymbol{c = -\frac35}$}

The expression \eqref{CovariantPFs} for the four conformal partition functions as sesquilinear forms in affine $u(1)$ characters gives
\begin{subequations}
\begin{alignat}{2}
\mathcal Z^{\textrm{\tiny$(0,0)$}}(3,5)
&=|\varkappa_0^{15}(q)|^2+2\varkappa_1^{15}(q)\varkappa_{11}^{15}(\qbar)+2\varkappa_2^{15}(q)\varkappa_{8}^{15}(\qbar)+2|\varkappa_3^{15}(q)|^2
+2\varkappa_4^{15}(q)\varkappa_{14}^{15}(\qbar)
\nonumber\\&
+2|\varkappa_5^{15}(q)|^2+2|\varkappa_6^{15}(q)|^2+2\varkappa_7^{15}(q)\varkappa_{13}^{15}(\qbar)+2\varkappa_8^{15}(q)\varkappa_{2}^{15}(\qbar)+2|\varkappa_9^{15}(q)|^2
\nonumber\\&
+2|\varkappa_{10}^{15}(q)|^2+2\varkappa_{11}^{15}(q)\varkappa_{1}^{15}(\qbar)+2|\varkappa_{12}^{15}(q)|^2+2\varkappa_{13}^{15}(q)\varkappa_{7}^{15}(\qbar)+2\varkappa_{14}^{15}(q)\varkappa_{4}^{15}(\qbar)+|\varkappa_{15}^{15}(q)|^2
\nonumber\\&
=|\varkappa_0^{60,+}(q)|^2+2\varkappa_2^{60,+}(q)\varkappa_{22}^{60,+}(\qbar)+2\varkappa_4^{60,+}(q)\varkappa_{16}^{60,+}(\qbar)+2|\varkappa_6^{60,+}(q)|^2
+2\varkappa_8^{60,+}(q)\varkappa_{28}^{60,+}(\qbar)
\nonumber\\&
+2|\varkappa_{10}^{60,+}(q)|^2+2|\varkappa_{12}^{60,+}(q)|^2+2\varkappa_{14}^{60,+}(q)\varkappa_{26}^{60,+}(\qbar)+2\varkappa_{16}^{60,+}(q)\varkappa_{4}^{60,+}(\qbar)
+2|\varkappa_{18}^{60,+}(q)|^2
\nonumber\\&
+2|\varkappa_{20}^{60,+}(q)|^2+2\varkappa_{22}^{60,+}(q)\varkappa_{2}^{60,+}(\qbar)+2|\varkappa_{24}^{60,+}(q)|^2+2\varkappa_{26}^{60,+}(q)\varkappa_{14}^{60,+}(\qbar)+2\varkappa_{28}^{60,+}(q)\varkappa_{8}^{60,+}(\qbar)
\nonumber\\&
+|\varkappa_{30}^{60,+}(q)|^2,
\\[4pt]
\mathcal Z^{\textrm{\tiny$(0,1)$}}(3,5)
&=|\varkappa_0^{15}(-1,q)|^2-2\varkappa_1^{15}(-1,q)\varkappa_{11}^{15}(-1,\qbar)-2\varkappa_2^{15}(-1,q)\varkappa_{8}^{15}(-1,\qbar)+2|\varkappa_3^{15}(-1,q)|^2
\nonumber\\&
-2\varkappa_4^{15}(-1,q)\varkappa_{14}^{15}(-1,\qbar)-2|\varkappa_5^{15}(-1,q)|^2+2|\varkappa_6^{15}(-1,q)|^2+2\varkappa_7^{15}(-1,q)\varkappa_{13}^{15}(-1,\qbar)
\nonumber\\&
-2\varkappa_8^{15}(-1,q)\varkappa_{2}^{15}(-1,\qbar)+2|\varkappa_9^{15}(-1,q)|^2+2|\varkappa_{10}^{15}(-1,q)|^2-2\varkappa_{11}^{15}(-1,q)\varkappa_{1}^{15}(-1,\qbar)
\nonumber\\&
+2|\varkappa_{12}^{15}(-1,q)|^2+2\varkappa_{13}^{15}(-1,q)\varkappa_{7}^{15}(-1,\qbar)-2\varkappa_{14}^{15}(-1,q)\varkappa_{4}^{15}(-1,\qbar)+|\varkappa_{15}^{15}(-1,q)|^2
\nonumber\\&
=|\varkappa_0^{60,-}(q)|^2-2\varkappa_2^{60,-}(q)\varkappa_{22}^{60,-}(\qbar)-2\varkappa_4^{60,-}(q)\varkappa_{16}^{60,-}(\qbar)+2|\varkappa_6^{60,-}(q)|^2
-2\varkappa_8^{60,-}(q)\varkappa_{28}^{60,-}(\qbar)
\nonumber\\&
-2|\varkappa_{10}^{60,-}(q)|^2+2|\varkappa_{12}^{60,-}(q)|^2+2\varkappa_{14}^{60,-}(q)\varkappa_{26}^{60,-}(\qbar)-2\varkappa_{16}^{60,-}(q)\varkappa_{4}^{60,-}(\qbar)
+2|\varkappa_{18}^{60,-}(q)|^2
\nonumber\\&
+2|\varkappa_{20}^{60,-}(q)|^2
-2\varkappa_{22}^{60,-}(q)\varkappa_{2}^{60,-}(\qbar)+2|\varkappa_{24}^{60,-}(q)|^2+2\varkappa_26^{60,-}(q)\varkappa_{14}^{60,-}(\qbar)
\nonumber\\&
-2\varkappa_{28}^{60,-}(q)\varkappa_{8}^{60,-}(\qbar)
+|\varkappa_{30}^{60,-}(q)|^2,&\\[4pt]
\mathcal Z^{\textrm{\tiny$(1,0)$}}(3,5)
&=2\varkappa_{1/2}^{15}(q)\varkappa_{19/2}^{15}(\qbar)+2|\varkappa_{3/2}^{15}(q)|^2+2\varkappa_{5/2}^{15}(q)\varkappa_{25/2}^{15}(\qbar)
+2\varkappa_{7/2}^{15}(q)\varkappa_{13/2}^{15}(\qbar)+2|\varkappa_{9/2}^{15}(q)|^2
\nonumber\\&
+2\varkappa_{11/2}^{15}(q)\varkappa_{29/2}^{15}(\qbar)\!+\!2\varkappa_{13/2}^{15}(q)\varkappa_{7/2}^{15}(\qbar)\!+\!2|\varkappa_{15/2}^{15}(q)|^2
\!+\!2\varkappa_{17/2}^{15}(q)\varkappa_{23/2}^{15}(\qbar)\!+\!2\varkappa_{19/2}^{15}(q)\varkappa_{1/2}^{15}(\qbar)
\nonumber\\&
+2|\varkappa_{21/2}^{15}(q)|^2+2\varkappa_{23/2}^{15}(q)\varkappa_{17/2}^{15}(\qbar)\nonumber
+2\varkappa_{25/2}^{15}(q)\varkappa_{5/2}^{15}(\qbar)+2|\varkappa_{27/2}^{15}(q)|^2+2\varkappa_{29/2}^{15}(q)\varkappa_{11/2}^{15}(\qbar)&
\\&
=2\varkappa_1^{60,+}(q)\varkappa_{19}^{60,+}(\qbar)+2|\varkappa_3^{60,+}(q)|^2+2\varkappa_5^{60,+}(q)\varkappa_{25}^{60,+}(\qbar)
+2\varkappa_7^{60,+}(q)\varkappa_{13}^{60,+}(\qbar)+2|\varkappa_9^{60,+}(q)|^2
\nonumber\\&
+2\varkappa_{11}^{60,+}(q)\varkappa_{29}^{60,+}(\qbar)+2\varkappa_{13}^{60,+}(q)\varkappa_{7}^{60,+}(\qbar)+2|\varkappa_{15}^{60,+}(q)|^2
+2\varkappa_{17}^{60,+}(q)\varkappa_{23}^{60,+}(\qbar)
\nonumber\\&
+2\varkappa_{19}^{60,+}(q)\varkappa_{1}^{60,+}(\qbar)
+2|\varkappa_{21}^{60,+}(q)|^2+2\varkappa_{23}^{60,+}(q)\varkappa_{17}^{60,+}(\qbar)
+2\varkappa_{25}^{60,+}(q)\varkappa_{5}^{60,+}(\qbar)+2|\varkappa_{27}^{60,+}(q)|^2
\nonumber\\&
+2\varkappa_{29}^{60,+}(q)\varkappa_{11}^{60,+}(\qbar),\hspace{-20pt}&\\[4pt]
\mathcal Z^{\textrm{\tiny$(1,1)$}}(3,5)
&=-2\varkappa_{1/2}^{15}(-1,q)\varkappa_{19/2}^{15}(-1,\qbar)+2|\varkappa_{3/2}^{15}(-1,q)|^2-2\varkappa_{5/2}^{15}(-1,q)\varkappa_{25/2}^{15}(-1,\qbar)
\nonumber\\&
-2\varkappa_{7/2}^{15}(-1,q)\varkappa_{13/2}^{15}(-1,\qbar)+2|\varkappa_{9/2}^{15}(-1,q)|^2+2\varkappa_{11/2}^{15}(-1,q)\varkappa_{29/2}^{15}(-1,\qbar)
\nonumber\\&
-2\varkappa_{13/2}^{15}(-1,q)\varkappa_{7/2}^{15}(-1,\qbar)+2|\varkappa_{15/2}^{15}(-1,q)|^2+2\varkappa_{17/2}^{15}(-1,q)\varkappa_{23/2}^{15}(-1,\qbar)
\nonumber\\&
-2\varkappa_{19/2}^{15}(-1,q)\varkappa_{1/2}^{15}(-1,\qbar)+2|\varkappa_{21/2}^{15}(-1,q)|^2+2\varkappa_{23/2}^{15}(-1,q)\varkappa_{17/2}^{15}(-1,\qbar)
\nonumber\\&
-2\varkappa_{25/2}^{15}(-1,q)\varkappa_{5/2}^{15}(-1,\qbar)+2|\varkappa_{27/2}^{15}(-1,q)|^2+2\varkappa_{29/2}^{15}(-1,q)\varkappa_{11/2}^{15}(-1,\qbar)
\nonumber\\&
=-2\varkappa_1^{60,-}(q)\varkappa_{19}^{60,-}(\qbar)+2|\varkappa_3^{60,-}(q)|^2-2\varkappa_5^{60,-}(q)\varkappa_{25}^{60,-}(\qbar)
-2\varkappa_7^{60,-}(q)\varkappa_{13}^{60,-}(\qbar)
\nonumber\\&
+2|\varkappa_9^{60,-}(q)|^2+2\varkappa_{11}^{60,-}(q)\varkappa_{29}^{60,-}(\qbar)-2\varkappa_{13}^{60,-}(q)\varkappa_{7}^{60,-}(\qbar)+2|\varkappa_{15}^{60,-}(q)|^2
\nonumber\\&
+2\varkappa_{17}^{60,-}(q)\varkappa_{23}^{60,-}(\qbar)
-2\varkappa_{19}^{60,-}(q)\varkappa_{1}^{60,-}(\qbar)
+2|\varkappa_{21}^{60,-}(q)|^2+2\varkappa_{23}^{60,-}(q)\varkappa_{17}^{60,-}(\qbar)
\nonumber\\&
-2\varkappa_{25}^{60,-}(q)\varkappa_{5}^{60,-}(\qbar)
+2|\varkappa_{27}^{60,-}(q)|^2
+2\varkappa_{29}^{60,-}(q)\varkappa_{11}^{60,-}(\qbar).
\end{alignat}
\end{subequations}

\subsection[$(p,p')=(4,5)$ with $c = -\frac7{10}$]{$\boldsymbol{(p,p')=(4,5)}$ with $\boldsymbol{c = -\frac7{10}}$}

This is the case of the dense and dilute logarithmic tricritical Ising model. 
{The expression \eqref{CovariantPFs} for the four conformal partition functions as sesquilinear forms in affine $u(1)$ characters gives}
\begin{subequations}
\begin{alignat}{2}
\mathcal Z^{\textrm{\tiny$(0,0)$}}(4,5)
&=|\varkappa_0^{20}(q)|^2+2\varkappa_1^{20}(q)\varkappa_9^{20}(\qbar)+2\varkappa_2^{20}(q)\varkappa_{18}^{20}(\qbar)+2\varkappa_3^{20}(q)\varkappa_{13}^{20}(\qbar)+2|\varkappa_4^{20}(q)|^2
+2|\varkappa_5^{20}(q)|^2
\nonumber\\&
+2\varkappa_6^{20}(q)\varkappa_{14}^{20}(\qbar)+2\varkappa_7^{20}(q)\varkappa_{17}^{20}(\qbar)+2|\varkappa_8^{20}(q)|^2+2\varkappa_9^{20}(q)\varkappa_1^{20}(\qbar)
+2|\varkappa_{10}^{20}(q)|^2+2\varkappa_{11}^{20}(q)\varkappa_{19}^{20}(\qbar)\nonumber\\
&+2|\varkappa_{12}^{20}(q)|^2+2\varkappa_{13}^{20}(q)\varkappa_{3}^{20}(\qbar)+2\varkappa_{14}^{20}(q)\varkappa_{6}^{20}(\qbar)+2|\varkappa_{15}^{20}(q)|^2+2|\varkappa_{16}^{20}(q)|^2
+2\varkappa_{17}^{20}(q)\varkappa_7^{20}(\qbar)
\nonumber\\&
+2\varkappa_{18}^{20}(q)\varkappa_{2}^{20}(\qbar)+2\varkappa_{19}^{20}(q)\varkappa_{11}^{20}(\qbar)+|\varkappa_{20}^{20}(q)|^2
\nonumber\\&
=|\varkappa_0^{80,+}(q)|^2+2\varkappa_2^{80,+}(q)\varkappa_{18}^{80,+}(\qbar)+2\varkappa_4^{80,+}(q)\varkappa_{36}^{80,+}(\qbar)+2\varkappa_6^{80,+}(q)\varkappa_{26}^{80,+}(\qbar)+2|\varkappa_8^{80,+}(q)|^2
\nonumber&\\
&+2|\varkappa_{10}^{80,+}(q)|^2+2\varkappa_{12}^{80,+}(q)\varkappa_{28}^{80,+}(\qbar)+2\varkappa_{14}^{80,+}(q)\varkappa_{34}^{80,+}(\qbar)+2|\varkappa_{16}^{80,+}(q)|^2+2\varkappa_{18}^{80,+}(q)\varkappa_2^{80,+}(\qbar)
\nonumber\\&
+2|\varkappa_{20}^{80,+}(q)|^2+2\varkappa_{22}^{80,+}(q)\varkappa_{38}^{80,+}(\qbar)+2|\varkappa_{24}^{80,+}(q)|^2+2\varkappa_{26}^{80,+}(q)\varkappa_{6}^{80,+}(\qbar)+2\varkappa_{28}^{80,+}(q)\varkappa_{12}^{80,+}(\qbar)
\nonumber\\&
+2|\varkappa_{30}^{80,+}(q)|^2+2|\varkappa_{32}^{80,+}(q)|^2+2\varkappa_{34}^{80,+}(q)\varkappa_{14}^{80,+}(\qbar)+2\varkappa_{36}^{80,+}(q)\varkappa_{4}^{80,+}(\qbar)+2\varkappa_{38}^{80,+}(q)\varkappa_{22}^{80,+}(\qbar)\nonumber\\
&+|\varkappa_{40}^{80,+}(q)|^2,
\\[4pt]
\mathcal Z^{\textrm{\tiny$(0,1)$}}(4,5)
&=|\varkappa_0^{20}(q)|^2-2\varkappa_1^{20}(q)\varkappa_9^{20}(\qbar)+2\varkappa_2^{20}(q)\varkappa_{18}^{20}(\qbar)-2\varkappa_3^{20}(q)\varkappa_{13}^{20}(\qbar)+2|\varkappa_4^{20}(q)|^2
-2|\varkappa_5^{20}(q)|^2
\nonumber\\&
+2\varkappa_6^{20}(q)\varkappa_{14}^{20}(\qbar)\!-\!2\varkappa_7^{20}(q)\varkappa_{17}^{20}(\qbar)\!+\!2|\varkappa_8^{20}(q)|^2\!-\!2\varkappa_9^{20}(q)\varkappa_1^{20}(\qbar)
\!+\!2|\varkappa_{10}^{20}(q)|^2\!-\!2\varkappa_{11}^{20}(q)\varkappa_{19}^{20}(\qbar)
\nonumber\\&
+2|\varkappa_{12}^{20}(q)|^2-2\varkappa_{13}^{20}(q)\varkappa_{3}^{20}(\qbar)+2\varkappa_{14}^{20}(q)\varkappa_{6}^{20}(\qbar)-2|\varkappa_{15}^{20}(q)|^2+2|\varkappa_{16}^{20}(q)|^2
-2\varkappa_{17}^{20}(q)\varkappa_7^{20}(\qbar)
\nonumber\\&
+2\varkappa_{18}^{20}(q)\varkappa_{2}^{20}(\qbar)-2\varkappa_{19}^{20}(q)\varkappa_{11}^{20}(\qbar)+|\varkappa_{20}^{20}(q)|^2
\nonumber\\&
=|\varkappa_0^{80,+}(q)|^2-2\varkappa_2^{80,+}(q)\varkappa_{18}^{80,+}(\qbar)+2\varkappa_4^{80,+}(q)\varkappa_{36}^{80,+}(\qbar)-2\varkappa_6^{80,+}(q)\varkappa_{26}^{80,+}(\qbar)+2|\varkappa_8^{80,+}(q)|^2
\nonumber\\&
-2|\varkappa_{10}^{80,+}(q)|^2+2\varkappa_{12}^{80,+}(q)\varkappa_{28}^{80,+}(\qbar)-2\varkappa_{14}^{80,+}(q)\varkappa_{34}^{80,+}(\qbar)+2|\varkappa_{16}^{80,+}(q)|^2-2\varkappa_{18}^{80,+}(q)\varkappa_2^{80,+}(\qbar)
\nonumber\\&
+2|\varkappa_{20}^{80,+}(q)|^2-2\varkappa_{22}^{80,+}(q)\varkappa_{38}^{80,+}(\qbar)+2|\varkappa_{24}^{80,+}(q)|^2-2\varkappa_{26}^{80,+}(q)\varkappa_{6}^{80,+}(\qbar)+2\varkappa_{28}^{80,+}(q)\varkappa_{12}^{80,+}(\qbar)
\nonumber\\&
-2|\varkappa_{30}^{80,+}(q)|^2+2|\varkappa_{32}^{80,+}(q)|^2-2\varkappa_{34}^{80,+}(q)\varkappa_{14}^{80,+}(\qbar)+2\varkappa_{36}^{80,+}(q)\varkappa_{4}^{80,+}(\qbar)-2\varkappa_{38}^{80,+}(q)\varkappa_{22}^{80,+}(\qbar)
\nonumber\\&
+|\varkappa_{40}^{80,+}(q)|^2,&\\[4pt]
\mathcal Z^{\textrm{\tiny$(1,0)$}}(4,5)
&=\varkappa_0^{20}(q)\varkappa_{20}^{20}(\qbar)\!+\!2\varkappa_1^{20}(q)\varkappa_{11}^{20}(\qbar)\!+\!2|\varkappa_2^{20}(q)|^2\!+\!2\varkappa_3^{20}(q)\varkappa_7^{20}(\qbar)
\!+\!2\varkappa_4^{20}(q)\varkappa_{16}^{20}(\qbar)\!+\!2\varkappa_5^{20}(q)\varkappa_{15}^{20}(\qbar)
\nonumber\\&
+2|\varkappa_6^{20}(q)|^2+2\varkappa_7^{20}(q)\varkappa_{3}^{20}(\qbar)+2\varkappa_8^{20}(q)\varkappa_{12}^{20}(\qbar)+2\varkappa_9^{20}(q)\varkappa_{19}^{20}(\qbar)+2|\varkappa_{10}^{20}(q)|^2
+2\varkappa_{11}^{20}(q)\varkappa_{1}^{20}(\qbar)
\nonumber\\&
+2\varkappa_{12}^{20}(q)\varkappa_{8}^{20}(\qbar)\!+\!2\varkappa_{13}^{20}(q)\varkappa_{17}^{20}(\qbar)\!+\!2|\varkappa_{14}^{20}(q)|^2\!+\!2\varkappa_{15}^{20}(q)\varkappa_{5}^{20}(\qbar)
\!+\!2\varkappa_{16}^{20}(q)\varkappa_{4}^{20}(\qbar)\!+\!2\varkappa_{17}^{20}(q)\varkappa_{13}^{20}(\qbar)
\nonumber\\&
+2|\varkappa_{18}^{20}(q)|^2+2\varkappa_{19}^{20}(q)\varkappa_{9}^{20}(\qbar)+\varkappa_{20}^{20}(q)\varkappa_{0}^{20}(\qbar)
\nonumber\\&
=\varkappa_0^{80,+}(q)\varkappa_{40}^{80,+}(\qbar)\!+\!2\varkappa_2^{80,+}(q)\varkappa_{22}^{80,+}(\qbar)\!+\!2|\varkappa_4^{80,+}(q)|^2\!+\!2\varkappa_6^{80,+}(q)\varkappa_{14}^{80,+}(\qbar)\!+\!2\varkappa_8^{80,+}(q)\varkappa_{32}^{80,+}(\qbar)
\nonumber\\&
+2\varkappa_{10}^{80,+}(q)\varkappa_{30}^{80,+}(\qbar)\!+\!2|\varkappa_{12}^{80,+}(q)|^2\!+\!2\varkappa_{14}^{80,+}(q)\varkappa_{6}^{80,+}(\qbar)\!+\!2\varkappa_{16}^{80,+}(q)\varkappa_{24}^{80,+}(\qbar)\!+\!2\varkappa_{18}^{80,+}(q)\varkappa_{38}^{80,+}(\qbar)
\nonumber\\&
+2|\varkappa_{20}^{80,+}(q)|^2+2\varkappa_{22}^{80,+}(q)\varkappa_{2}^{80,+}(\qbar)+2\varkappa_{24}^{80,+}(q)\varkappa_{16}^{80,+}(\qbar)+2\varkappa_{26}^{80,+}(q)\varkappa_{34}^{80,+}(\qbar)+2|\varkappa_{28}^{80,+}(q)|^2
\nonumber\\&
+2\varkappa_{30}^{80,+}(q)\varkappa_{10}^{80,+}(\qbar)+2\varkappa_{32}^{80,+}(q)\varkappa_{8}^{80,+}(\qbar)+2\varkappa_{34}^{80,+}(q)\varkappa_{26}^{80,+}(\qbar)+2|\varkappa_{36}^{80,+}(q)|^2
\nonumber\\&
+2\varkappa_{38}^{80,+}(q)\varkappa_{18}^{80,+}(\qbar)+\varkappa_{40}^{80,+}(q)\varkappa_{0}^{80,+}(\qbar),\\
\mathcal Z^{\textrm{\tiny$(1,1)$}}(4,5)
&=\varkappa_0^{20}(q)\varkappa_{20}^{20}(\qbar)\!-\!2\varkappa_1^{20}(q)\varkappa_{11}^{20}(\qbar)\!+\!2|\varkappa_2^{20}(q)|^2-2\varkappa_3^{20}(q)\varkappa_7^{20}(\qbar)
\!+\!2\varkappa_4^{20}(q)\varkappa_{16}^{20}(\qbar)\!-\!2\varkappa_5^{20}(q)\varkappa_{15}^{20}(\qbar)
\nonumber\\&
+2|\varkappa_6^{20}(q)|^2-2\varkappa_7^{20}(q)\varkappa_{3}^{20}(\qbar)+2\varkappa_8^{20}(q)\varkappa_{12}^{20}(\qbar)-2\varkappa_9^{20}(q)\varkappa_{19}^{20}(\qbar)+2|\varkappa_{10}^{20}(q)|^2
-2\varkappa_{11}^{20}(q)\varkappa_{1}^{20}(\qbar)
\nonumber\\&
+2\varkappa_{12}^{20}(q)\varkappa_{8}^{20}(\qbar)\!-\!2\varkappa_{13}^{20}(q)\varkappa_{17}^{20}(\qbar)\!+\!2|\varkappa_{14}^{20}(q)|^2\!-\!2\varkappa_{15}^{20}(q)\varkappa_{5}^{20}(\qbar)
\!+\!2\varkappa_{16}^{20}(q)\varkappa_{4}^{20}(\qbar)\!-\!2\varkappa_{17}^{20}(q)\varkappa_{13}^{20}(\qbar)
\nonumber\\&
+2|\varkappa_{18}^{20}(q)|^2-2\varkappa_{19}^{20}(q)\varkappa_{9}^{20}(\qbar)+\varkappa_{20}^{20}(q)\varkappa_{0}^{20}(\qbar)
\nonumber\\&
=\varkappa_0^{80,+}(q)\varkappa_{40}^{80,+}(\qbar)\!-\!2\varkappa_2^{80,+}(q)\varkappa_{22}^{80,+}(\qbar)\!+\!2|\varkappa_4^{80,+}(q)|^2\!-\!2\varkappa_6^{80,+}(q)\varkappa_{14}^{80,+}(\qbar)\!+\!2\varkappa_8^{80,+}(q)\varkappa_{32}^{80,+}(\qbar)
\nonumber\\&
-2\varkappa_{10}^{80,+}(q)\varkappa_{30}^{80,+}(\qbar)\!+\!2|\varkappa_{12}^{80,+}(q)|^2\!-\!2\varkappa_{14}^{80,+}(q)\varkappa_{6}^{80,+}(\qbar)\!+\!2\varkappa_{16}^{80,+}(q)\varkappa_{24}^{80,+}(\qbar)\!-\!2\varkappa_{18}^{80,+}(q)\varkappa_{38}^{80,+}(\qbar)
\nonumber\\&
+2|\varkappa_{20}^{80,+}(q)|^2-2\varkappa_{22}^{80,+}(q)\varkappa_{2}^{80,+}(\qbar)+2\varkappa_{24}^{80,+}(q)\varkappa_{16}^{80,+}(\qbar)-2\varkappa_{26}^{80,+}(q)\varkappa_{34}^{80,+}(\qbar)
+2|\varkappa_{28}^{80,+}(q)|^2
\nonumber\\&
-2\varkappa_{30}^{80,+}(q)\varkappa_{10}^{80,+}(\qbar)+2\varkappa_{32}^{80,+}(q)\varkappa_{8}^{80,+}(\qbar)-2\varkappa_{34}^{80,+}(q)\varkappa_{26}^{80,+}(\qbar)+2|\varkappa_{36}^{80,+}(q)|^2
\nonumber\\&
-2\varkappa_{38}^{80,+}(q)\varkappa_{18}^{80,+}(\qbar)+\varkappa_{40}^{80,+}(q)\varkappa_{0}^{80,+}(\qbar).
\end{alignat}
\end{subequations}

%

\end{document}